\documentclass[a4paper,12pt]{article}

\usepackage{jheppub} 
   \usepackage[T1]{fontenc}                  
\usepackage{hyperref}
\hypersetup{
    bookmarks=true,         
    unicode=false,          
    pdftoolbar=true,        
    pdfmenubar=true,        
    pdffitwindow=false,     
    pdfstartview={FitH},    
    pdftitle={My title},    
    pdfauthor={Author},     
    pdfsubject={Subject},   
    pdfcreator={Creator},   
    pdfproducer={Producer}, 
    pdfkeywords={keyword1} {key2} {key3}, 
    pdfnewwindow=true,      
    colorlinks=true,       
    linkcolor=red,          
    citecolor=violet,        
    filecolor=magenta,      
    urlcolor=blue,           
    linktocpage=true
}
\usepackage{graphics}
\usepackage{rotating}
\usepackage{subfigure}
\usepackage{pstricks}
\usepackage{color}
\usepackage{amsfonts}
\usepackage{mathrsfs}
\usepackage{epsfig}
\usepackage{amsmath,amssymb,amsthm,graphicx,latexsym}
\usepackage{ulem}

\newcommand{\be}{\begin{equation}}
\newcommand{\ee}{\end{equation}}
\newcommand{\bea}{\begin{eqnarray}}
\newcommand{\eea}{\end{eqnarray}}
\newcommand{\bml}{\begin{subequations}}
\newcommand{\eml}{\end{subequations}}
\newcommand{\bfig}{\begin{figure}}
\newcommand{\efig}{\end{figure}}

\newcommand{\slashed}{\hspace{-1.1ex}/}

\begin{document}
$~~~~~~~~~~~~~~~~~~~~~~~~~~~~~~~~~~~~~~~~~~~~~~~~~~~~~~~~~~~~~~~~~~~~~~~~~~~~~~~~~~~~$\textcolor{red}{\bf TIFR/TH/15-22}
\title{\textsc{\fontsize{32.5}{90}\selectfont \sffamily \bfseries Reconstructing inflationary paradigm
within Effective Field Theory framework}}

\author[a]{Sayantan Choudhury
\footnote{\textcolor{purple}{\bf Presently working as a Visiting (Post-Doctoral) fellow at DTP, TIFR, Mumbai, \\$~~~~~$Alternative
 E-mail: sayanphysicsisi@gmail.com}. ${}^{}$}}

\affiliation[a]{Department of Theoretical Physics, Tata Institute of Fundamental Research, Colaba, Mumbai - 400005, India
}

\emailAdd{sayantan@theory.tifr.res.in}

\abstract{In this paper my prime objective is to analyze the constraints on a sub-Planckian excursion of a single inflaton field within Effective Field Theory framework
in a model independent fashion. For a generic single field inflationary potential, using the various parameterization of the primordial power spectrum 
I have derived the most general expression for the field excursion in terms of various inflationary observables, applying the observational constraints obtained from recent Planck 2015 and Planck 2015 +BICEP2/Keck Array data. 
By explicit computation I have reconstructed the structural form of the 
inflationary potential by constraining the Taylor expansion co-efficients appearing in the generic expansion of the potential within the Effective
Field Theory. Next I have explicitly derived, a set of higher order inflationary {\it consistency} relationships,
which would help us to break the degeneracy between various class of inflationary models by differentiating them. 
I also provided two simple examples of Effective Theory of inflation- {\it inflection-point} model and {\it saddle-point} model to check the compatibility of the prescribed methodology
in the light of Planck 2015 and Planck 2015 +BICEP2/Keck Array data. Finally, I have also checked the validity of the prescription by estimating the cosmological parameters and 
fitting the theoretical CMB TT, TE and EE angular power spectra with the observed data within the multipole range $2<l<2500$.
}

\maketitle
\flushbottom

\section{Introduction}

The primordial inflation~\cite{Guth:1980zm,Linde,Albrecht} has two {\it key} predictions - creating the scalar density perturbations and the tensor perturbations during the accelerated phase of expansion~\cite{Mukhanov:1981xt}, for a review, see~\cite{Mukhanov:1990me}. One of the predictions, namely the temperature anisotropy due to the 
scalar density fluctuations has now been tested very accurately by the observations from the temperature anisotropy and polarization 
in the cosmic microwave background (CMB) radiation~\cite{Hinshaw:2012aka,Planck-1,Planck-infl}. 
In the last year the detection of tensor modes has been initially confirmed by the ground based BICEP2 
experiment~\cite{Ade:2014xna}, which according to the initial claim has detected
for the first time a non-zero value of the tensor-to-scalar ratio at $7\sigma$ C.L. The value obtained by the BICEP2 team in conjunction with
\textcolor{red}{ Planck (2014)+WMAP-9+high~L+BICEP2 (dust)}~\footnote{Throughout the article we use \textcolor{red}{red} for the obsoleted \textcolor{red}{ Planck (2014)+WMAP-9+high~L+BICEP2 (dust)}
data.} to put a bound on the primordial gravitational waves, via tensor-to-scalar ratio, within a window: 
\be 0.15 \leq r(k_\star)\equiv P_T(k_\star)/P_S(k_\star)\leq  0.27,\ee at the pivot scale, $k_\star = 0.002 {\rm Mpc^{-1}}$~\cite{Ade:2014xna},
where
$P_T$ and $P_S$ denote the power spectrum for the tensor and scalar modes, respectively. But just after releasing this result
BICEP2 analysis was put
into question by several works \cite{Liu:2014mpa,Mortonson:2014bja,Flauger:2014qra,Adam:2014gaa} on its correctness in the physical ground.
Most importantly it accounting for the
contribution of foreground dust which will shift the value of tensor-to-scalar ratio $r$ downward by an amount and further better
constrained by the joint analysis performed by Planck and BICEP2/Keck Array team \cite{Ade:2015tva}. The final result 
is expressed as a likelihood curve for $r$, and yields an upper limit: \be r(k_\star)\equiv P_T(k_\star)/P_S(k_\star)\leq 0.12 \ee at the pivot scale, $k_\star = 0.05 {\rm Mpc^{-1}}$
with $2\sigma$ confidence. Marginalizing over dust contribution and $r$, finally it is reported that the lensing B-modes
are detected at $7\sigma$ significance. Very recently in ref.~\cite{Ade:2015lrj} the Planck 
team in 2015 data release also fixed the upper bound on the tensor-to-scalar ratio as:
\be r(k_\star)\equiv P_T(k_\star)/P_S(k_\star)\leq 0.11 \ee at the pivot scale, $k_\star = 0.002 {\rm Mpc^{-1}}$ with $2\sigma$ C.L. and perfectly consistent with the joint analysis performed by Planck and BICEP2/Keck Array team.

 Note that large $r(k_\star)$ is possible if the 
initial conditions for gravitational waves is quantum Bunch-Davis vacuum~\cite{BD}~\footnote{ Apart from the correctness of this argument, it is additionally importnat to 
mention here that, still it doesn't require the initial conditions for the quantum vacuum to be Bunch-Davis strictly.}, for a classical initial condition the amplitude of the gravitational 
waves would be very tiny and undetectable, therefore this can be treated as the first observable proof of quantum gravity. 
However, apart from the importance and applicability of quantum Bunch-Davis vacuum on its theoretical and observational ground it is still not at all clear from the previous works in this area that whether the quantum Bunch-Davies vacuum is the only source of
generating large value of $r(k_{\star})$ during inflationary epoch or not. One of the prime possibilities comes from the  
deviation from quantum Bunch-Davies vacuum {\it aka} consideration of quantum non-Bunch-Davies
or arbitrary vacuum in this picture which may also responsible for the generation of large $r(k_{\star})$ during inflation~\footnote{In this article I have not explored the possibility of non Bunch Davies vacuum.} .

In this paper, our aim will be to illustrate that it is possible to explain the current data sets 
within a sub-Planckian model of inflation, where:
\begin{itemize}

\item{$\phi_0\ll M_p$ - vev of the inflaton must be bounded by the cut-off of the particle theory,  where $M_p=2.4\times 10^{18}$~GeV. We are assuming that $4$ dimensions $M_p$ puts a natural cut-off here for any physics beyond the Standard Model.}

\item{ $|\Delta \phi |\approx |\phi_\star-\phi_e| \lesssim M_p$ - the inflaton potential has to be flat enough  during which a successful
inflation can occur. Here \be \phi_\star \geq \phi_0\geq \phi_e\ee represents the field VEV, and $\Delta\phi$ denotes the range of the field values around which all 
the relevant inflation occurs, $\phi_\star$ corresponds to the pivot scale and $\phi_e$ denotes the end of inflation. Note that the
flatness of the potential has to be fine tuned and also there is no particle physics symmetry which can maintain the
flatness. We will assume \be V''(\phi_0)\approx 0,\ee where $V(\phi)$ denotes the inflaton potential,
and prime denotes derivative w.r.t. the $\phi$ field. Naturally, the potential has to be flat enough within $\Delta \phi$ to support slow roll inflation. }

\end{itemize}

The above requirements are important if the origin of the inflaton has to be embedded within a particle theory, where inflaton is part of
a {\it visible sector} gauge group, i.e. Standard Model gauge group, instead of an arbitrary gauge singlet. 
If the inflaton is {\it gauged} under some gauge group,
as in the case of a minimal supersymmetric Standard Model (MSSM), Ref.~\cite{Allahverdi:2006iq}, then the inflaton VEV must be 
bounded by $M_p$, in order to keep the sanctity of an effective field theory description~\footnote{ An arbitrary moduli or a gauge singlet inflaton can 
take large VEVs ( super-Planckian ) as in the case of a
chaotic inflation~\cite{Linde}. Although, in the case of {\it assisted inflation}~\cite{assisted}, see {\it chaotic assisted inflation}~\cite{kanti},  the individual VEVs of the inflatons are sub-Planckian.}.

Our prescribed methodology will be very generic, with a Taylor expanded potential around VEV $\phi_{0}$ is given by~\footnote{For the sake of completeness I suggest the readers to see ref.~\cite{Choudhury:2014sua}, where
I have explicitly studied the inflationary reconstruction technique within the framework of Randall Sundrum single brane set up, 
using the observational constraints obtained from 
Planck 2015 and BICPE2/Keck Array joint constraints.}:
\begin{eqnarray}\label{rt10a}
V(\phi)&=&\sum^{\infty}_{n=0}\frac{(\phi-\phi_{0})^{n}}{n!}\left(\frac{d^{n}V(\phi)}{d\phi^{n}}\right)_{\phi=\phi_{0}},\nonumber\\
&=&V(\phi_0)+V^{\prime}(\phi_0)(\phi-\phi_{0})+\frac{V^{\prime\prime}(\phi_0)}{2}(\phi-\phi_{0})^{2}+\frac{V^{\prime\prime\prime}(\phi_0)}{6}(\phi-\phi_{0})^{3}\,\nonumber\\
&&~~~~~~~~~~~~~~~~~~~~~~~~~~~~~~~~~~~~~~~~~~~~~~~~+\frac{V^{\prime\prime\prime\prime}(\phi_0)}{24}(\phi-\phi_{0})^{4}+\cdots\,,
\end{eqnarray}
where the expansion co-efficients are characterized as follows:
\begin{itemize}
 \item The first term: \be V(\phi_0)\ll M_p^4\ee denotes the height of the potential. Also this term will play the most significant
 role in fixing the scale of inflation within the present framework.
 \item The Taylor expansion coefficients of the effective potential: 
 \bea V^{\prime}(\phi_0) &\leq& M_p^3,\\ 
 V^{\prime\prime}(\phi_0)&\leq& M_p^2,\\ 
 V^{\prime\prime\prime}(\phi_0)&\leq& M_p,\\
 V^{\prime\prime\prime\prime}(\phi_0)&\leq& {\cal O}(1),\eea determine the shape of the potential
 in terms of the model parameters.
 \item The {\it prime} denotes the derivative w.r.t. $\phi$. In particular, some specific choices of the potential
would be a {\it saddle point}, when  \be V^{\prime}(\phi_0)=0=V^{\prime\prime}(\phi_0),\ee
an  {\it inflection point}, when \be V^{\prime\prime}(\phi_0)=0.\ee
\end{itemize}

Previous studies regarding obtaining large $r(k_\star)$ within sub-Planckian VEV models of inflation have been studied in 
Refs.~\cite{BenDayan:2009kv,Shafi,Choudhury:2013iaa}. In Ref.~\cite{BenDayan:2009kv},
the authors could match the amplitude of the power spectrum, $P_S$, at the pivot point, but not at the entire range of $\Delta\phi$ for the observable window of $\Delta N$, where
$N$ is the number of e-foldings of inflation. In Ref.~\cite{Hotchkiss:2011gz}, the authors have looked into higher order slow roll corrections by expanding the  potential 
around $\phi_0$. They pointed out that large \be r\sim 0.05\ee could be obtained in an {\it inflection-point} model of inflation where the slow roll parameter, $\epsilon_V$, changes 
non-monotonically (for a definition of $\epsilon_V$, see Eq.~(\ref{ra1})). The $\epsilon_V$ parameter first increases within the observational window of $\Delta N$ and then decreases before increasing to exit the slow roll inflation by violating the slow roll condition, i.e. $\epsilon_V\approx 1$
~\footnote{The
smallness of the observational window on $\Delta N$ also allows us to exploit another loophole
in the derivation of the Lyth bound \cite{Lyth:1996im}, namely the assumption that $\epsilon_{V}$
increases monotonically. This seems a natural assumption given that during
inflation \be \epsilon_{V} \ll 1\ee and at the end of inflation \be \epsilon_{V} \sim 1,\ee
current observational constraints also strongly favour $\epsilon_{V}$ increasing
during the $\Delta N\sim 17$ $e$-folds of the window achieved by the CMB
distortion observation. However, outside
this window the behaviour of $\epsilon_{V}$ is not constrained and the assumption of
monotonicity is not strictly necessary. Relaxing this assumption makes it
possible to construct a scalar potential for a single field that violates the
Lyth bound. For more details see Ref.~\cite{Hotchkiss:2011gz}.}.
The value of $r$ was still small in order to accommodate the WMAP data, which had probed roughly $ \Delta N\approx 8$ as compared to the Planck,
 which has now probed $\Delta N\approx 17$ e-foldings of inflation \cite{Choudhury:2013iaa,Choudhury:2014kma,Choudhury:2014wsa,Khatri:2013xwa,Clesse:2014pna} can be achieved by the CMB
distortion observation. A generic bound on tensor-to-scalar ratio was explicitly derived in Ref.~\cite{Choudhury:2013iaa,Choudhury:2014kma,Choudhury:2014wsa} for a generic type of {\it inflection-point} inflationary model,
where I set $V^{\prime\prime}(\phi_0)=0$ in the Taylor expanded form of the potential. 

In this paper I will consider the full potential of Eq.~(\ref{rt10a}), and our prime objective is to determine the values 
of the Taylor expansion co-efficients $V(\phi_0),~V^{\prime}(\phi_0)$, $V^{\prime\prime}(\phi_0),~V^{\prime\prime\prime}(\phi_0)$
and $V^{\prime\prime\prime\prime}(\phi_0)$ from the latest Planck 2015 and BICEP2/Keck Array+Planck 2015 joint data~\footnote{
Additionally I have also mentioned the results using the obsoleted Planck+WAMP-9+high L+BICEP2(dust) data throughout the paper to explicitly show 
the validity of our prescribed methodology for 
any kind of observed data set.}. In this respect I will be reconstructing the inflationary potential around $\phi_0$ and at the pivot
scale, $\phi_\star=\phi(k_\star)$, 
where I fix $k_\star = 0.002 {\rm Mpc^{-1}}$. We will also provide for the second order consistency relations for a sub-Planckian excursion of the inflaton field. This will be treated as
an observational discriminator which could rule out various sub-Planckian models of inflation in future.

Within the region of  $\Delta {N}\approx {\cal O}(8-17)$ e-foldings \cite{Choudhury:2013iaa,Choudhury:2014kma,Choudhury:2014wsa,Khatri:2013xwa,Clesse:2014pna}, I will be able to constrain the power spectrum: $P_S$, spectral tilt: $n_S$,
running of the spectral tilt: $\alpha_S$, and running of running of the spectral tilt: $\kappa_S$, 
in the background of $\Lambda$CDM model for:\\
\underline{\bf Planck (2013)+WMAP-9+high~L data sets:\cite{Planck-1,Planck-infl}}
 \begin{eqnarray}\label{obscons1a}
  r(k_\star)&\leq&  0.12~~ ({\rm within}~ 2\sigma ~C.L.),\\
 \label{obscons1b}\ln(10^{10}P_{S})&=&3.089^{+0.024}_{-0.027}~~ ({\rm within}~ 2\sigma ~C.L.),\\
 \label{obscons1c} n_{S}&=&0.9600 \pm 0.0071~~ ({\rm within}~ 3\sigma ~C.L.),\\
 \label{obscons1d}\alpha_{S}&=&dn_{S}/d\ln k=-0.013\pm 0.009~~({\rm within}~1.5\sigma~C.L.),\\
 \label{obscons1e}\kappa_{S}&=&d^{2}n_{S}/d\ln k^{2}=0.020^{+0.016}_{-0.015}~~({\rm within}~1.5\sigma~C.L.)\,.
 \end{eqnarray}
 \underline{\bf\textcolor{red}{ Planck (2014)+WMAP-9+high~L+BICEP2 (dust)} data sets:\cite{Ade:2014xna}}
 \begin{eqnarray}
 \label{obscons2a}0.15 &\leq & r(k_\star)\leq  0.27\\
 \label{obscons2b}\ln(10^{10}P_{S})&=&3.089^{+0.024}_{-0.027}~~ ({\rm within}~ 2\sigma ~C.L.),\\
 \label{obscons2c}n_{S}&=&0.9600 \pm 0.0071~~ ({\rm within}~ 3\sigma ~C.L.),\\
 \label{obscons2d}\alpha_{S}&=&dn_{S}/d\ln k=-0.022\pm 0.010~~({\rm within}~1.5\sigma~C.L.),\\
 \label{obscons2e}\kappa_{S}&=&d^{2}n_{S}/d\ln k^{2}=0.020^{+0.016}_{-0.015}~~({\rm within}~1.5\sigma~C.L.)\,.
 \end{eqnarray}
\underline{\bf Planck (2015)+WMAP-9+high~L(TT) data sets:\cite{Ade:2015lrj}}
 \begin{eqnarray}
  \label{obscons3a}r(k_\star)&\leq&  0.11~~ ({\rm within}~ 2\sigma ~C.L.),\\
 \label{obscons3b}\ln(10^{10}P_{S})&=&3.089\pm 0.036~~ ({\rm within}~ 2\sigma ~C.L.),\\
 \label{obscons3c}n_{S}&=&0.9569 \pm 0.0077~~ ({\rm within}~ 3\sigma ~C.L.),\\
 \label{obscons3d}\alpha_{S}&=&dn_{S}/d\ln k=0.011^{+ 0.014}_{-0.013}~~({\rm within}~1.5\sigma~C.L.),\\
 \label{obscons3e}\kappa_{S}&=&d^{2}n_{S}/d\ln k^{2}=0.029^{+0.015}_{-0.016}~~({\rm within}~1.5\sigma~C.L.)\,.
 \end{eqnarray}
\underline{\bf Planck (2015)+BICEP2/Keck Array joint data sets:\cite{Ade:2015tva}}
 \begin{eqnarray}
  \label{obscons4a}r(k_\star)&\leq&  0.12~~ ({\rm within}~ 2\sigma ~C.L.),\\
 \label{obscons4b}\ln(10^{10}P_{S})&=&3.089^{+0.024}_{-0.027}~~ ({\rm within}~ 2\sigma ~C.L.),\\
 \label{obscons4c}n_{S}&=&0.9600 \pm 0.0071~~ ({\rm within}~ 3\sigma ~C.L.),\\
 \label{obscons4d}\alpha_{S}&=&dn_{S}/d\ln k=-0.022\pm 0.010~~({\rm within}~1.5\sigma~C.L.),\\
 \label{obscons4e}\kappa_{S}&=&d^{2}n_{S}/d\ln k^{2}=0.020^{+0.016}_{-0.015}~~({\rm within}~1.5\sigma~C.L.)\,.
 \end{eqnarray}
In this paper-
\begin{itemize}
 \item We will briefly recap the key equations for the inflationary tensor-to-scalar ratio
 in the most generalized case by taking into account of the effect of higher order slow-roll corrections. Throughout the analysis of the paper
 I assume:
 \begin{enumerate}
  \item Inflaton 
 field $\phi$ is minimally coupled to the Einstein gravity sector.
 \item Slow-roll prescription perfectly holds good after considering higher order slow-roll corrections.
 \item Convergence of the Taylor expanded potential.
 \item For the numerical estimations the generic form of Taylor expanded potential is truncated at the fourth power of inflaton field. 
 \item We also assume that the major contribution for the generic version of inflationary potential comes from the first term $V(\phi_{0})$ of the Taylor series for which convergence of the taylor sereies holds good perfectly within the 
 present framework. We will explicitly show in the next sections, within the slow-roll regime of inflation this assumption is compatible with the results obtained by applying the recent observational constraints.
 \item As the contribution from all the non-renormalizable effective field theory operators are highly suppressed by the various powers of the 
 UV cut-off scale of the effective field theory $\Lambda_{UV}$, in the present context I neglect all such contribution due to its smallness.
\item Initial condition for inflation is fixed via Bunch-Davies vacuum. 
\item Sound speed is fixed at $c_{S}=1$.
\item UV cut-off of the effective theory is fixed at $\Lambda_{UV}=M_{p}$, where $M_{p}$ is the reduced Planck mass. But in principle one can fix the 
scale between GUT scale and reduced Planck scale i.e. $\Lambda_{GUT}<\Lambda_{UV}\leq M_{p}$. But in such a situation $\Lambda_{UV}$ acts as a regulating 
parameter in the effective field theory. To avoid all such complications I fix it at $M_{p}$.
 \end{enumerate}
 
 \item We will then derive the {\it most general} bound on $r(k_\star)$ for
 a generic sub-Planckian VEV inflation by considering the effect of running and running of the running in primordial scalar and tensor power spectrum, and the corresponding values of $H_\star$ and $V(\phi_\star)$
 within the framework of effective field theory. For completeness of the presented analysis, in the Appendix of this paper I give also the expression for the bound obtained from the  various parameterization in the primordial 
 power spectrum.
 \item We will then discuss the non-monotonic behaviour of slow-roll parameter and its effectiveness to evade the {\it Lyth bound} \cite{Lyth:1996im} in section \ref{w3}.
 \item Further I reconstruct the shape of the potential in section \ref{w4}, by providing
 the observational constraints from Planck (2013)+WMAP-9+high-L \cite{Planck-1,Planck-infl}, Planck (2014)+WMAP-9+high-L+BICEP2 (dust) \cite{Ade:2014xna}, Planck (2015)+WMAP-9+high-L(TT) \cite{Ade:2015lrj}
 and Planck (2015)+BICEP2/Keck Array joint constraints \cite{Ade:2015tva} on the various Taylor expansion co-efficients 
 $V^{\prime}(\phi_0),~V^{\prime\prime}(\phi_0),~V^{\prime\prime\prime}(\phi_0)$
 and $V^{\prime\prime\prime\prime}(\phi_0)$ at VEV $\phi_{0}$, which are expressed in terms of the Taylor expansion co-efficients 
 $V^{\prime}(\phi_{\star}),~V^{\prime\prime}(\phi_{\star})$, $V^{\prime\prime\prime}(\phi_{\star})$
 and $V^{\prime\prime\prime\prime}(\phi_{\star})$ at pivot/CMB scale $\phi_{\star}$.
 \item In section \ref{w5}, I will discuss the inflationary consistency relationships
 considering up to second order correction in slow-roll parameters.
 \item In section \ref{w6}, I will consider a specific case of inflection point
 inflation for the purpose of illustration.
 \item In section \ref{w6a}, I will consider a specific case of saddle point
 inflation for the purpose of illustration.
 \item Finally in section \ref{w7} I have discussed the multipole scanning of CMB TT,
 TE, EE and BB spectra via reconstructed potential by applying the constraints from Planck (2013)+WMAP-9+high-L, Planck (2014)+WMAP-9+high-L+BICEP2 (dust), Planck (2015)+WMAP-9+high-L(TT)
 and Planck (2015)+BICEP2/Keck Array data.
 \item Additionally, in the Appendix of our paper for completeness I will provide all
 the key equations and detailed derivations of main results 
 which are used frequently throughout the prescribed analysis of the paper.
\end{itemize}


\section{Brief introduction to tensor to scalar ratio in inflation} 
\label{w1}
The tensor to scalar ratio can be  defined by taking into account of
 the higher order corrections, see Refs.~\cite{Choudhury:2013iaa,Choudhury:2013jya,Easther:2006tv}:
\be\label{para 21ea} 
r=16\epsilon_{H}\frac{\left[1-({\cal C}_{E}+1)\epsilon_{H}\right]^{2}}{\left[1-(2{\cal C}_{E}+1)\epsilon_{H}
+{\cal C}_{E}\eta_{H}\right]^{2}}\,,\ee
where \be {\cal C}_{E}=4(\ln 2+\gamma_{E})-5\ee  with $\gamma_{E}=0.5772$ is the {\it Euler-Mascheroni constant} \cite{Easther:2006tv}.
In Eq~(\ref{para 21ea}) the Hubble slow roll parameters $(\epsilon_{H},\eta_{H})$ are defined as:
\begin{eqnarray}\label{hub1}
    \epsilon_{H}&=&-\frac{d\ln H}{d\ln a}=-\frac{\dot{H}}{H^{2}}\,,\\
    \eta_{H}&=&-\frac{d\ln \dot{\phi}}{d\ln a}=-\frac{\ddot{\phi}}{H\dot{\phi}}\,,
   \end{eqnarray}
where dot denotes time derivative with respect to the physical time.
Now considering  the effect from the leading order dominant contributions from the slow-roll parameters, the Hubble slow-roll parameters
can be expressed in terms of the potential dependent slow-roll parameters, $(\epsilon_{V},~\eta_{V})$, as:
\bea \epsilon_{H}&\approx&\epsilon_{V}+\cdots,\\  \eta_{H}&\approx&\eta_{V}-\epsilon_{V}+\cdots,\eea
where $\cdots$ comes from the higher order contributions of $(\epsilon_{V},~\eta_{V})$.

The tensor to scalar ratio can be re-expressed in terms of inflationary potential as:
\be\label{para 21e} 
r\approx16\epsilon_{V}\frac{\left[1-({\cal C}_{E}+1)\epsilon_{V}\right]^{2}}{\left[1-(3{\cal C}_{E}+1)\epsilon_{V}
+{\cal C}_{E}\eta_{V}\right]^{2}}\,
\ee
where 
slow-roll parameters  $(\epsilon_{V},~\eta_{V})$ are given by in terms of the inflationary potential $V(\phi)$, which can be expressed as:
\begin{eqnarray}\label{ra1}
    \epsilon_{V}&=&\frac{M^{2}_{p}}{2}\left(\frac{V^{\prime}}{V}\right)^{2}\,,\\
    \label{ra2} \eta_{V}&=&{M^{2}_{P}}\left(\frac{V^{\prime\prime}}{V}\right)\,.
   \end{eqnarray}
We would also require two other slow-roll parameters, $(\xi^{2}_{V},\sigma^{3}_{V})$, in our analysis, which are given by:
\begin{eqnarray}\label{ja1}
    \xi^{2}_{V}&=&M^{4}_{p}\left(\frac{V^{\prime}V^{\prime\prime\prime}}{V^{2}}\right)\,,\\
    \label{ja2} \sigma^{3}_{V}&=&M^{6}_{p}\left(\frac{V^{\prime 2}V^{\prime\prime\prime\prime}}{V^{3}}\right)\,.
   \end{eqnarray}
Note that I have neglected the contributions from the higher order slow-roll terms, as they are sub-dominant at the leading order.
With the help of 
\be\label{con1}
 \frac{d}{d\ln k}=-M_p\frac{\sqrt{2\epsilon_{H}}}{1-\epsilon_{H}}\frac{d}{d\phi}\,\approx -M_p\frac{\sqrt{2\epsilon_{V}}}{1-\epsilon_{V}}\frac{d}{d\phi}\,,
\ee
I can derive a simple expression for the tensor-to-scalar ratio, $r$, as:~\footnote{ We have derived some of the key expressions in an Appendix, see for instance, Eq.~(\ref{para 21e}), which I would require to derive the above expression, Eq.~(\ref{con2}). }
\be\label{con2}
 r=\frac{8}{M^{2}_{p}}\frac{(1-\epsilon_{V})^{2}\left[1-({\cal C}_{E}+1)\epsilon_{V}\right]^{2}}{\left[1-(3{\cal C}_{E}+1)\epsilon_{V}
+{\cal C}_{E}\eta_{V}\right]^{2}}\left(\frac{d\phi}{d{\ln k}}\right)^{2}. 
 \ee
We can now derive a bound on $r(k)$  in terms of the momentum scale:
\be\begin{array}{llll}\label{con4}
    \displaystyle \int^{{ k}_{\star}}_{{k}_{e}}\frac{dk}{k}\sqrt{\frac{r({k})}{8}} \\ \displaystyle =
\frac{1}{M_p}\int^{{\phi}_{\star}}_{{\phi}_{e}}d {\phi}\frac{(1-\epsilon_{V})\left[1-({\cal C}_{E}+1)\epsilon_{V}\right]}{\left[1-(3{\cal C}_{E}+1)\epsilon_{V}
+{\cal C}_{E}\eta_{V}\right]},\\
 \displaystyle \approx \frac{1}{M_p}\int^{{\phi}_{\star}}_{{\phi}_{e}}d {\phi}(1-\epsilon_{V})\left[1+{\cal C}_{E}(2\epsilon_{V}-\eta_{V})+....\right],\\
\displaystyle \approx \frac{\Delta\phi}{M_p} \left\{ 1+\frac{1}{\Delta\phi}\left[(2{\cal C}_{E}-1)\int^{{\phi}_{\star}}_{{\phi}_{e}}d {\phi}~\epsilon_{V}
-{\cal C}_{E}\int^{{\phi}_{\star}}_{{\phi}_{e}}d {\phi}~\eta_{V}\right]+....\right\}\,,
   \end{array}\ee
where note that \be \Delta\phi \approx \phi_{\star}-\phi_{e}>0\ee is positive in Eq.~(\ref{con4}), and $\phi_e$ denotes the inflaton VEV at the end of inflation,
and $\phi_{\star}$ denote the field VEV when the corresponding mode $k_\star$ is leaving the Hubble patch during inflation. Here I have used the
slow-roll approximation \be \dot{\phi}/H\simeq\sqrt{2\epsilon_{V}}.\ee The physical significance of the Eq~(\ref{con4}) are appended below:
\begin{itemize}
 \item This gives the the analytical expression for the field excursion during inflation in terms of tensor-to-scalar ratio and other inflationary observables.
 \item This relation can be treated as a discriminator between sub-Planckian and super-Planckian inflationary models, depending on the value of field excursion.
 \item This relation also justifies the validity and correctness of Effective Field Theory framework, depending on the value of field excursion. 
 \item Also this relation can be used to break the degeneracy between various class of inflationary models.
\end{itemize}

Note that $\Delta\phi>0$ implies that the left 
hand side of the integration over momentum within an interval, $k_{e}<k<k_{\star}$, is also positive, where
$k_e$ represents corresponding momentum scale at the end of inflation.
 Individual integrals 
involving $\epsilon_V$ and $\eta_V$ are estimated in an Appendix, see Eqs.~(\ref{hj1}) and (\ref{hj2}).

In order to perform the momentum integration in the left hand side of Eq~(\ref{con4}), I have used the running of
 $r(k)$, which can be expressed as:
\be\label{con5}
 r(k)=r(k_{\star})\left(\frac{k}{k_{\star}}\right)^{a+\frac{b}{2}\ln\left(\frac{k}{k_{\star}}\right)
+\frac{c}{6}\ln^{2}\left(\frac{k}{k_{\star}}\right)+....}\,,
\ee
where 
\bea
a&=&n_{T}-n_{S}+1,\\ b&=&\left(\alpha_{T}-\alpha_{S}\right),\\ c&=&\left(\kappa_{T}-\kappa_{S}\right)\,
\eea
defined at the momentum pivot scale $k_{\star}$.
 These parameterization characterizes the spectral indices, $n_S,~n_T$, running 
of the spectral indices, $\alpha_S,~\alpha_T$, and running of the running of the spectral indices, $\kappa_S,~\kappa_T$. 
Here the subscripts, $(S,~T)$, represent the scalar and tensor modes. Now substituting the explicit form of the 
potential stated in Eq.~(\ref{rt1a}), I can evaluate the crucial integrals of the first and second slow-roll parameters ($\epsilon_{V},~\eta_{V}$)
appearing in the right hand side of Eq.~(\ref{con4}). For the details of the computation, see appendix.

It was earlier confirmed by the WMAP9+high-{\it l}+BAO+$H_{0}$ combined constraints that \cite{Hinshaw:2012aka}:
\bea \alpha_{S}&=&-0.023 \pm 0.011,\\ \kappa_{S}&=&0\eea within less than
$1\sigma$ C.L. . After the Planck release it is important to see the impact on $r(k_\star)$ due  to running, and running of the running of 
the spectral tilt by modifying the generic power law form of the parameterization of tensor-to-scalar ratio.
The combined Planck (2013)+WMAP-9 constraint 
confirms that \cite{Planck-infl}: \bea \alpha_{S}&=&-0.0134\pm 0.0090,\\ \kappa_{S}&=&0.020^{+0.016}_{-0.015}\eea 
within $1.5\sigma$ statistical accuracy, which additionally includes \be \kappa_{S}\neq 0\ee possibility for the first time. Also 
the recent combined Planck (2015)+WMAP-9 constraint 
confirms that \cite{Ade:2015lrj}: \bea \alpha_{S}&=&0.011^{+ 0.014}_{-0.013},\\ \kappa_{S}&=&0.029^{+0.015}_{-0.016},\eea 
within $1.5\sigma$ statistical accuracy, which first additionally includes \be \alpha_{S}>0\ee possibility alongwith large running compared to the Planck (2013) data.
 
 At the next to leading order, the simplest way to modify the power law parameterization is to incorporate the effects of 
higher order Logarithmic corrections in terms of the presence of non-negligible running, and running of the running
of the spectral tilt as shown in Eq~(\ref{con5}), which involves higher order slow-roll
corrections~\footnote{It is important to note that when Ref.~\cite{Lyth:1996im}  first derived a bound on large tensor-to-scalar ratio 
for super-Planckian inflationary models (with $\Delta\phi >M_{p}$), the above mentioned constraints on $\alpha_S,
\kappa_S$ were not taken into account due to lack of observational constraints.}.

After substituting Eq~(\ref{con5}) in Eq~(\ref{con4}), I will show that  additional information can be gained from our analysis:
first of all it provides more accurate and improved bound on tensor-to-scalar ratio in presence of non-negligible running and
 running of the running of the  spectral tilt. In our analysis super-Planckian physics doesn't play any role as the effective theory 
puts naturally an upper cut-off set by the Planck scale. Consequently the prescription only holds good for:
\begin{enumerate}
 \item  sub-Planckian VEVs, $\phi_{0}<M_{p}$,
 \item field excursion, 
$\Delta\phi<M_{p}$ for inflation.
\end{enumerate}
  Both of these outcomes open a completely new insight into the particle physics
motivated models of inflation, which are valid below the Planck scale.

Further note that the momentum integral has non-monotonous behaviour of the slow-roll parameters ($\epsilon_{V},\eta_{V}$)
 within the interval, $k_{e}<k<k_{cmb}$, which implies that 
$\epsilon_{V}$ and $\eta_{V}$ initially increase within an observable window of e-foldings 
(which I will define in the next section, see Eq~(\ref{efold})), and then decrease at some point during the inflationary epoch when the observable 
scales had left the Hubble patch, and  eventually increase again to end inflation \cite{BenDayan:2009kv,Hotchkiss:2011gz}. 

After substituting Eq~(\ref{con5}) in the left hand side of I Eq~(\ref{con4}), I obtain:
\be\begin{array}{llll}\label{eq6}
    \displaystyle \int^{{ k}_{\star}}_{{k}_{e}}\frac{dk}{k}\sqrt{\frac{r({k})}{8}}
\displaystyle =\sqrt{\frac{r(k_{\star})}{8}}\int^{{ k}_{\star}}_{{k}_{e}}\frac{dk}{k}\sqrt{\left(\frac{k}{k_{\star}}\right)^{a+\frac{b}{2}\ln\left(\frac{k}{k_{\star}}\right)
+\frac{c}{6}\ln^{2}\left(\frac{k}{k_{\star}}\right)}},\\
\displaystyle ~~~~~~~~~~~~~~~~~~~=\sqrt{\frac{r(k_{\star})}{8}}\int^{{ k}_{\star}}_{{k}_{e}}\frac{dk}{k_{\star}\left(\frac{k}{k_{\star}}\right)}\left(\frac{k}{k_{\star}}\right)^{A+B\ln\left(\frac{k}{k_{\star}}\right)
+C\ln^{2}\left(\frac{k}{k_{\star}}\right)},
   \end{array}\ee
where $$A=\frac{a}{2},~B=\frac{b}{4},~C=\frac{c}{12}.$$
Let us substitute, \be k/k_{\star}=\ln y,\ee to simplify the mathematical form of the above Eq~(\ref{eq6}). Consequently, I get:
\be\begin{array}{llll}\label{eq7}
    \displaystyle \int^{{ k}_{\star}}_{{k}_{e}}\frac{dk}{k}\sqrt{\frac{r({k})}{8}}
\displaystyle =\sqrt{\frac{r(k_{\star})}{8}}\int^{e^{1}}_{e^{{k}_{e}/k_{\star}}}\frac{dy}{y\ln y}\left(\ln y\right)^{A+B\ln\left(\ln y\right)
+C\ln^{2}\left(\ln y\right)},\end{array}\ee
To evaluate the integral analytically, I apply the following technique.
Let us consider:
\be\begin{array}{lll}\label{f1}
(\ln y)^{\alpha},~~~ {\rm where} ~~~\alpha<<1
\end{array}\ee
where the exponent $\alpha$ is defined as:
\be\label{f2}
\alpha=A+B\ln\left(\ln y\right)
+C\ln^{2}\left(\ln y\right)
\ee
where \be |A|,|B|,|C| \ll 1\ee with \be |A|>|B|>|C|.\ee Now, for $\alpha<<1$, which is  typically the case, one can expand the 
function mentioned in Eq~(\ref{f1}) as~\footnote{One can verify the validity of $\alpha << 1$ for a generic single field slow roll inflation, within the interval $8.2\times 10^{-11}~{\rm Mpc}^{-1}
\leq k\leq 0.056~{\rm Mpc}^{-1}$.}: 
\be\begin{array}{lll}\label{f3}
(\ln y)^{\alpha}=1+\alpha\ln(\ln y)+\cdots
\end{array}\ee

 Let us take first two terms in the right hand side of the series expansion.
This finally results in:
\be\begin{array}{llll}\label{eq7a}
\displaystyle \int^{{ k}_{\star}}_{{k}_{e}}\frac{dk}{k}\sqrt{\frac{r({k})}{8}}
\displaystyle \approx\sqrt{\frac{r(k_{\star})}{8}}\int^{e^{1}}_{e^{{k}_{e}/k_{\star}}}\frac{dy}{y\ln y}\left\{1+\left[A+B\ln\left(\ln y\right)
+C\ln^{2}\left(\ln y\right)\right]\ln (\ln y)\right\},\\
\displaystyle~~~~~~~~~~~~~~~~~~~ =\sqrt{\frac{r(k_{\star})}{8}}\left[\left(1-A+2B-6C\right)\ln y
+\left(A-2B+6C\right)(\ln y)\ln(\ln y) 
\right.\\ \left.\displaystyle ~~~~~~~~~~~~~~~~~~~~~~~~~~~~~~~~~~~~~ +\left(B-3C\right)(\ln y)\ln^{2}(\ln y) 
+C(\ln y)(\ln(\ln y))^{3}\right]^{e^{1}}_{e^{{k}_{e}/k_{\star}}},\\
\displaystyle~~~~~~~~~~~~~~~~~~ =\sqrt{\frac{r(k_{\star})}{8}}\left[\left(1-A+2B-6C\right)\left[1-\frac{k_{e}}{k_{\star}}\right]
-\left(A-2B+6C\right)\frac{k_{e}}{k_{\star}}\ln\left(\frac{k_{e}}{k_{\star}}\right) 
\right.\\ \left.\displaystyle ~~~~~~~~~~~~~~~~~~~~~~~~~~~~~~~~~~~~~~~~~~~~~~~ -\left(B-3C\right)\frac{k_{e}}{k_{\star}}\ln^{2}\left(\frac{k_{e}}{k_{\star}}\right) 
-C\frac{k_{e}}{k_{\star}}\ln^{3}\left(\frac{k_{e}}{k_{\star}}\right)\right],\\
\displaystyle~~~~~~~~~~~~~~~~~~ =\sqrt{\frac{r(k_{\star})}{8}}\left[\left(2-\frac{a}{2}+\frac{b}{2}-\frac{c}{2}\right)\left[1-\frac{k_{e}}{k_{\star}}\right]
-\left(\frac{a}{2}-\frac{b}{2}+\frac{c}{2}-1\right)\frac{k_{e}}{k_{\star}}\ln\left(\frac{k_{e}}{k_{\star}}\right) 
\right.\\ \left.\displaystyle ~~~~~~~~~~~~~~~~~~~~~~~~~~~~~~ -\left(\frac{b}{4}-\frac{c}{4}\right)\frac{k_{e}}{k_{\star}}\ln^{2}\left(\frac{k_{e}}{k_{\star}}\right) 
-\frac{c}{12}\frac{k_{e}}{k_{\star}}\ln^{3}\left(\frac{k_{e}}{k_{\star}}\right)\right].
   \end{array}\ee

In any arbitrary momentum scale $k$ spectral tilt, running of the tilt, and running of the running of the tilt for the scalar and tensor perturbations can be written
as:
\begin{eqnarray}
\label{ns1} n_{S}(k)-1&\equiv& \frac{d\ln P_{S}(k)}{d\ln k}\nonumber \\
&=& n_{S}(k_{\star})-1+\alpha_{S}(k_{\star})\ln\left(\frac{k}{k_{\star}}\right)
 +\frac{\kappa_{S}(k_{\star})}{2}\ln^{2}\left(\frac{k}{k_{\star}}\right)+\cdots\,\\
\label{nt1} n_{T}(k)&\equiv& \frac{d\ln P_{T}(k)}{d\ln k}=n_{T}(k_{\star})+\alpha_{T}(k_{\star})\ln\left(\frac{k}{k_{\star}}\right)
 +\frac{\kappa_{T}(k_{\star})}{2}\ln^{2}\left(\frac{k}{k_{\star}}\right)+\cdots\,\\
\label{asat1} \alpha_{S,T}(k)&\equiv&\frac{dn_{S,T}(k)}{d\ln k}=\frac{d^{2}\ln P_{S,T}(k)}{d\ln k^{2}}=\alpha_{S,T}(k_{\star})
+\kappa_{S,T}(k_{\star})\ln\left(\frac{k}{k_{\star}}\right)+\cdots\,\\
\label{ksat1} \kappa_{S,T}(k)&\equiv&\frac{d\alpha_{S,T}(k)}{d\ln k}=\frac{d^2n_{S,T}(k)}{d\ln k^{2}}
=\frac{d^3\ln P_{S,T}(k)}{d\ln k^{3}}\approx\kappa_{S,T}(k_{\star})+\cdots\,.
\end{eqnarray}

Also at scale $k$ tilt, running ang running of the running in tensor-to-scalar ratio can be expressed as:
\begin{eqnarray}
\label{nr1} n_{r}(k)&\equiv& \frac{dr(k)}{d\ln k}
= r(k)\left[a+b\ln\left(\frac{k}{k_{\star}}\right)
 +\frac{c}{2}\ln^{2}\left(\frac{k}{k_{\star}}\right)+\cdots\right]\,\\
\label{asat1} \alpha_{r}(k)&\equiv&\frac{dn_{r}(k)}{d\ln k}=\frac{d^{2}r(k)}{d\ln k^{2}}\nonumber \\
&=&\left[an_{r}(k)+br(k)\right]+\left[bn_{r}(k)+cr(k)\right]\ln\left(\frac{k}{k_{\star}}\right)
 +\frac{cn_{r}(k)}{2}\ln^{2}\left(\frac{k}{k_{\star}}\right)+\cdots\,~~~~~~~~~~\\
\label{ksat1} \kappa_{r}(k)&\equiv&\frac{d\alpha_{r}(k)}{d\ln k}=\frac{d^2n_{r}(k)}{d\ln k^{2}}
=\frac{d^3 r(k)}{d\ln k^{3}}\nonumber\\
&=&\left[a\alpha_{r}(k)+2bn_{r}(k)+cr(k)\right]+\left[b\alpha_{r}(k)+2cn_{r}(k)\right]\ln\left(\frac{k}{k_{\star}}\right)+\cdots\,.
\end{eqnarray}

Here at an arbitrary scale $k$, the parameters $a,b$ and $c$ defined as:
\begin{eqnarray}
\label{a} a(k)&\equiv& \frac{d\ln r(k)}{d\ln k}
= \left[a+b\ln\left(\frac{k}{k_{\star}}\right)
 +\frac{c}{2}\ln^{2}\left(\frac{k}{k_{\star}}\right)+\cdots\right]\,\\
\label{b} b(k)&\equiv&\frac{da(k)}{d\ln k}=\frac{d^{2}\ln r(k)}{d\ln k^{2}}=\left[b
 +c\ln\left(\frac{k}{k_{\star}}\right)+\cdots\right]\,\\
\label{c} c(k)&\equiv&\frac{d\alpha_{r}(k)}{d\ln k}=\frac{d^2n_{r}(k)}{d\ln k^{2}}
=\frac{d^3 \ln r(k)}{d\ln k^{3}}\approx\left[ c+\cdots\right]\,.
\end{eqnarray}

In the present context the potential dependent slow-roll parameters:
($\epsilon_{V},~\eta_{V},\cdots$), satisfy the joint Planck (2013)+WMAP-9 constraints, which imply that~\cite{Planck-infl}:
\bea
\epsilon_V<10^{-2}\,~~ ({\rm within}~ 1.5\sigma ~C.L.),\\ 5\times 10^{-3}<|\eta_{V}|<0.021\,~~ ({\rm within}~ 1.5\sigma ~C.L.),
\eea
and also satisfy the joint Planck (2015)+WMAP-9+high-L (TT) constraints, which imply that~\cite{Ade:2015lrj}:
\bea
\epsilon_V<0.011\,~~ ({\rm within}~ 2\sigma ~C.L.),\\ 8\times 10^{-3}<|\eta_{V}|<0.021\,~~ ({\rm within}~ 1.5\sigma ~C.L.),\\
8\times 10^{-3}<|\xi^2_{V}|<0.021\,~~ ({\rm within}~ 1.5\sigma ~C.L.),
\eea
for which the inflationary potential is concave in nature for both the cases. 
In the next section, I will discuss model independent bounds on the coefficients ($V(\phi_\star),V^{\prime}(\phi_\star),\cdots$) for a generic sub-Planckian 
VEV inflationary setup, for which I will satisfy the joint constraints from: Planck(2013 \& 2015)+WMAP-9+high-L, Planck (2015)+BICEP2/Keck Array and 
Planck (2014)+WMAP-9+high~L+BICEP2 (dust) data sets.


\section{Constraining the scale of effective field theory inflation via field excursion}
\label{w2}
The number of
e-foldings, ${N}(k)$,  can be expressed as~\cite{Burgess:2005sb}:
%
\be\begin{array}{llll}\label{efold}
\displaystyle {N}(k) \approx  71.21 - \ln \left(\frac{k}{k_{\star}}\right)  
+  \frac{1}{4}\ln{\left( \frac{V_{\star}}{M^4_{P}}\right) }
+  \frac{1}{4}\ln{\left( \frac{V_{\star}}{\rho_{e}}\right) }  
+ \frac{1-3w_{int}}{12(1+w_{int})} 
\ln{\left(\frac{\rho_{rh}}{\rho_{e}} \right)},
\end{array}\ee
%
where $\rho_{e}$ is the energy density at the end of inflation, 
$\rho_{rh}$ is an energy scale during reheating, 
$k_{\star}=a_\star H_\star$ is the present Hubble scale, 
$V_{\star}$ corresponds to the potential energy when the relevant modes left the Hubble patch 
during inflation corresponding to the momentum scale $k_{\star}$, and $w_{int}$ characterises the effective equation of state 
parameter between the end of inflation, and the energy scale during reheating. 

Within the momentum interval,
$k_{e}<k<k_{\star}$, the corresponding number of e-foldings is given by, $\Delta {N}$, as:
\be\begin{array}{lll}\label{intnk}
    \displaystyle \Delta {N} =N_{\star}-N_{e}\approx \ln\left(\frac{k_{\star}}{k_{e}}\right)
=\ln\left(\frac{a_{\star}}{a_{e}}\right)+\ln\left(\frac{H_{\star}}{H_{e}}\right)
\approx \ln\left(\frac{a_{\star}}{a_{e}}\right)+\frac{1}{2}\ln\left(\frac{V_{\star}}{V_{e}}\right)\,
   \end{array}\ee
where $(a_{\star},H_{\star})$ and $(a_{e}H_{e})$
represent the scale factor and the Hubble parameter at the pivot scale and end of inflation, and I have used the fact that 
\be H^{2}=\frac{V(\phi)}{3M^{2}_{p}}\ee
which is true within the framework of Einstein's General Relativity.
We can estimate the contribution of the last term of the right hand side by using Eq~(\ref{rt1a}) as follows:
\be\begin{array}{llll}\label{espot}
    \displaystyle \ln\left(\frac{V_{\star}}{V_{e}}\right)=\ln\left(
\frac{\alpha+\beta(\phi_{\star}-\phi_{0})+\gamma(\phi_{\star}-\phi_{0})^{3}+\kappa(\phi_{\star}-\phi_{0})^{4}
+\cdots\,}{\alpha+\beta(\phi_{e}-\phi_{0})+\gamma(\phi_{e}-\phi_{0})^{3}+\kappa(\phi_{e}-\phi_{0})^{4}+\cdots\,}\right),\\
\displaystyle ~~~~~~~~~~~\approx\ln\left(1+\underbrace{M_{p}\frac{\beta}{\alpha}\left(\frac{\Delta\phi}{M_{p}}\right)}_{<< 1}[1+\underbrace{\cdots}_{<< 1}]\right),\\
\displaystyle ~~~~~~~~~~~\approx \ln(1+\underbrace{\cdots}_{<< 1}),\\ 
   \end{array}\ee
 where \be (\Delta\phi/M_{p})<<1,\ee and additionally I assume that \be (\beta M_{p}/\alpha)<<1.\ee
 Consequently, Eq~(\ref{intnk}) reduces to the following simplified expression:
\be\begin{array}{lll}\label{intnk1}
    \displaystyle \Delta {N} \approx \ln\left(\frac{k_{\star}}{k_{e}}\right)
\approx\ln\left(\frac{a_{\star}}{a_{e}}\right) \approx {\cal O}(8-17)~{\rm efolds}\,.
   \end{array}\ee
Within the observed limit of CMB distortion, i.e. $\Delta N\approx 17$, the slow-roll parameters, see  Eqs.~(\ref{hj1},~\ref{hj2}) of Appendix, show non-monotonic behaviour, 
where the corresponding scalar and tensor amplitude of the power spectrum remains almost unchanged
~\footnote{In this paper I fix $\Delta {\cal N}\approx {\cal O}(8-17)$ e-foldings as within this interval the combined constraints from 
Planck (2013 \& 2015)+WMAP-9+high-L and Planck (2015)+BICEP2/Keck Array are satisfied. Additionally I have also mentioned the results by applying 
the constraints from \textcolor{red}{ Planck (2014)+WMAP-9+high~L+BICEP2 (dust)} data to explicitly show that our prescribed methodology also holds
good for $r(k_{\star})>0.12$.}.
Substituting the results obtained from Eq.~(\ref{hj1}) and Eq.~(\ref{hj2}) (see Appendix), and with the help of Eq.~(\ref{intnk}), 
up to the leading order, I obtain:
%
\be\begin{array}{llll}\label{con9}
    \displaystyle \sum^{\infty}_{n=0}{\bf G}_{n}\left(\frac{|\Delta\phi|}{M_p}\right)^{n}\approx\displaystyle\sqrt{\frac{r(k_{\star})}{8}}\times
\left|\frac{a}{2}-\frac{b}{2}+\frac{c}{2}-2\right|+\cdots\end{array}\ee
where for $\Delta N= 17$ and $\Delta N=8$ I have used \bea (k_{e}/k_{\star})&\approx& \exp(-\Delta{N})=\exp(-17)\approx 4.13\times 10^{-8},\\
(k_{e}/k_{\star})&\approx& \exp(-\Delta{N})=\exp(-8)\approx 3.35\times 10^{-4}.\eea 
 Here I will concentrate on \be a,~b,~c \neq 0\ee  where additionally  \be a>>b>>c\ee  case is satisfied (for the details, 
 see~\cite{Choudhury:2013iaa,Choudhury:2014kma,Choudhury:2014wsa,Choudhury:2013woa}). 
 In Eq~(\ref{con9}) the series appearing in the left side of the above expression is convergent, since the expansion coefficients can be expressed as:
\be\begin{array}{lll}\label{expco}
\displaystyle {\bf G}_{n}=\left\{\begin{array}{ll}
                    \displaystyle  \left(1 +\underbrace{\sum^{\infty}_{m=0}{\bf A}_{m}
\left(\frac{\phi_{e}-\phi_{0}}{M_p}\right)^{m}}_{\ll 1} \right)\sim 1 &
 \mbox{ {\bf for} $n=1$}  \\ 
         \displaystyle  << 1 & \mbox{ {\bf for} $n\geq 2$}\,,
  \end{array}
\right. \end{array}\ee
%
and I have defined a new dimensionless 
binomial expansion co-efficient (${\bf A}_{m}$), defined as:
\be\label{cond}
{\bf A}_{m}=M^{m+2}_{p}\left[\left({\cal C}_{E}-\frac{1}{2}\right){\bf C}_{m}-{\cal C}_{E}{\bf D}_{m}\right]~~~~(\forall m=0,1,2,....)\,,
 \ee
which is obtained from the binomial series expansion  from the
leading order results of the slow-roll integrals stated in the Appendix~\footnote{In Eq.~(\ref{cond}), and 
Eqs.~(\ref{hj1},~\ref{hj2}) (see Appendix), ${\bf C}_{p}$ and ${\bf D}_{q}$ are Planck suppressed dimensionful (mass dimension 
$\left[M^{-(m+2)}_{p}\right]$) binomial series expansion coefficients, which are expressed in terms of the generic model 
parameters $(V(\phi_\star),V^{\prime}(\phi_\star),\cdots)$ as presented
in Eq~(\ref{rt1a}).}.

Note that the expansion co-efficient ${\bf A}_{m}(\forall m)$ are suppressed by, 
$V(\phi_0)$, which is the leading order term in a generic expansion of the 
inflationary potential as shown in Eq~(\ref{rt10a}) and Eq~(\ref{rt1a})(see also Eq~(\ref{ceff}) in the Appendix). 
We can expand the left side of Eq.~(\ref{con9}) 
in the powers of ${\Delta\phi}/{M_p}$, using the additional constraint 
\be \Delta\phi<(\phi_{e}-\phi_{0})<M_p,\ee and I keep the leading order terms in $\Delta\phi/M_p$.


To the  first order approximation - I can 
neglect all the higher powers of $k_{e}/k_{\star}\approx {\cal O}(10^{-4}-10^{-8})$ from the left hand side of Eq~(\ref{con9}), within ${\cal O}(8-17)$ e-foldings of inflation \cite{Choudhury:2013iaa,Khatri:2013xwa,Clesse:2014pna}.
Consequently, Eq.~(\ref{con9}) reduces to the following compact form for $r(k_\star)$:
%
\be\begin{array}{llll}\label{con10sd}
    \displaystyle \frac{9}{25}\sqrt{\frac{r(k_{\star})}{0.27}}\left|\frac{27}{1600}\left(\frac{r(k_{\star})}{0.27}\right)
-\frac{\eta_{V}(k_{\star})}{2}-1+\cdots\,\right|
\displaystyle \approx \frac{\left |\Delta\phi\right|}{M_p}\leq 1\,,
   \end{array}\ee
%
provided at the pivot scale, $k=k_{\star}>>k_{e}$, in this regime 
\be |\eta_{V}|>>\left\{\epsilon^{2}_{V},\eta^{2}_{V},\xi^{2}_{V},\sigma^{3}_{V},\cdots\right\}\ee
approximation is valid.  Our expression, Eq.~(\ref{con10sd}),
shows that large value of $r(k_\star)$ can be obtained for models of inflation where
 inflation occurs below the Planck cut-off. Once the field excursion, $|\Delta\phi|/M_p$, and $\eta_{V}$, are known from any type of sub-Planckian inflationary 
setup, one can easily compute the tensor-to-scalar ratio by finding the roots. 

The expression, as stated in Eq~(\ref{con10sd}), is in the form of a simple algebraic (cubic) equation.  In order to find the roots of 
tensor-to-scalar ratio $r$ in terms of the field excursion $|\Delta\phi|/M_p$, one has to solve a cubic equation:
\bea\label{rooteq}
 x^{3}+Ux-W=0,
\eea
where 
\bea\label{cont}
 x:&=&\sqrt{\frac{r(k_{\star})}{0.27}},\\
 U:&=&\frac{800}{27}\left(\eta_{V}+2\right),\\ 
 W:&=&\frac{40000}{243}\frac{\left |\Delta\phi\right|}{M_p}.
\eea
The three roots  $x_{1},x_{2},x_{3}$ are explicitly given by:
\bea\label{root1}
x_{1}&=&-\frac{\left(\frac{2}{3}\right)^{1/3} U}{Y}
+\frac{Y}{2^{1/3} 3^{2/3}},\\
x_{2}&=&\frac{\left(1+i \sqrt{3}\right) U}{2^{2/3} 3^{1/3} Y}
-\frac{\left(1-i \sqrt{3}\right) Y}{ 2^{4/3} 3^{2/3}},\\
x_{3}&=&\frac{\left(1-i \sqrt{3}\right) U}{2^{2/3} 3^{1/3} Y}
-\frac{\left(1+i \sqrt{3}\right) Y}{ 2^{4/3} 3^{2/3}}.
\eea
where the symbol, \be Y=\left(9 W+\sqrt{3} \sqrt{4 U^3+27 W^2}\right)^{1/3}.\ee Here the complex roots $x_{2}$ and $x_{3}$ are physically redundant.
The only acceptable root is the real one, i.e. $x_{1}$ which will finally contribute for the estimation of field excursion
for a specified value of tensor-to-scalar ratio $r(k_{\star})$.

Further note that our formulation will also hold true if inflation were to start at the hill-top, such as \be \phi=0.\ee However in this case one would have 
to proceed similarly by expanding the potential around \be \phi_0=0,\ee and then follow the algorithm I have provided here.

Now, it is also possible to recast $a(k),~b(k),~c(k)$, in terms of $r(k)$, and the slow roll parameters by using the 
relation, Eq.~(\ref{para 21e}) (see Appendix):
\be\begin{array}{lll}\label{apara}
    \displaystyle a(k_{\star})\approx \left[\frac{r(k_{\star})}{4}-2\eta_{V}(k_{\star})
+\cdots\right],\\
    \displaystyle b(k_{\star})\approx \left[16\epsilon^{2}_{V}(k_{\star})-12\epsilon_{V}(k_{\star})\eta_{V}(k_{\star})+2\xi^{2}_{V}(k_{\star})+\cdots\right],\\
\displaystyle c(k_{\star})\approx \left[-2\sigma^{3}_{V}+\cdots\right],
   \end{array}\ee
where $``\cdots''$ involve the higher order slow roll contributions, which are negligibly small in the leading order approximation.

Now at the pivot scale, $k_{\star}=0.002~Mpc^{-1}$, the scale of inflation $V^{1/4}(\phi_\star)$ can be expressed as:
\begin{equation}\label{scale}
     V^{1/4}(\phi_{\star})=\left(\frac{3}{2}P_{S}(k_{\star})r(k_{\star})\right)^{1/4}\times\sqrt{\pi}M_{p}\leq (2.40\times 10^{16}{\rm GeV})\times~\left(\frac{r(k_{\star})}{0.27}\right)^{1/4}.
   \end{equation}
The equivalent statement can be made in terms of the upper bound on the
numerical value of the Hubble parameter at the exit of the relevant modes:
\begin{equation}\label{hubinfla}
     H_{\star}\leq 1.38\times 10^{14}\times\sqrt{\frac{r(k_{\star})}{0.27}}~{\rm GeV}.
   \end{equation}
 Combining Eqs.~(\ref{con10sd}), (\ref{hubinfla}) and (\ref{scale}),
I can now obtain a closed relationships:
\begin{equation}\label{con13}
\frac{|\Delta\phi|}{M_p}  \leq  \frac{\sqrt{V_{\star}}}{(1.55\times 10^{-2}~M_p)^{2}}\left|\frac{V_{\star}}{(2.78\times 10^{-2}M_p)^{4}}-\frac{\eta_{V}(k_{\star})}{2}-1\right|\,.
\end{equation}

\begin{equation}\label{con13xc}
\frac{|\Delta\phi|}{M_p}  \leq  \frac{H_{\star}}{(1.39\times 10^{-4}~M_p)}\left|\frac{H^{2}_{\star}}{(1.99\times 10^{-7}~M^{2}_{p})}-\frac{\eta_{V}(k_{\star})}{2}-1 \right|\,.
\end{equation}
where additionally
\be |\eta_{V}|>>\left\{\epsilon^{2}_{V},\eta^{2}_{V},\xi^{2}_{V},\sigma^{3}_{V},\cdots\right\}\ee are satisfied. Similar expressions were derived in Ref.~\cite{Choudhury:2013iaa}
for {\it inflection-point} model of inflation, where an additional constraint \be V^{\prime\prime}(\phi_0)=0\ee is considered for the computation.
The above Eqs.~(\ref{con13},~\ref{con13xc}) characterize the bounds on $\Delta \phi$ for: \bea \phi_0 &<& M_p,\\ \Delta \phi &\lesssim& M_p.\eea

Our conditions, Eqs.~(\ref{con10sd},~\ref{con13}), provide 
new constraints on model building for inflation within  particle theory, where the inflaton potential is always constructed 
within an effective field theory with a cut-off. Note that \be |\eta_V(k_{\star})|> 0 \ee can provide the largest contribution, 
in order to satisfy the latest bound on the tensor-to-scalar ratio,
the shape of the potential has to be {\it concave} in nature. Further applying this input in Eq~(\ref{scale}), one can get a preferred bound 
on the scale of sub-Planckian VEV inflation as:\\
 \underline{\bf Planck (2013)+WMAP-9+high~L:}\\
  \bea
  \label{gutr1}\sqrt[4]{V_{\star}}&\leq& 1.96\times 10^{16}~{\rm GeV}~~~{\bf  for}~~~~ r_{\star}\leq 0.12,\\
   \label{gutr2}H_{\star}&\leq& 9.20\times 10^{13}~{\rm GeV}~~~{\bf  for}~~~~ r_{\star}\leq 0.12,\eea
  \underline{\bf\textcolor{red}{ Planck (2014)+WMAP-9+high~L+BICEP2 (dust)}:}
  \bea
  \label{gutr3}2.07\times10^{16}~{\rm GeV}&\leq&\sqrt[4]{V_{\star}}\leq 2.40\times 10^{16}~{\rm GeV}~~~{\bf  for}~~~~ 0.15\leq r_{\star}\leq 0.27,\\
  \label{gutr4}1.03\times10^{14}~{\rm GeV}&\leq&H_{\star}\leq 1.38\times 10^{14}~{\rm GeV}~~~{\bf  for}~~~~ 0.15\leq r_{\star}\leq 0.27.\eea
  \underline{\bf Planck (2015)+WMAP-9+high~L(TT):}
  \bea
  \label{gutr5}\sqrt[4]{V_{\star}}&\leq& 1.92\times 10^{16}~{\rm GeV}~~~{\bf  for}~~~~ r_{\star}\leq 0.11,\\ 
  \label{gutr6}H_{\star}&\leq& 8.80\times 10^{13}~{\rm GeV}~~~{\bf  for}~~~~ r_{\star}\leq 0.11,\eea
  \underline{\bf Planck (2015)+BICEP2/Keck Array :}
  \bea
  \label{gutr7}\sqrt[4]{V_{\star}}&\leq& 1.96\times 10^{16}~{\rm GeV}~~~{\bf  for}~~~~ r_{\star}\leq 0.12,\\
  \label{gutr8}H_{\star}&\leq& 9.20\times 10^{13}~{\rm GeV}~~~{\bf  for}~~~~ r_{\star}\leq 0.12.\eea
%


\begin{figure}[t]
\centering
\subfigure[$\epsilon_{V}$~vs~$\phi-\phi_{0}$]{
    \includegraphics[width=7.1cm, height=9.1cm] {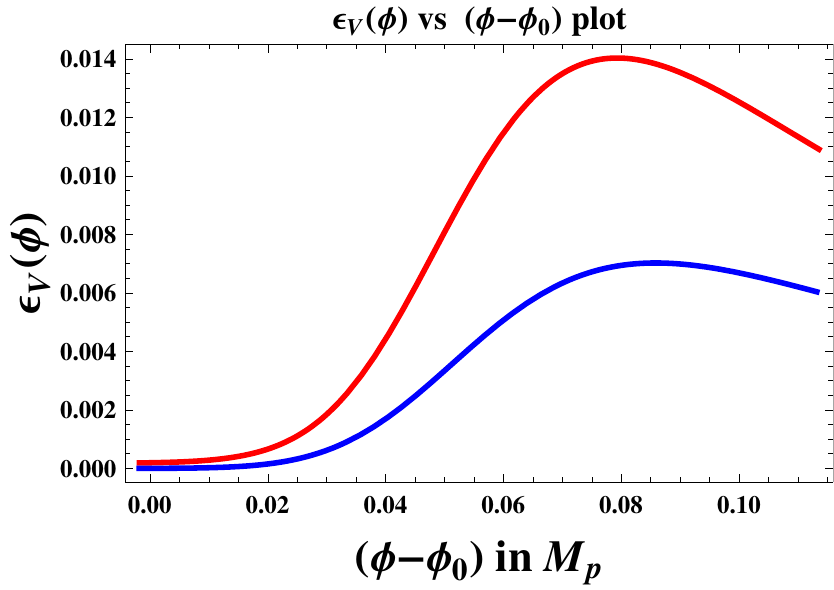}
    \label{fig:subfig1}
}
\subfigure[$|\eta_{V}|$~vs~$\phi-\phi_{0}$]{
    \includegraphics[width=7.1cm, height=9.1cm] {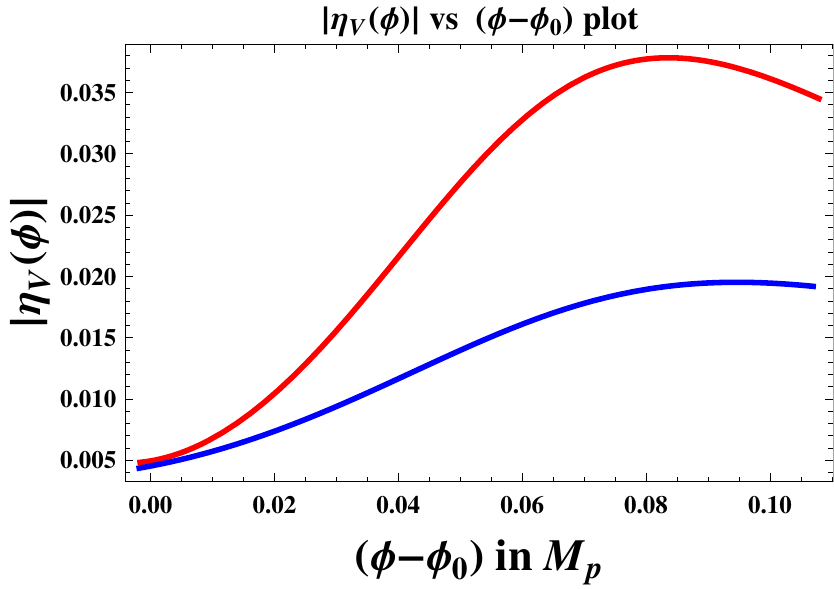}
    \label{fig:subfig2}
}
\subfigure[$|\xi^{2}_{V}|$~vs~$\phi-\phi_{0}$]{
    \includegraphics[width=7.1cm, height=9.1cm] {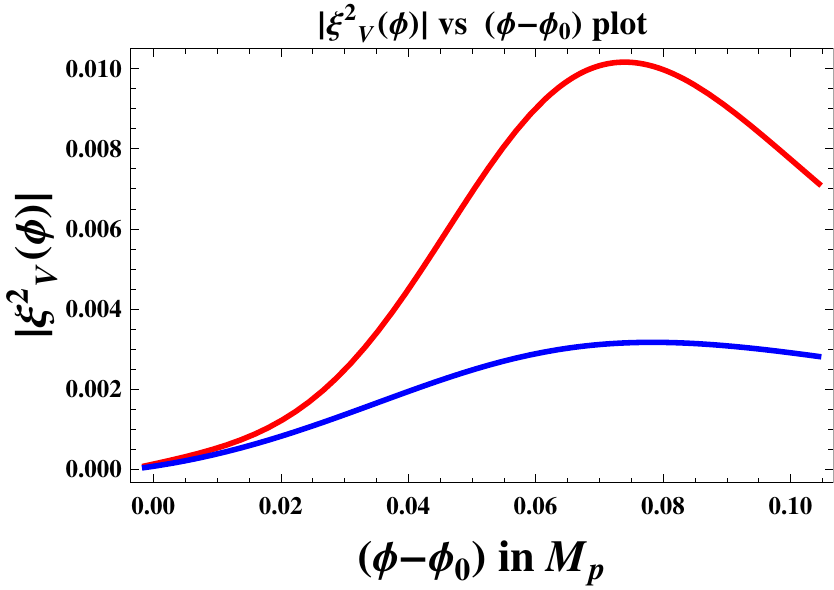}
    \label{fig:subfig3}
}
\subfigure[$|\sigma^{3}_{V}|$~vs~$\phi-\phi_{0}$]{
    \includegraphics[width=7.1cm, height=9.1cm] {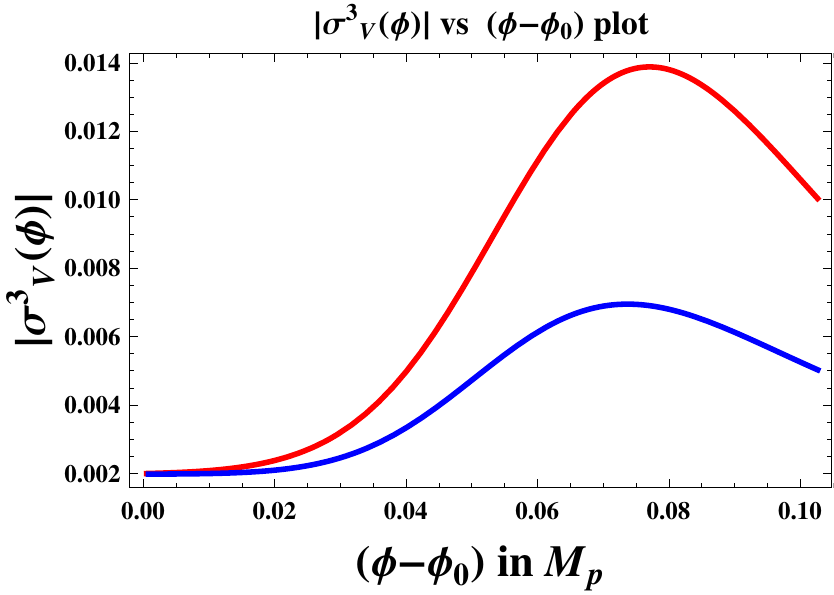}
    \label{fig:subfig4}
}
\caption[Optional caption for list of figures]{Non-monotonous evolution of the slow roll parameters are shown with respect to $\phi-\phi_0$. The upper and lower bounds are set by
Eqs.~(\ref{constraint6}-\ref{constraint10}).  
}
\label{fig3}
\end{figure}



\begin{figure}[t]
\centering
\subfigure[$\epsilon_{V}$~vs~$N$]{
    \includegraphics[width=7.1cm, height=9.1cm] {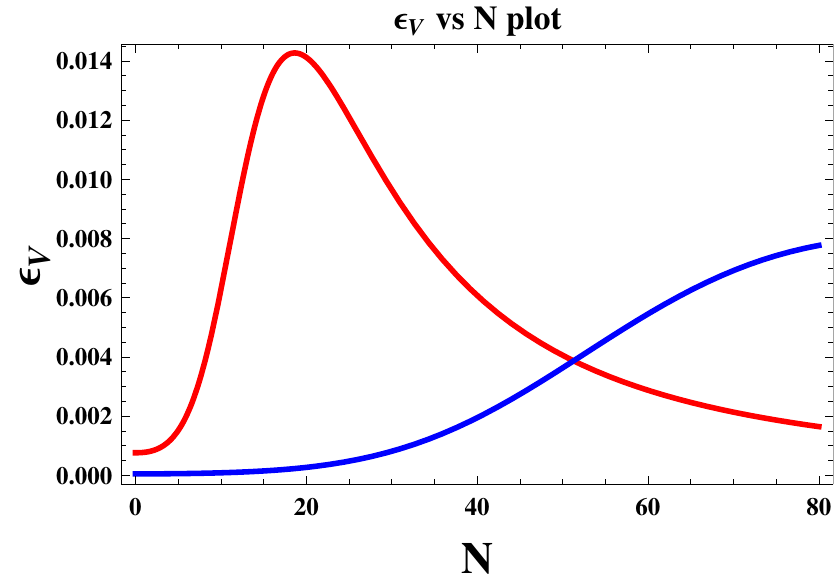}
    \label{fig:subvv1}
}
\subfigure[$|\eta_{V}|$~vs~$N$]{
    \includegraphics[width=7.1cm, height=9.1cm] {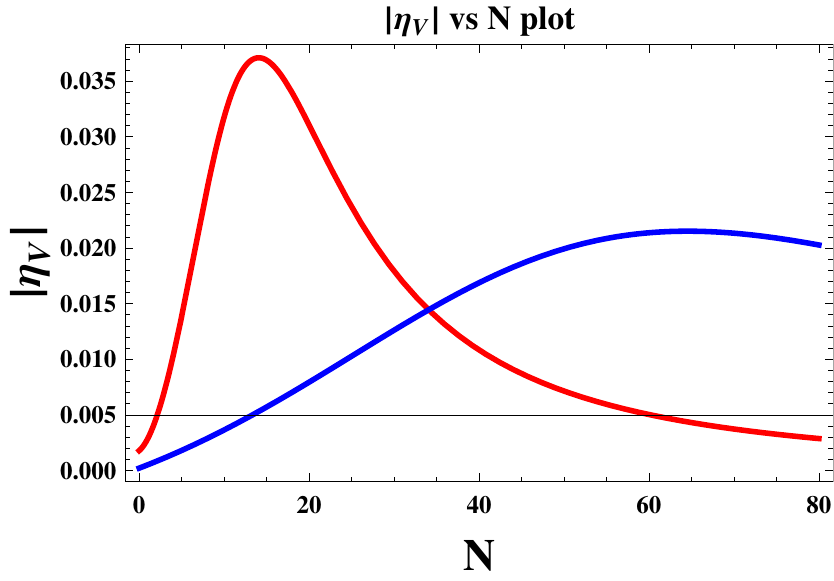}
    \label{fig:subvv2}
}
\subfigure[$|\xi^{2}_{V}|$~vs~$N$]{
    \includegraphics[width=7.1cm, height=9.1cm] {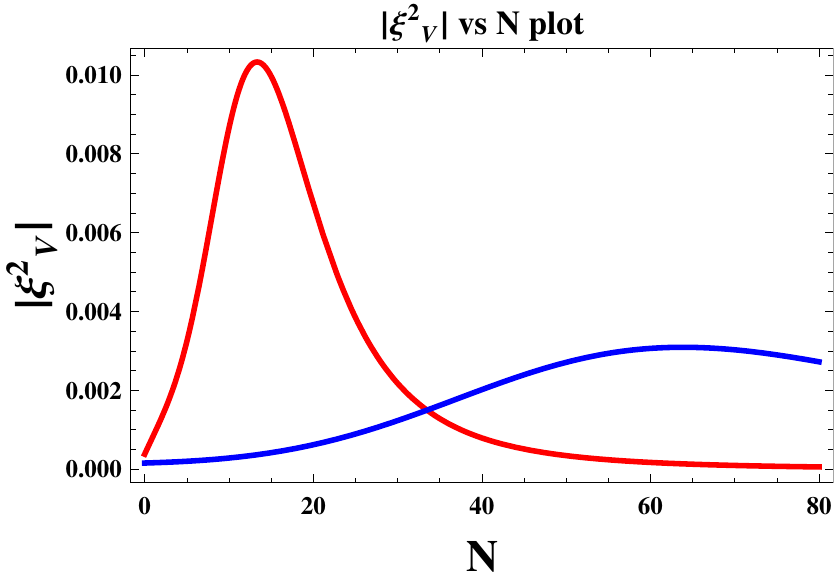}
    \label{fig:subvv3}
}
\subfigure[$|\sigma^{3}_{V}|$~vs~$N$]{
    \includegraphics[width=7.1cm, height=9.1cm] {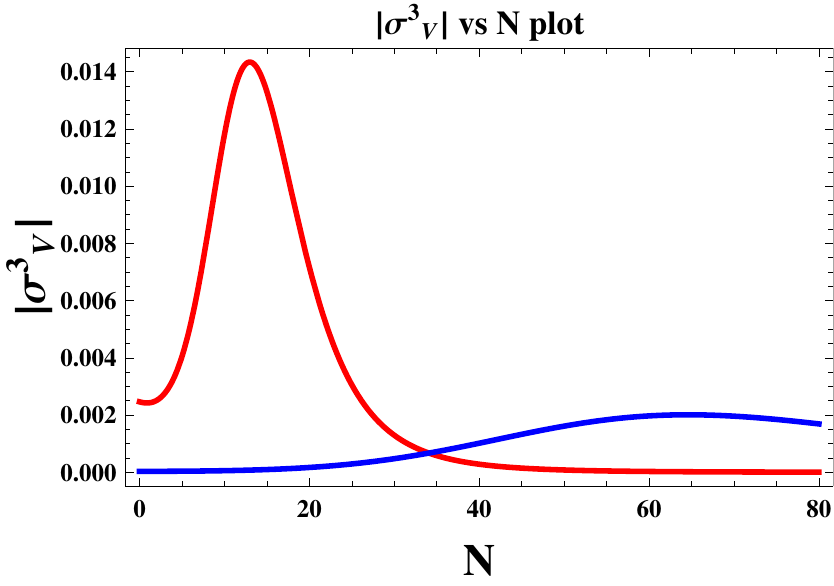}
    \label{fig:subvv4}
}
\caption[Optional caption for list of figures]{Non-monotonous evolution of the slow roll parameters are shown with respect to number of e-foldings $N$. The upper and lower bounds are set by
Eqs.~(\ref{constraint6}-\ref{constraint10}).  
}
\label{fig3new}
\end{figure}
\section{Non-monotonic behaviour of slow-roll parameters within effective field theory}
\label{w3}

Let us now discuss the non-monotonous features of slow-roll parameters $\epsilon_{V},\eta_{V},\xi^{2}_{V}, \sigma^{3}_{V}$ appearing in the present context of the paper:-
\begin{itemize}
 \item To violate the Lyth bound mainly $\epsilon_{V}$ and also the other slow-roll parameters $\eta_{V},\xi^{2}_{V}, \sigma^{3}_{V}$ must decrease at
some point during the inflationary epoch. This is evident from
the definition of no. of e-foldings, \bea N&=&\frac{1}{M_p}\int_{\phi_{cmb}}^{\phi_{\rm e}}
\frac{d\phi}{\sqrt{2\epsilon_{V}}}\nonumber\\
&\approx&\frac{1}{M_p}\int_{\phi_{cmb}}^{\phi_{\rm e}}
\sqrt{\frac{8}{r}}d\phi\;,\eea where, if $\epsilon_{V}$ decreases, $\Delta N$ will increase
for the same field excursion, $\Delta \phi$. Unfortunately this alone is not sufficient
enough to successfully evade the Lyth bound and match all observations.

\item  At CMB scale, \be k=k_{cmb}(=aH),\ee at least $\epsilon_{V}$ must be large enough to generate an observable value of tensor-to-scalar ratio $r$.
In our methodology I have considered that the other slow-roll parameters $\eta_{V},\xi^{2}_{V}, \sigma^{3}_{V}$ also be sufficiently large to confront observation 
at CMB scale.

\item  Also $\epsilon_{V}$ and the other slow-roll parameters must increase over the $\Delta N\approx $ 17 e-fold observational
 window \cite{Choudhury:2013iaa,Khatri:2013xwa,Clesse:2014pna}. This is dictated by a combination of the spectral index constraint and
the observed value of $\sigma_{8}$ from large-scale structure (LSS) \cite{Mantz:2009fw}, which means the
spectrum must decrease over the observational window. When running of the spectral
index is allowed, the best fit value for the spectral index is indeed \be n_S>1,\ee
but the running is \be \alpha_{S}<0,\ee therefore $\epsilon_{V}$ must increase
eventually. Also, the value of $\sigma_8$ measured independently from LSS
strongly favours a primordial spectrum that decreases amplitude between Planck
and LSS scales.

\item After observable scales have left the horizon ($k>aH$), $\epsilon_{V}$ and the other slow-roll parameters must quickly decrease.
The quick decrease of $\epsilon_{V}$ and as well as other slow-roll parameters are necessary to generate enough e-folds required for inflation.
 If instead $\epsilon_{V}$ decreases gradually within the present context, then it will need
to eventually decrease to a much smaller value because in such a situation 
\bea\epsilon_{V} &\propto& (\Delta \phi/2M_p\Delta N)^2,\\ \eta_{V}&\propto& 2\epsilon_{V}+(1/\sqrt{2}\Delta N),\\
\xi^{2}_{V}&\propto& 2\epsilon_{V}\left[1+2\epsilon_{V}+(\sqrt{2}/\Delta N)\right]+(1/\sqrt{2}\Delta N)^2,\\
\sigma^{3}_{V}&\propto& 2\epsilon_{V}\left\{4\epsilon_{V}\left[1+\epsilon_{V}+(\sqrt{2}/\Delta N)\right]+(1/\sqrt{2}\Delta N)
\left[\sqrt{2}+(1/2\epsilon_{V}\Delta N)^2\right]+3(1/\sqrt{2}\Delta N)^2\right\},\nonumber\\ \eea and I finally 
require \be \Delta \phi \leq M_p\ee to violate the Lyth bound in the present context. Consequently, the slope of $\epsilon_{V},\eta_{V},\xi^{2}_{V},\sigma^{3}_{V}$ 
in fig~(\ref{fig3}) and fig~(\ref{fig3new}) can be computed as:
\bea\Delta\epsilon_{V}/\Delta\phi &\propto& (\sqrt{\epsilon_{V}}/M_p\Delta N),\\
 \Delta\eta_{V}/\Delta\phi &\propto& 2(\sqrt{\epsilon_{V}}/M_p\Delta N)+(1/2M_p\sqrt{2\epsilon_{V}}(\Delta N)^2),\\
\Delta\xi^{2}_{V}/\Delta\phi&\propto& 2(\sqrt{\epsilon_{V}}/M_p\Delta N)\left[1+4\epsilon_{V}+(\sqrt{2}/\Delta N)\right]+(\sqrt{2\epsilon_{V}}/M_p(\Delta N)^2)
\nonumber\\&&~~~~~~~~~~~~~~~~~~~~~~~~~~~~~~~~~~~~~~~+(1/4M_p\sqrt{\epsilon_{V}}(\Delta N)^3),\\
\Delta\sigma^{3}_{V}/\Delta\phi&\propto& (\sigma^3_{V}/M_p\sqrt{\epsilon_{V}}\Delta N)
+(8\epsilon_{V}\sqrt{\epsilon_{V}}/M_p\Delta N)\left[1+\epsilon_{V}+(\sqrt{2}/\Delta N)\right]\nonumber\\
&&+8\epsilon^2_{V}\left[(\sqrt{\epsilon_{V}}/M_p\Delta N)+(1/M_p\sqrt{2\epsilon_{V}}(\Delta N)^2)\right]\nonumber\\ 
&& +(\sqrt{\epsilon_{V}}/M_p\sqrt{2}(\Delta N)^2)
\left[\sqrt{2}+(1/2\epsilon_{V}\Delta N)^2\right]\nonumber\\&&
+(1/2M_p(2\epsilon_{V})^{3/2}(\Delta N)^4)+(6\sqrt{\epsilon_{V}}/4M_p(\Delta N)^3),\eea
and 
\bea|\Delta\epsilon_{V}/\Delta N| &\propto& (4\epsilon_{V}/\Delta N),\\
 |\Delta\eta_{V}/\Delta N| &\propto& (8\epsilon_{V}/\Delta N)+(1/\sqrt{2}(\Delta N)^2),\\
|\Delta\xi^{2}_{V}/\Delta N|&\propto& (8\epsilon_{V}/\Delta N)\left[1+2\epsilon_{V}+(\sqrt{2}/\Delta N)\right]
+2\epsilon_{V}\left[(8\epsilon_{V}/\Delta N)+(\sqrt{2}/(\Delta N)^2)\right]
\nonumber\\&&~~~~~~~~~~~~~~~~~~~~~~~~~~~~~~~~~~~~~~~~~~~~~~~~~~~~~~~+(1/\Delta N)^3,\\
|\Delta\sigma^{3}_{V}/\Delta N|&\propto& (2\sigma^3_{V}/\Delta N)+(32\epsilon^2_{V}/\Delta N)\left[1+\epsilon_{V}+(\sqrt{2}/\Delta N)\right]\nonumber\\
&& +8\epsilon^2_{V}\left[(4\epsilon_{V}/\Delta N)+(\sqrt{2}/(\Delta N)^2)\right]\nonumber\\ && +(\sqrt{2}\epsilon_{V}/(\Delta N)^2)
\left[\sqrt{2}+(1/2\epsilon_{V}\Delta N)^2\right]\nonumber\\
&& +(\sqrt{2}/(\Delta N)^4)+(6\sqrt{2}\epsilon_{V}/(\Delta N)^3).\eea
\item Just before the end of inflation $\epsilon_{V}$ and the other slow-roll parameters must eventually increase again.

\item  The change in $\epsilon_{V}$ and other slow-roll parameters must also not be too sharp
in order not to violate slow-roll condition within the prescribed setup.

\item In case of usual monotonic case the end of inflation ($\phi_{e}$) is fixed by the condition:
      \be {\rm max}_{\phi=\phi_{e}}\left[\epsilon_{V},|\eta_{V}|,|\xi^{2}_{V}|,|\sigma^{3}_{V}|\right]\equiv 1.\ee
But in the present case the non-monotonicity of the slow-roll parameters cannot be able to push the maximum field value upto unity at the end of inflation. In this case fig~(\ref{fig3}) suggest that 
\be\label{r1} {\rm max}_{\phi=\phi_{e}}\left[|\eta_{V}|\right]> {\rm max}_{\phi=\phi_{e}}\left[\epsilon_{V}\right]={\rm max}_{\phi=\phi_{e}}\left[|\sigma^{3}_{V}|\right]
>{\rm max}_{\phi=\phi_{e}}
\left[|\xi^{2}_{V}|\right]<<1\ee 
and this also implies that from the maximization of $|\eta_{V}|$ one can compute the field value corresponding to the end of inflation. Eq~(\ref{r1}) can 
also be represented in the from the following constraint conditions:
\bea |V^{''}(\phi_{e})|&<<&\left|\frac{V(\phi_e)}{M^2_{p}}\right|,\\
V^{'}(\phi_{e})&<<&\frac{\sqrt{2}V(\phi_e)}{M_{p}},\\ 
|V^{'}(\phi_{e})|&<<&\sqrt{2|V(\phi_e)V^{''}(\phi_e)|},\\ 
\label{eqas}V(\phi_e)&=&2M^{4}_{p}V^{''''}(\phi_e),\\ 
|V^{''''}(\phi_e)|&>>&\left|\frac{\sqrt{V^{'}(\phi_e)V^{'''}(\phi_e)}}{2M^2_p}\right|,\\ 
|V^{'}(\phi_e)|&>>& 2M^2_p |V^{'''}(\phi_e)|,\\ 
|V(\phi_e)|&>>& M^2_p |V^{'''}(\phi_e)|. \eea
Further substituting Eq~(\ref{rt10a}) in Eq~(\ref{eqas}) finally one can find the field value $(\phi_e-\phi_0)$. 
\end{itemize}

 The most demanding aspect of the conditions is that
the single scalar potential in a non-trivial fashion reduces $\epsilon_{V},\eta_{V},\xi^{2}_{V},\sigma^{3}_{V}$ after
observable scales leaving the horizon at $k>aH$ and subsequently in a non-standard way increases
$\epsilon_{V},\eta_{V},\xi^{2}_{V},\sigma^{3}_{V}$ just before the end of inflation. If inflation is, instead, brought to an end by
another non-trivial mechanism, such as a hybrid transition \cite{Shafi}, then one can relax the fifth condition in the 
present context. This makes the task of reconstructing the inflationary potential simpler. However, in this work I have utilized almost
all of the conditions to reconstruct the structure of inflationary potential.

Most importantly, the condition that $\epsilon_{V}$ first increases and then decreases is not
very uncommon in the context of inflationary model building and can easily be achieved by simply adding a constant vacuum energy correction term to 
a potential that already supports inflation and has increased $\epsilon_{V}$ as well. Before
the constant vacuum energy correction term becomes completely dominant, $\epsilon_{V}$ and the slow-roll parameters will continue to increase as before. 
Then, when the constant term does come to completely dominate,
$\epsilon_{V}$ as well as the other slow-roll parameters approach towards zero. If the contribution from such vacuum energy dominated 
correction term is sufficiently large enough, this transition will occur before the end of inflation. This also suggests that to successfully evade 
the Lyth bound in the present context, the occurrence of this special feature must coincide with the end of 
the observational window and occur with the right magnitude, which is not guaranteed. However, the non-monotonic variation of 
$\epsilon_{V}$ as well as the other slow-roll parameters are the examples of
this physical mechanism, which will take care of all these issues successfully.

 Additionally it is important to note that with non-monotonic evolution of
$\epsilon_{V}$ will modify the power-law feature in the primordial density
perturbations. Apart from an observable tensor-to-scalar ratio, such non-montonicity criteria of the slow-roll parameters generically
 exhibit the following unconventional features in the present context:
\begin{itemize}
 \item Existence of a non-negligible, scale-dependent running $(\alpha_{S})$ and running of the running $(\kappa_{S})$, of the scalar spectral index $(n_{S})$.
 \item Appearance of a significant increase in the primordial power spectrum on very small momentum
scales.
\end{itemize}
  
The non-negligible running and running of the running arises because there must be significant evolution of
$\epsilon_{V}$ as well as the other slow-roll parameters while observable momentum scales are crossing the horizon at $k=aH$. This is
possible in principle for all the evolution of $\epsilon_{V}$ to occur only after
the observable scales have crossed the horizon, it requires sharp
features in $\epsilon_{V}$ and after that the decrement in
$\epsilon_{V}$ and other slow-roll parameters begin, the greater the decrement must be in order to
get sufficient $e$-folds to achieve inflation. If this decrement is huge and very faster,
then the perturbative approach of the scalar fluctuations, and consequently the consistency
of inflation, breaks down in the context of inflationary reconstruction of potential. This problem can only be addressed by a concurrent
sharp fall in the energy scale associated with inflation. For details see ref.~\cite{Adams:1997de} where such a sharp change would be
 described as an inflationary model with multiple periods of inflation, rather than one single period with an evolving $\epsilon_{V}$. Also to explain the 
non-monotonicity of slow-roll parameters the reconstruction technique is implemented in order to impose a sufficient running
and running of the running at the pivot scale. The
necessary evolution of $\epsilon_{V}$ required for consistency of inflation pointing towards the fact that the running and running of the running
will be scale-dependent. This implies that the description of the scalar primordial power spectrum using the simple
power-law parameterization in momentum scale is not sufficient enough to explain non-monotonic evolution of
$\epsilon_{V}$ and as well as other slow-roll parameters to evade the Lyth bound. 
Finally, the increase in the primordial power spectrum at very small momentum scales is an outcome of the
necessary decrement of $\epsilon_{V}$ and other slow-roll parameters outside the observational window. If the
amplitude of the primordial power spectrum on these scales is sufficiently large enough then primordial black
holes (PBHs) will be formed~\footnote{Present cosmological constraints on PBH's restrict
$P_{S}\lesssim 10^{-2}$. See 
refs.~\cite{Carr:2009jm,Alabidi:2009bk,Drees:2011hb} for details.}.


\section{Reconstruction technique of the structure of inflationary effective potential }
\label{w4}

Let us now mention the crucial steps followed during the reconstruction of inflationary potential:-
\begin{itemize}
 \item \underline{\bf STEP I}:\\First of all I need the required data for inflationary observables from an observational
 probe at CMB scale. In our case I use Planck (2013)+WMAP-9+high-L, \textcolor{red}{ Planck (2014)+WMAP-9+high~L+BICEP2 (dust)}, Planck (2015) +WMAP-9+high-L (TT) and 
 Planck (2015)+BICEP2/Keck Array data sets for the numerical estimations.
       Most importantly, I need to implement the reconstruction technique in such a way that I can check the validity and convergence of the proposed methodology using any observational 
       probe. 
 \item \underline{\bf STEP II}:\\Next I assume the slow-roll paradigm within our prescription. Also I demand from the field theoretic requirement that the reconstructed potential is renormalizable
and for this reason the Taylor series expansion of the potential is truncated at the fourth order derivative term.

\item \underline{\bf STEP III}:\\Further I construct the scale and the Taylor series expansion co-efficients of inflationary potential from the derivatives upto fourth order at the CMB scale. 

\item \underline{\bf STEP IV}:\\Hence I use matrix inversion technique to determine the Taylor series expansion co-efficients of the potential around the VEV $\phi_0$ of the inflationary potential.
This is only possible when the system (square) matrix computed from the difference in the field value of the potential at the CMB scale and at VEV is non-singular. 

\item \underline{\bf STEP V}:\\Finally, to determine the various cosmological parameters and to check the validity of the reconstruction technique I fit the reconstructed potential with the observed 
CMB angular power spectra from TT, TE, EE and BB mode obtained from WMAP9, 
Planck (2013, 2014, 2015) alog with BICEP2/Keck Array joint data.
\end{itemize}

Let us now discuss observational constraints on the Taylor expansion coefficients $(V(\phi_0),V^{'}(\phi_0),\cdots)$ as appearing in
Eq~(\ref{rt10a}). 

Let us first write down 
$V(\phi_{\star}),V^{\prime}(\phi_{\star}),V^{\prime\prime}(\phi_{\star}),\cdots$ in terms of the inflationary observables (see Appendix, 
Eqs.~(\ref{ps},~\ref{para 21c},~\ref{para 21e})):
\be\begin{array}{llll}\label{aq1}
\displaystyle V(\phi_{\star})= \frac{3}{2}P_{S}(k_{\star})r(k_{\star})\pi^{2}M^{4}_{p},\\
\displaystyle V^{'}(\phi_{\star})= \frac{3}{2}P_{S}(k_{\star})r(k_{\star})\pi^{2}\sqrt{\frac{r(k_{\star})}{8}}M^{3}_{p},\\
\displaystyle V^{''}(\phi_{\star})= \frac{3}{4}P_{S}(k_{\star})r(k_{\star})\pi^{2}\left(n_{S}(k_{\star})-1+\frac{3r(k_{\star})}{8}\right)M^{2}_{p},\\
\displaystyle V^{'''}(\phi_{\star})= \frac{3}{2}P_{S}(k_{\star})r(k_{\star})\pi^{2}\left[\sqrt{2r(k_{\star})}\left(n_{S}(k_{\star})-1+\frac{3r(k_{\star})}{8}\right)
\right.\\ \left.~~~~~~~~~~~~~~~~~~~~~~~~~~~~~~~~~~~~~~~~~~~~~~~~~~~~~~~~~~~~\displaystyle -\frac{1}{2}
\left(\frac{r(k_{\star})}{8}\right)^{\frac{3}{2}}-\alpha_{S}(k_{\star})\sqrt{\frac{2}{r(k_{\star})}}\right]M_{p},\\
\displaystyle V^{''''}(\phi_{\star})=12P_{S}(k_{\star})\pi^{2}\left\{\frac{\kappa_{S}(k_{\star})}{2}-
\frac{1}{2}\left(\frac{r(k_{\star})}{8}\right)^{2}\left(n_{S}(k_{\star})-1+\frac{3r(k_{\star})}{8}\right)\right.\\ \left.\displaystyle 
~~~~~~~~~~~~~~~~~~~~~~~~~~~~~~~~~~+12\left(\frac{r(k_{\star})}{8}\right)^{3}+r(k_{\star})\left(n_{S}(k_{\star})-1+\frac{3r(k_{\star})}{8}\right)^{2}
\right.\\ \left.\displaystyle 
~~~~~~~~~~~~~~~~~~~~~~~~~~~~~~~~~~+\left[\sqrt{2r(k_{\star})}\left(n_{S}(k_{\star})-1+\frac{3r(k_{\star})}{8}\right)
\right.\right.\\ \left.\left.~~~~~~~~~~~~~~~~~~~~~~~~~~~~~~~~~~~~~~~~~~~~~~~~~~~~~~~~~~~~\displaystyle -\frac{1}{2}
\left(\frac{r(k_{\star})}{8}\right)^{\frac{3}{2}}-\alpha_{S}(k_{\star})\sqrt{\frac{2}{r(k_{\star})}}\right]\right.\\ \left.
\displaystyle 
~~~~~~~~~~~~~~~~~~~~~~~~~~~~~~~~~~\times \left[\sqrt{\frac{r(k_{\star})}{8}}\left(n_{S}(k_{\star})-1+\frac{3r(k_{\star})}{8}\right)
-6\left(\frac{r(k_{\star})}{8}\right)^{\frac{3}{2}}\right]\right\}
\end{array}\ee

Therefore, for any chosen sub-Planckian VEV of $\phi_{\star}$, I can obtain a matrix  equation characterizing the coefficients:
 $V(\phi_0),~V^{\prime}(\phi_0),~V^{\prime\prime}(\phi_0),\cdots$:
\begin{equation}\label{mat1}\underbrace{\left(\begin{tabular}{cccccc}
 $1$ & $\vartheta_{\star}$ & $\frac{\vartheta^{2}_{\star}}{2}$& $\frac{\vartheta^{3}_{\star}}{6}$ & $\frac{\vartheta^{4}_{\star}}{24}$\\
$0$ & $1$ & $\vartheta_{\star}$& $\frac{\vartheta^{2}_{\star}}{2}$ & $\frac{\vartheta^{3}_{\star}}{6}$\\
$0$ & 0 & 1& $\vartheta_{\star}$ & $\frac{\vartheta^{2}_{\star}}{2}$\\
$0$ & $0$ & $0$& $1$ & $\vartheta_{\star}$\\
$0$ & $0$ & $0$& $0$ & $1$
      \end{tabular}\right)}
\left(\begin{tabular}{c}
      $V(\phi_{0})$\\
      $V^{'}(\phi_0)$\\
      $V^{''}(\phi_0)$\\
      $V^{'''}(\phi_0)$\\
      $V^{''''}(\phi_0)$\\
      \end{tabular}\right)
\begin{tabular}{c}
      \\
      =\\
      \\
      \end{tabular}
\left(\begin{tabular}{c}
      $V(\phi_{\star})$\\
      $V^{'}(\phi_{\star})$\\
      $V^{''}(\phi_{\star})$\\
     $V^{'''}(\phi_{\star})$\\
      $V^{''''}(\phi_{\star})$\\
      \end{tabular}\right)\,,
\end{equation}
where I have define $$\vartheta_{\star}:=(\phi_{\star}-\phi_0)< M_p.$$
The square matrix marked by the symbol $\underbrace{\cdots}$ is nonsingular,
since its determinant is nonzero, for which the matrix inversion technique is applicable in
the present context.
Finally, I get the following physical solution to the problem:
\begin{equation}\label{mat2}
\left(\begin{tabular}{c}
      $V(\phi_{0})$\\
      $V^{'}(\phi_0)$\\
      $V^{''}(\phi_0)$\\
      $V^{'''}(\phi_0)$\\
      $V^{''''}(\phi_0)$\\
      \end{tabular}\right)
\begin{tabular}{c}
      \\
      =\\
      \\
      \end{tabular}
\left(\begin{tabular}{cccccc}
 $1$ & $-\vartheta_{\star}$ & $\frac{\vartheta^{2}_{\star}}{2}$& $-\frac{\vartheta^{3}_{\star}}{6}$ & $\frac{\vartheta^{4}_{\star}}{24}$\\
$0$ & $1$ & $-\vartheta_{\star}$& $\frac{\vartheta^{2}_{\star}}{2}$ & $-\frac{\vartheta^{3}_{\star}}{6}$\\
$0$ & 0 & 1& $-\vartheta_{\star}$ & $\frac{\vartheta^{2}_{\star}}{2}$\\
$0$ & $0$ & $0$& $1$ & $-\vartheta_{\star}$\\
$0$ & $0$ & $0$& $0$ & $1$
      \end{tabular}\right)
\left(\begin{tabular}{c}
      $V(\phi_{\star})$\\
      $V^{'}(\phi_{\star})$\\
      $V^{''}(\phi_{\star})$\\
     $V^{'''}(\phi_{\star})$\\
      $V^{''''}(\phi_{\star})$\\
      \end{tabular}\right).
\end{equation}
For a model independent constraint on the shape of the potential, the parameter, \be \vartheta_{\star}\sim {\cal O}(10^{-1}M_{p}),\ee as \be \Delta\phi\sim {\cal O}(10^{-1}M_p),\ee
 which is applicable 
to a large class of sub-Planckian inflationary models to generate sufficient amount of inflation within $50<N_{total}<70$. This criteria holds good for high scale inflation, but within the regime of sub-Planckian cut-off.

In order to satisfy the preferred bounds, see Eq.~(\ref{obscons1a})-Eq.~(\ref{obscons4e}),
 the following model independent theoretical constraints 
on $V(\phi_{\star}),V^{'}(\phi_{\star}),\cdots$ have to be imposed:
\\
\underline{\bf Planck (2013)+WMAP-9+high~L:}
 \bea\label{constraint1}
 V(\phi_{\star})\leq {\cal O}(3.79-3.99)\times 10^{-9}M^{4}_{p},\\  
   \label{constraint2}
  V^{'}(\phi_{\star})\leq {\cal O}(4.65-4.89)\times 10^{-10}M^{3}_{p}, \\
\label{constraint3}
   V^{''}(\phi_{\star})\leq {\cal O}((-0.41)-2.42)\times 10^{-11}M^{2}_{p}, \\
\label{constraint4}
  V^{'''}(\phi_{\star})\leq {\cal O}(8.52-35.1)\times 10^{-11}M_{p}, \\
\label{constraint5}
    V^{''''}(\phi_{\star})\leq {\cal O}(0.39-4.76)\times 10^{-9},
   \eea
 \underline{\bf\textcolor{red}{ Planck (2014)+WMAP-9+high~L+BICEP2 (dust)}:}
 \bea\label{constraint6}
   5.27\times 10^{-9}M^{4}_{p}\leq V(\phi_{\star})\leq 9.52\times 10^{-9}M^{4}_{p},\\  
   \label{constraint7}
   2.45\times 10^{-10}M^{3}_{p}\leq V^{'}(\phi_{\star})\leq 1.75\times 10^{-9}M^{3}_{p}, \\
\label{constraint8}
   2.41\times 10^{-11}M^{2}_{p}\leq V^{''}(\phi_{\star})\leq 3.25\times 10^{-10}M^{2}_{p}, \\
\label{constraint9}
   6.35\times 10^{-10}M_{p}\leq V^{'''}(\phi_{\star})\leq 7.56\times 10^{-10}M_{p}, \\
\label{constraint10}
   5.56\times 10^{-10}\leq V^{''''}(\phi_{\star})\leq 4.82\times 10^{-9},
   \eea
\underline{\bf Planck (2015)+WMAP-9+high~L(TT):}
 \bea\label{constraint11}
   V(\phi_{\star})\leq {\cal O}(3.45-3.71)\times 10^{-9}M^{4}_{p},\\  
   \label{constraint12}
   V^{'}(\phi_{\star})\leq {\cal O}(4.04-4.35)\times 10^{-10}M^{3}_{p}, \\
\label{constraint13}
   V^{''}(\phi_{\star})\leq {\cal O}((-2.93)-1.08)\times 10^{-11}M^{2}_{p}, \\
\label{constraint14}
   V^{'''}(\phi_{\star})\leq {\cal O}((-0.48)-(-3.88))\times 10^{-10}M_{p}, \\
\label{constraint15}
   V^{''''}(\phi_{\star})\leq {\cal O}(5.52-5.76)\times 10^{-9},
   \eea
\underline{\bf Planck (2015)+BICEP2/Keck Array:}
 \bea\label{constraint16}
  V(\phi_{\star})\leq {\cal O}(3.80-3.99)\times 10^{-9}M^{4}_{p},\\  
   \label{constraint17}
   V^{'}(\phi_{\star})\leq {\cal O}(4.65-4.89)\times 10^{-10}M^{3}_{p}, \\
\label{constraint18}
    V^{''}(\phi_{\star})\leq {\cal O}((-0.41)-2.42)\times 10^{-11}M^{2}_{p}, \\
\label{constraint19}
    V^{'''}(\phi_{\star})\leq {\cal O}(8.52-35.1)\times 10^{-10}M_{p}, \\
\label{constraint20}
    V^{''''}(\phi_{\star})\leq {\cal O}(0.39-4.76)\times 10^{-9},
   \eea
Now, substituting the above expressions in Eq~(\ref{mat2}), I obtain model independent constraints on the Taylor expansion 
co-efficients at $\phi=\phi_{0}$ i.e. $V(\phi_0),V^{\prime}(\phi_0),\cdots$ as:
\\
\underline{\bf Planck (2013)+WMAP-9+high~L:}
 \bea\label{constraint1}
 V(\phi_{0})\leq {\cal O}(3.79-3.94)\times 10^{-9}M^{4}_{p},\\  
   \label{constraint2}
  V^{'}(\phi_{0})\leq {\cal O}(4.66-4.88)\times 10^{-10}M^{3}_{p}, \\
\label{constraint3}
   V^{''}(\phi_{0})\leq {\cal O}((-1.07)-1.29)\times 10^{-11}M^{2}_{p}, \\
\label{constraint4}
  V^{'''}(\phi_{0})\leq {\cal O}(8.13-(-1.25))\times 10^{-10}M_{p}, \\
\label{constraint5}
    V^{''''}(\phi_{0})\leq {\cal O}(0.39-4.76)\times 10^{-9},
   \eea
 \underline{\bf\textcolor{red}{ Planck (2014)+WMAP-9+high~L+BICEP2 (dust)}:}
 \bea\label{constraint25}
   5.26\times 10^{-9}M^{4}_{p}\leq V(\phi_0)\leq 9.50\times 10^{-9}M^{4}_{p},\\  
   \label{constraint26}
   2.44\times 10^{-10}M^{3}_{p}\leq V^{'}(\phi_0)\leq 1.74\times 10^{-9}M^{3}_{p}, \\
\label{constraint27}
   2.10\times 10^{-11}M^{2}_{p}\leq V^{''}(\phi_0)\leq 3.22\times 10^{-10}M^{2}_{p}, \\
\label{constraint28}
   6.29\times 10^{-10}M_{p}\leq V^{'''}(\phi_0)\leq 7.08\times 10^{-10}M_{p}, \\
\label{constraint29}
   5.56\times 10^{-10}\leq V^{''''}(\phi_0)\leq 4.82\times 10^{-9},
   \eea
\underline{\bf Planck (2015)+WMAP-9+high~L(TT):}
 \bea\label{constraint11}
   V(\phi_{0})\leq {\cal O}(3.41-3.67)\times 10^{-9}M^{4}_{p},\\  
   \label{constraint12}
   V^{'}(\phi_{0})\leq {\cal O}(4.06-4.31)\times 10^{-10}M^{3}_{p}, \\
\label{constraint13}
   V^{''}(\phi_{0})\leq {\cal O}(0.31-7.84)\times 10^{-11}M^{2}_{p}, \\
\label{constraint14}
   V^{'''}(\phi_{0})\leq {\cal O}((-6.00)-(-9.64))\times 10^{-10}M_{p}, \\
\label{constraint15}
   V^{''''}(\phi_{0})\leq {\cal O}(5.52-5.76)\times 10^{-9},
   \eea
\underline{\bf Planck (2015)+BICEP2/Keck Array:}
 \bea\label{constraint16}
  V(\phi_{0})\leq {\cal O}(3.75-3.95)\times 10^{-9}M^{4}_{p},\\  
   \label{constraint17}
   V^{'}(\phi_{0})\leq {\cal O}(4.70-5.03)\times 10^{-10}M^{3}_{p}, \\
\label{constraint18}
    V^{''}(\phi_{0})\leq {\cal O}((-8.74)-30.30)\times 10^{-11}M^{2}_{p}, \\
\label{constraint19}
    V^{'''}(\phi_{0})\leq {\cal O}(8.13-30.34)\times 10^{-10}M_{p}, \\
\label{constraint20}
    V^{''''}(\phi_{0})\leq {\cal O}(0.39-4.76)\times 10^{-9},
   \eea
Consequently, the slow-roll parameters $(\epsilon_{V},\eta_{V},\xi^{2}_{V},\sigma^{3}_{V})$ are constrained by:
\\
\underline{\bf Planck (2013)+WMAP-9+high~L:}
 \bea\label{constraint21}
   \epsilon_{V}\lesssim {\cal O}(7.66-7.72)\times 10^{-3},\\  
   \label{constraint22}
   |\eta_{V}|\lesssim{\cal O}(6.14\times 10^{-3}-0.019), \\
\label{constraint23}
   |\xi^{2}_{V}|\lesssim{\cal O}(2.34\times 10^{-6}-0.027), \\
\label{constraint24}
   |\sigma^{3}_{V}|\lesssim {\cal O}(1.58\times 10^{-3}-0.019).
\eea
 \underline{\bf\textcolor{red}{ Planck (2014)+WMAP-9+high~L+BICEP2 (dust)}:}
 \bea\label{constraint25}
   \epsilon_{V}\sim {\cal O}(0.10-1.69)\times 10^{-2},\\  
   \label{constraint26}
   |\eta_{V}|\sim{\cal O}(4.57\times 10^{-3}-0.030), \\
\label{constraint27}
   |\xi^{2}_{V}|\sim{\cal O}(5.60\times 10^{-3}-0.014), \\
\label{constraint28}
   |\sigma^{3}_{V}|\sim {\cal O}(2.28\times 10^{-4}-0.017).
\eea
\underline{\bf Planck (2015)+WMAP-9+high~L(TT):}
 \bea\label{constraint29}
   \epsilon_{V}\lesssim {\cal O}(7.02-7.03)\times 10^{-3},\\  
   \label{constraint30}
   |\eta_{V}|\lesssim{\cal O}(2.94-8.59)\times 10^{-3}, \\
\label{constraint31}
   |\xi^{2}_{V}|\lesssim{\cal O}(1.52\times 10^{-3}-0.012), \\
\label{constraint32}
   |\sigma^{3}_{V}|\lesssim {\cal O}(0.022-0.023).
\eea
\underline{\bf Planck (2015)+BICEP2/Keck Array:}
 \bea\label{constraint33}
   \epsilon_{V}\lesssim {\cal O}(7.70-9.67)\times 10^{-3},\\  
   \label{constraint34}
   |\eta_{V}|\lesssim{\cal O}(5.20\times 10^{-4}-0.160), \\
\label{constraint35}
   |\xi^{2}_{V}|\lesssim{\cal O}(5.60\times 10^{-3}-0.123), \\
\label{constraint36}
   |\sigma^{3}_{V}|\lesssim {\cal O}(1.60\times 10^{-3}-0.023).
\eea
Further, by applying the joint constraints from Planck (2013)+WMAP-9+high~L, \textcolor{red}{ Planck (2014)+WMAP-9+high~L+BICEP2 (dust)}, Planck (2015)+WMAP-9+high~L(TT) and Planck (2015)+BICEP2/Keck Array I obtain the following model
 independent bound on field excursion $|\Delta\phi|/M_p$ by using Eq~(\ref{con10sd}) or Eq~(\ref{con13}):  
\\
\underline{\bf Planck (2013)+WMAP-9+high~L:}
 \be\begin{array}{llll}\label{con10sdqx1}
    \displaystyle \frac{\left |\Delta\phi\right|}{M_p}\,\leq {\cal O}(0.239-0.241)~~~{\bf  for}~~~~ r_{\star}\leq 0.12.
   \end{array}\ee
 \underline{\bf\textcolor{red}{ Planck (2014)+WMAP-9+high~L+BICEP2 (dust)}:}
 \be\begin{array}{llll}\label{con10sdqx2}
    \displaystyle 0.242 \leq\frac{\left |\Delta\phi\right|}{M_p}\,\leq 0.354~~~{\bf  for}~~~~ 0.15\leq r_{\star}\leq 0.27.
   \end{array}\ee
\underline{\bf Planck (2015)+WMAP-9+high~L(TT):}
   \be\begin{array}{llll}\label{con10sdqx3}
    \displaystyle \frac{\left |\Delta\phi\right|}{M_p}\,\leq {\cal O}(0.230-0.231)~~~{\bf  for}~~~~ r_{\star}\leq 0.11.
   \end{array}\ee
\underline{\bf Planck (2015)+BICEP2/Keck Array:}
 \be\begin{array}{llll}\label{con10sdqx4}
    \displaystyle \frac{\left |\Delta\phi\right|}{M_p}\,\leq {\cal O}(0.223-0.242)~~~{\bf  for}~~~~ r_{\star}\leq 0.12.
   \end{array}\ee

Finally following the present analysis I get the following
 constraints on the tensor spectral tilt, $n_{T}$, running of the tensor spectral tilt, $\alpha_{T}$, the
running of the tensor-to-scalar ratio $n_{r}$ and running of the running of tensor spectral tilt $\kappa_{T}$
and tensor-to-scalar ratio $\kappa_{r}$ as:
\\
\underline{\bf Planck (2013)+WMAP-9+high~L:}
 \bea\label{wq1}
n_{T}<{\cal O}((-0.0151)-(-0.0154)),\\
\label{wq2}
\alpha_{T}<{\cal O}((-2.69)-(-4.815))\times 10^{-4},\\
\label{wq3}
|n_{r}|<{\cal O}(3.00-3.23)\times 10^{-4} ,\\
\label{wq4}
\kappa_{T}<{\cal O}((-5.92)-(-32.20))\times 10^{-5},\\
\label{wq5}
\kappa_{r}<{\cal O}((-6.86)-14.53)\times 10^{-4},
\eea
 \underline{\bf\textcolor{red}{ Planck (2014)+WMAP-9+high~L+BICEP2 (dust)}:}
 \bea\label{wq6}
-0.019<n_{T}<-0.033,\\
\label{wq7}
-2.97\times 10^{-4}<\alpha_{T}<2.86\times 10^{-5},\\
\label{wq8}
2.28\times 10^{-4}<|n_{r}|<0.010 ,\\
\label{wq9}
-0.11\times 10^{-4}<\kappa_{T}<-3.58\times 10^{-4},\\
\label{wq10}
-5.25\times 10^{-3}<\kappa_{r}<-6.27\times 10^{-3},
\eea
\underline{\bf Planck (2015)+WMAP-9+high~L(TT):}
   \bea\label{wq11}
n_{T}<{\cal O}((-0.0140)-(-0.0142)),\\
\label{wq12}
\alpha_{T}<{\cal O}((-2.98)-(-5.09))\times 10^{-4},\\
\label{wq13}
|n_{r}|<{\cal O}(2.47\times 10^{-3}) ,\\
\label{wq14}
\kappa_{T}<{\cal O}((-0.07)-3.46)\times 10^{-4},\\
\label{wq15}
\kappa_{r}<{\cal O}((-0.14)-2.89)\times 10^{-3},
\eea
\underline{\bf Planck (2015)+BICEP2/Keck Array:}
 \bea\label{wq16}
n_{T}<{\cal O}((-0.0153)-(-0.0154)),\\
\label{wq17}
\alpha_{T}<{\cal O}((-2.69)-(-4.82))\times 10^{-4},\\
\label{wq18}
|n_{r}|<{\cal O}(2.11-3.75)\times 10^{-3} ,\\
\label{wq19}
\kappa_{T}<{\cal O}((-1.79)-(-4.72))\times 10^{-4},\\
\label{wq20}
\kappa_{r}<{\cal O}((-0.20)-(-4.49))\times 10^{-3},
\eea
%
 Further, if I had set $\vartheta$ to a slightly
larger value, $\vartheta\sim {\cal O}(10^{-1}~M_{p})$ as $\Delta\phi\sim {\cal O}(10^{-1}~M_{p})$, then the order of magnitude
 of the numerics would not change, but the numerical pre-factors would slightly change.


\section{Higher order consistency relationships in effective theory} 
\label{w5}
Let us now provide the new set of consistency relationships {\it between slow roll parameters} for sub-Planckian models of inflation:  
\begin{eqnarray}\label{wq6}n_{T}&=&-\frac{r}{8}\left(2-\frac{r}{8}-n_{S}\right)+\cdots,\\
\label{wq7}\alpha_{T}&=&\frac{dn_{T}}{d\ln k}=\frac{r}{8}\left(\frac{r}{8}+n_{S}-1\right)+\cdots,\\
\label{wq8} n_{r}&=&\frac{dr}{d\ln k}=\frac{16}{9}\left(n_{S}-1+\frac{3r}{4}\right)\left(2n_{S}-2+\frac{3r}{8}\right)+\cdots,\\
\label{wq9} \kappa_{T}&=&\frac{d^{2}n_{T}}{d\ln k^{2}}=\frac{2}{9}\left(n_{S}-1+\frac{3r}{4}\right)\left(2n_{S}-2+\frac{3r}{8}\right)\left(\frac{r}{8}+n_{S}-1\right)\\
&&~~~~~~~~~~~~~~~~~~+\frac{r}{8}\left[\alpha_{S}+\frac{2}{9}\left(n_{S}-1+\frac{3r}{4}\right)\left(2n_{S}-2+\frac{3r}{8}\right)\right]+\cdots,\\
\label{wq10} \kappa_{r}&=&\frac{d^{2}r}{d\ln k^{2}}\nonumber\\
&=&\frac{16}{9}\left(2n_{S}-2+\frac{3r}{8}\right)\left\{\alpha_{S}+\frac{4}{3}\left(n_{S}-1+\frac{3r}{4}\right)\left(2n_{S}-2+\frac{3r}{8}\right)\right\}\\
&&~~~~~~+\frac{16}{9}\left(n_{S}-1+\frac{3r}{4}\right)\left\{2\alpha_{S}+\frac{2}{3}\left(n_{S}-1+\frac{3r}{4}\right)\left(2n_{S}-2+\frac{3r}{8}\right)\right\}+\cdots.\nonumber
\end{eqnarray}
One can compare these relationships with respect to super-Planckian models of inflation where the slow roll parameters vary monotonically, see Re.~\cite{Lyth}.
Observationally, now one can differentiate sub versus super Planckian excursion models of inflation with the help of the above consistency relationships.
In particular, the slope of the tensor modes, see Eq.~(\ref{wq6}),  will play a crucial role in deciding the fate of the sub-primordial inflation in the early universe.

In Fig.~(\ref{fig3}) and Fig.~(\ref{fig3new}), I have shown the evolution of $\epsilon_V,~|\eta_V|,~|\xi^2_V|,~|\sigma_V^3|$ (see Eqs.~(\ref{slpara})) with respect to 
$|\phi-\phi_0|$ and number of e-foldings $N$ respectively, where the upper and lower bounds are given by Eqs.~(\ref{constraint1}-\ref{constraint20}). In particular, note that the evolution of $\epsilon_V$ is
non-monotonic for sub-Planckian inflation, which is in stark contrast 
with the Lyth-bound for the super-Planckian models of inflation, where $\epsilon_V$ can evolve monotonically for polynomial potentials~\cite{Lyth}. In future 
the data would be  would be sufficiently good to compare the running of the gravitational tensor perturbations, $n_T$, for sub-vs-super-Planckian 
excursions of the inflaton.

\begin{figure}[t]
\centering
\subfigure[$P_{T}$~vs~$k$]{
    \includegraphics[width=7cm, height=6.7cm] {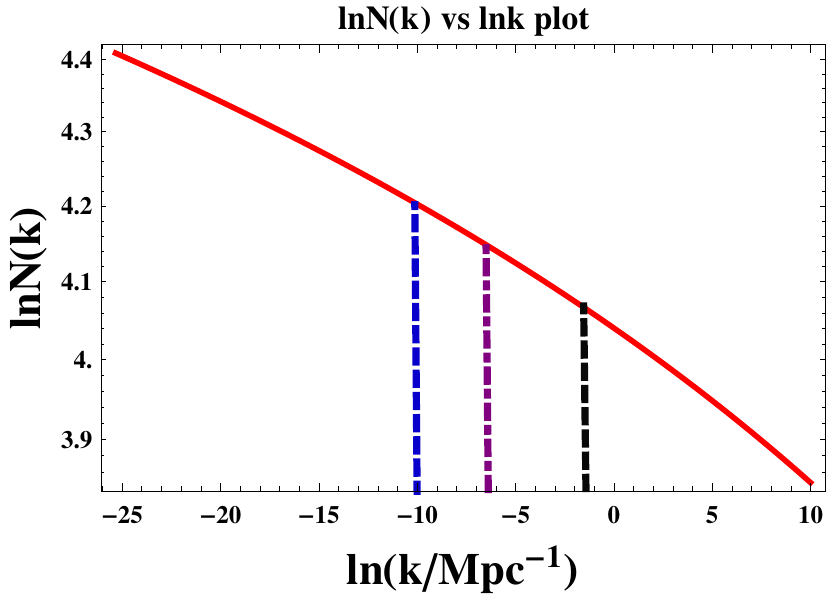}
    \label{fig:subfigz1a}
}
\subfigure[$P_{T}$~vs~$k$]{
    \includegraphics[width=7cm, height=6.7cm] {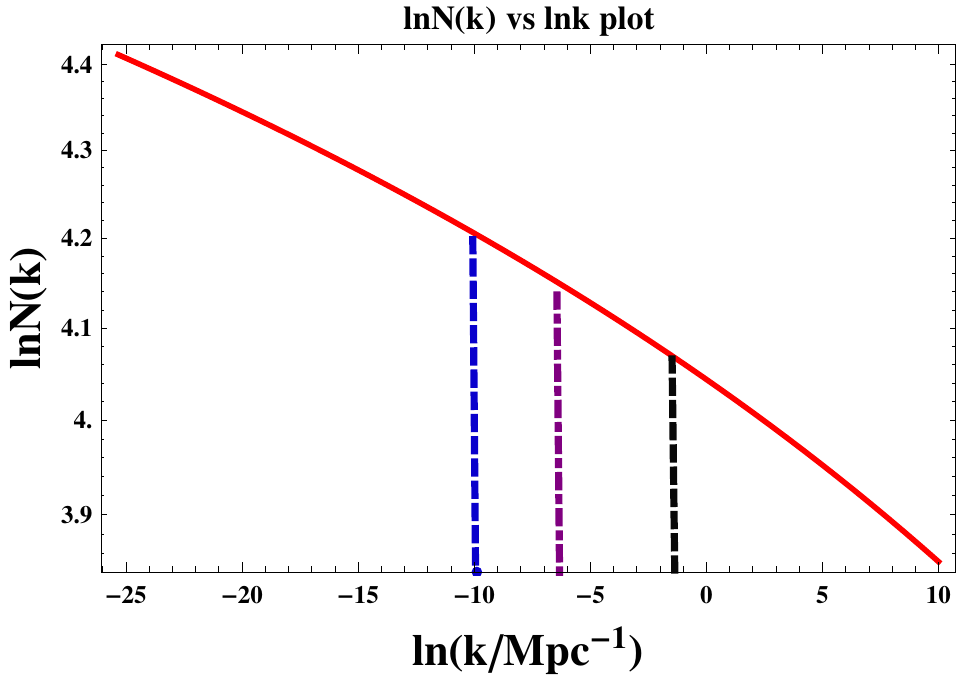}
    \label{fig:subfigz2a}
}
\subfigure[$r(k)$~vs~$k$]{
    \includegraphics[width=7cm, height=6.7cm] {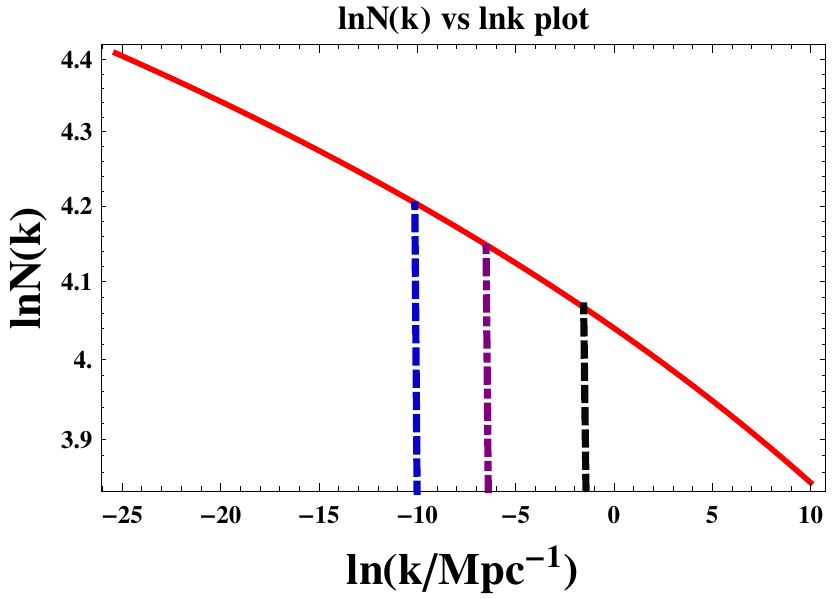}
    \label{fig:subfigz3a}
}
\subfigure[$r(k)$~vs~$k$]{
    \includegraphics[width=7cm, height=6.7cm] {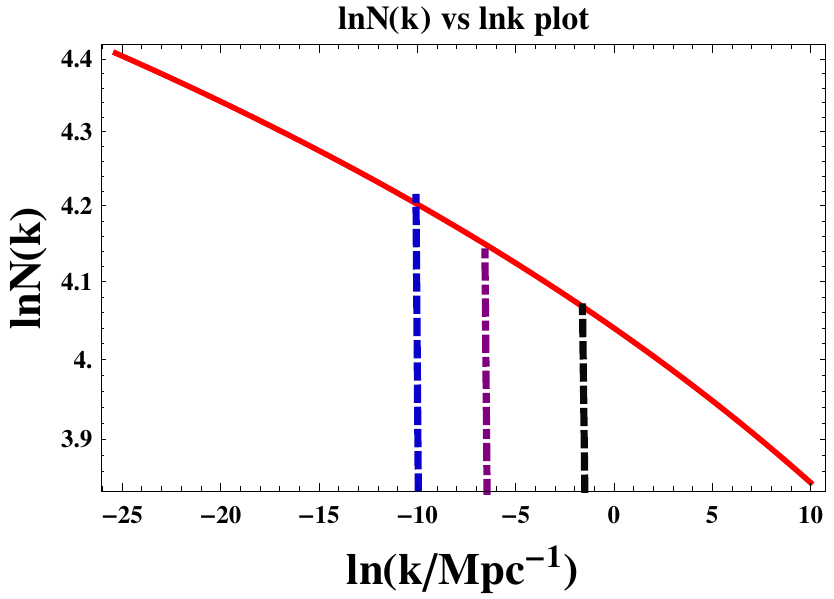}
    \label{fig:subfigz4a}
}
\caption[Optional caption for list of figures]{ Here I show the variation of the total number of e-foldings $N(k)$, with respect to the momentum scale $k$.
The {\bf black} dotted line corresponds to $k_{max}=0.3~{\rm Mpc}^{-1}$ for $l_{max}=2500$, the \textcolor{blue}{\bf blue} dotted line corresponds to
$k_{min}=4.488\times 10^{-5}~{\rm Mpc}^{-1}$ for $l_{min}=2$,  and in all the plots \textcolor{violet}{\bf violet} dashed dotted line
 represents the pivot scale of momentum at $k_{\star}=0.002~{\rm Mpc}^{-1}$ for $l_{\star}\sim 80$, at which
 for \ref{fig:subfigz1a} $P_{S}(k_{\star})=2.2\times 10^{-9}$, $n_{S}=0.9600$, $\alpha_{S}=-0.013$ and $N(k_{\star})=63.26$ (Planck (2013)+WMAP-9+high-L), 
\ref{fig:subfigz2a} $P_{S}(k_{\star})=2.2\times 10^{-9}$, $n_{S}=0.9600$, $\alpha_{S}=-0.022$ and $N(k_{\star})=63.26$ (Planck (2014)+WMAP-9+high-L+BICEP2 (dust)),
\ref{fig:subfigz3a} $P_{S}(k_{\star})=2.2\times 10^{-9}$, $n_{S}=0.9569$, $\alpha_{S}=0.011$ and $N(k_{\star})=63.26$ (Planck (2015)+WMAP-9+high-L(TT)),
\ref{fig:subfigz4a} $P_{S}(k_{\star})=2.2\times 10^{-9}$, $n_{S}=0.9600$, $\alpha_{S}=-0.022$ and $N(k_{\star})=63.26$ (Planck (2015)+BICEP2/Keck Array). Within $2<l<2500$ the
 value of the required momentum scale is determined by the relation,
$k_{reqd}\sim \frac{l_{reqd}}{\eta_{0}\pi}$ \cite{Choudhury:2013jya}, where the conformal time at the present epoch is $\eta_{0}\sim 14000~{\rm Mpc}$.
}
\label{figv1}
\end{figure}

\begin{figure}[t]
\centering
\subfigure[$P_{S}$~vs~$k$]{
    \includegraphics[width=10.5cm, height=7.8cm] {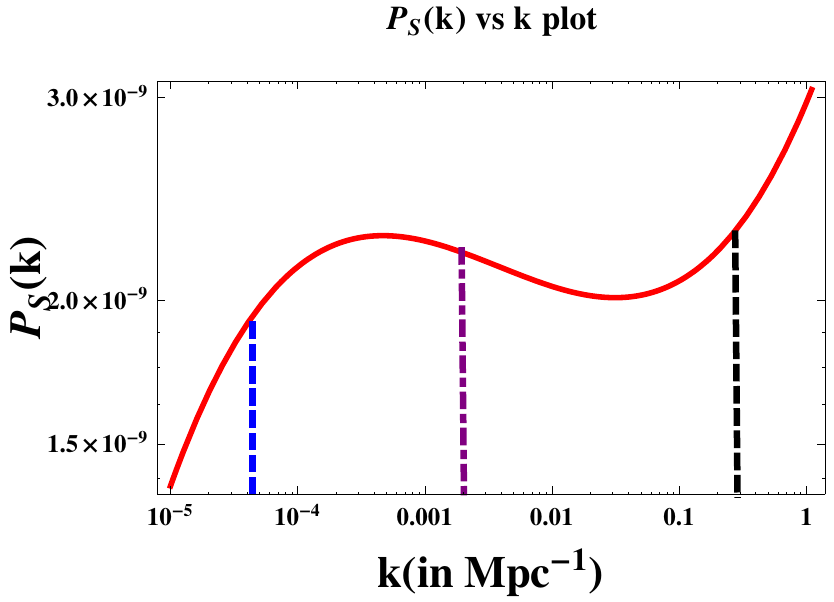}
    \label{fig:subfigv5a}
}
\subfigure[$n_{S}$~vs~$k$]{
    \includegraphics[width=7cm, height=7cm] {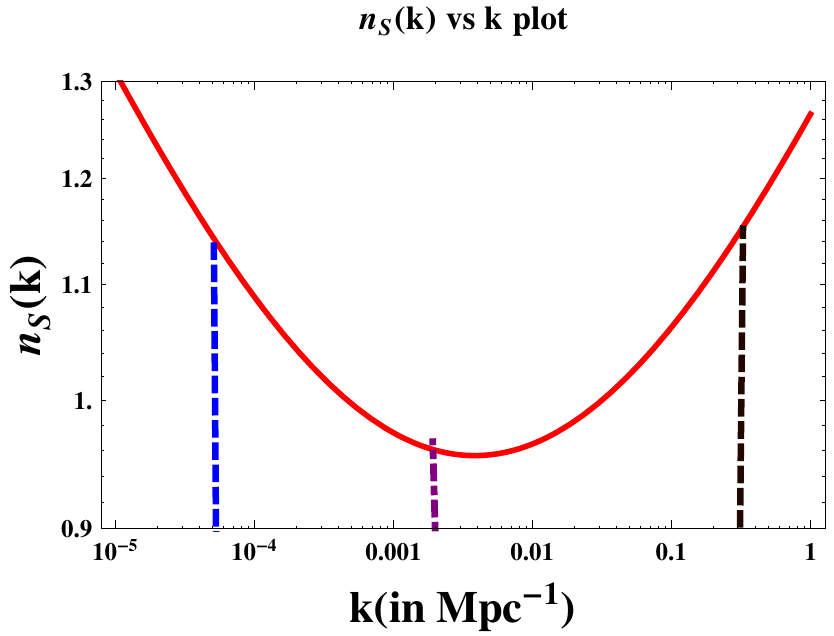}
    \label{fig:subfigv6a}
}
\subfigure[$\alpha_{S}$~vs~$k$]{
    \includegraphics[width=7cm, height=7cm] {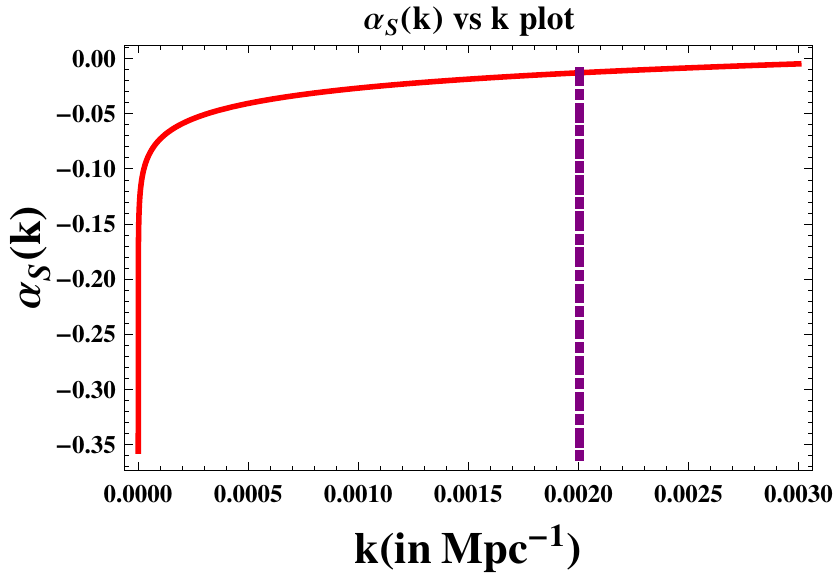}
    \label{fig:subfigv7a}
}
\caption[Optional caption for list of figures]{ In \ref{fig:subfigv5a}, I show the scalar Power spectrum $P_{s}(k)$ \ref{fig:subfigv6a}, I show the scalar spectral index $n_{s}(k)$,
 and in \ref{fig:subfigv7a}, I show the running of the scalar spectral index $\alpha_{s}(k)$, with respect to the momentum scale $k$.
The {\bf black} dotted line corresponds to $k_{max}=0.3~{\rm Mpc}^{-1}$ for $l_{max}=2500$, the \textcolor{blue}{\bf blue} dotted line corresponds to
$k_{min}=4.488\times 10^{-5}~{\rm Mpc}^{-1}$ for $l_{min}=2$,  and in all the plots \textcolor{violet}{\bf violet} dashed dotted line
 represents the pivot scale of momentum at $k_{\star}=0.002~{\rm Mpc}^{-1}$ for $l_{\star}\sim 80$ at which Planck (2013)+WMAP-9+high L constraints
 $P_{S}(k_{\star})=2.2\times 10^{-9}$, $n_{S}=0.9600$, $\alpha_{S}=-0.02$ and $N(k_{\star})=63.26$ are satisfied. Within $2<l<2500$ the
 value of the required momentum scale is determined by the relation,
$k_{reqd}\sim \frac{l_{reqd}}{\eta_{0}\pi}$ \cite{Choudhury:2013jya}, where the conformal time at the present epoch is $\eta_{0}\sim 14000~{\rm Mpc}$.
}
\label{figv2a}
\end{figure}

\begin{figure}[t]
\centering
\subfigure[$P_{S}$~vs~$k$]{
    \includegraphics[width=10.5cm, height=7.8cm] {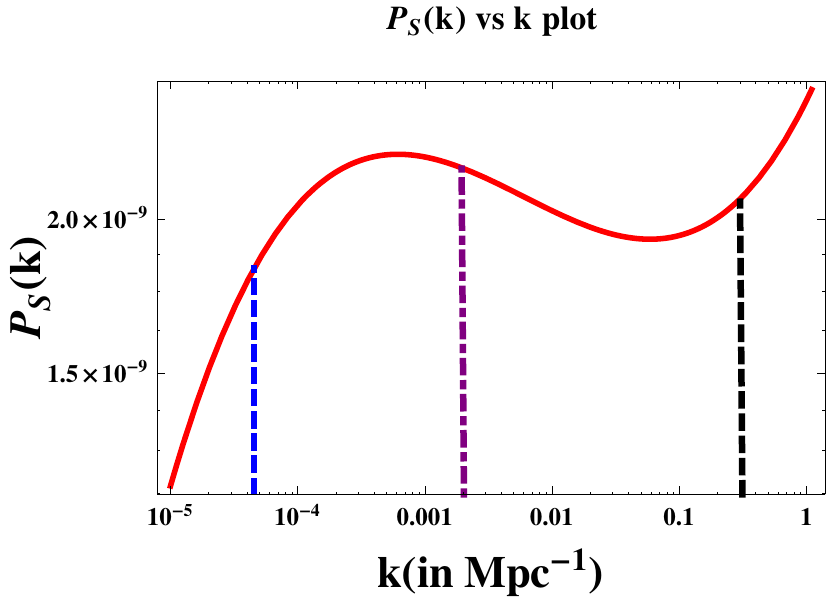}
    \label{fig:subfigv5}
}
\subfigure[$n_{S}$~vs~$k$]{
    \includegraphics[width=7cm, height=7cm] {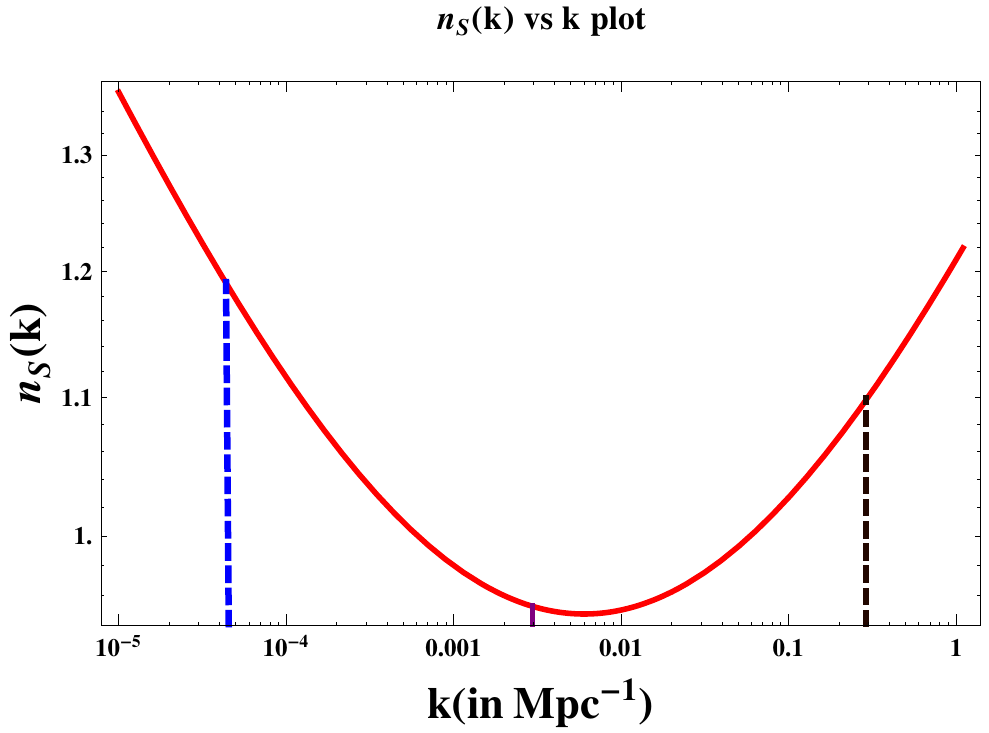}
    \label{fig:subfigv6}
}
\subfigure[$\alpha_{S}$~vs~$k$]{
    \includegraphics[width=7cm, height=7cm] {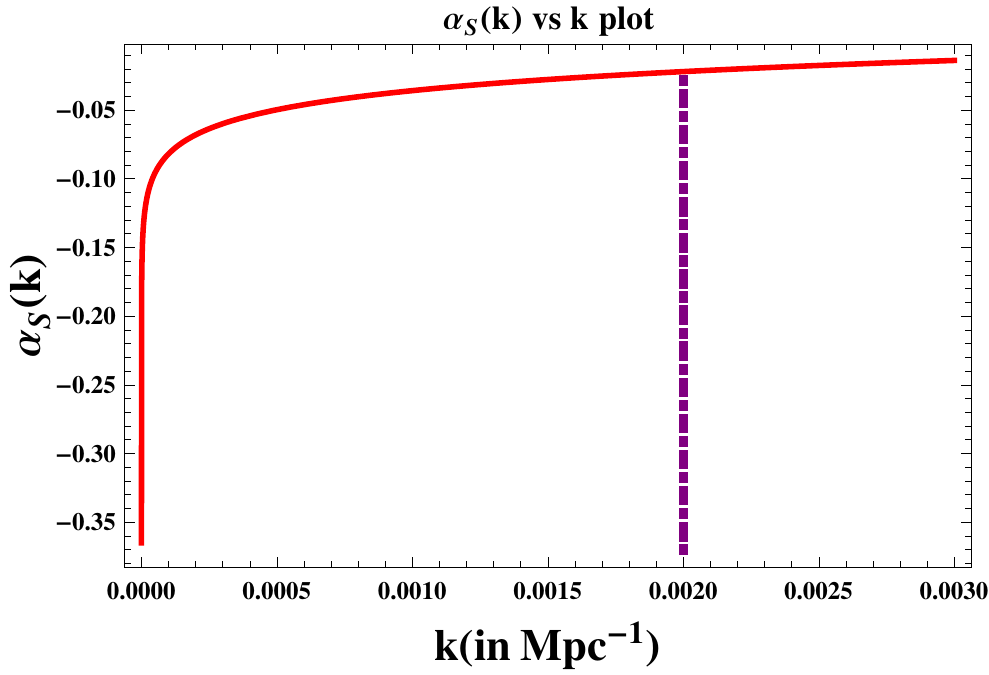}
    \label{fig:subfigv7}
}
\caption[Optional caption for list of figures]{ In \ref{fig:subfigv5}, I show the scalar Power spectrum $P_{s}(k)$ \ref{fig:subfigv6}, I show the scalar spectral index $n_{s}(k)$,
 and in \ref{fig:subfigv7}, I show the running of the scalar spectral index $\alpha_{s}(k)$, with respect to the momentum scale $k$.
The {\bf black} dotted line corresponds to $k_{max}=0.3~{\rm Mpc}^{-1}$ for $l_{max}=2500$, the \textcolor{blue}{\bf blue} dotted line corresponds to
$k_{min}=4.488\times 10^{-5}~{\rm Mpc}^{-1}$ for $l_{min}=2$,  and in all the plots \textcolor{violet}{\bf violet} dashed dotted line
 represents the pivot scale of momentum at $k_{\star}=0.002~{\rm Mpc}^{-1}$ for $l_{\star}\sim 80$ at which Planck (2014)+WMAP-9+high-L+BICEP2 (dust)
 $P_{S}(k_{\star})=2.2\times 10^{-9}$, $n_{S}=0.9600$, $\alpha_{S}=-0.02$ and $N(k_{\star})=63.26$ are satisfied. Within $2<l<2500$ the
 value of the required momentum scale is determined by the relation,
$k_{reqd}\sim \frac{l_{reqd}}{\eta_{0}\pi}$ \cite{Choudhury:2013jya}, where the conformal time at the present epoch is $\eta_{0}\sim 14000~{\rm Mpc}$.
}
\label{figv2}
\end{figure}

\begin{figure}[t]
\centering
\subfigure[$P_{S}$~vs~$k$]{
    \includegraphics[width=10.5cm, height=7.8cm] {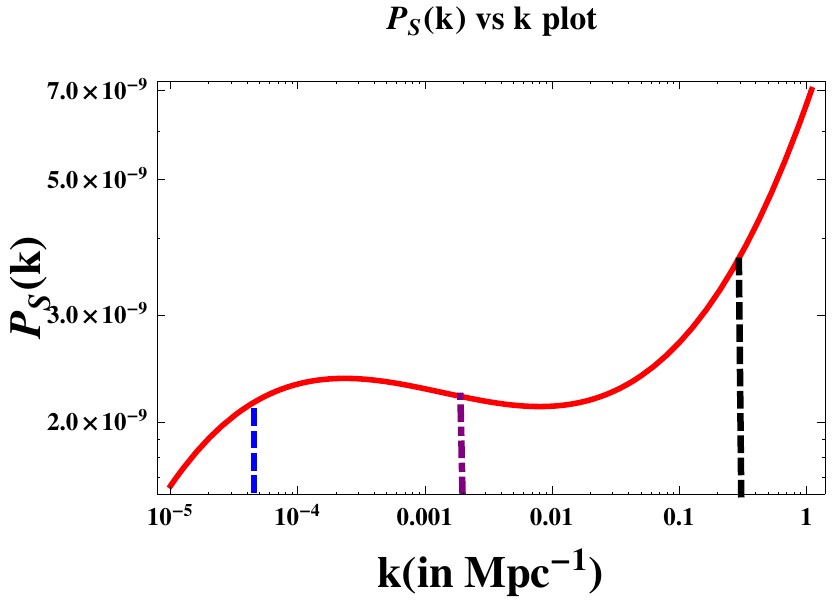}
    \label{fig:subfigv5b}
}
\subfigure[$n_{S}$~vs~$k$]{
    \includegraphics[width=7cm, height=7cm] {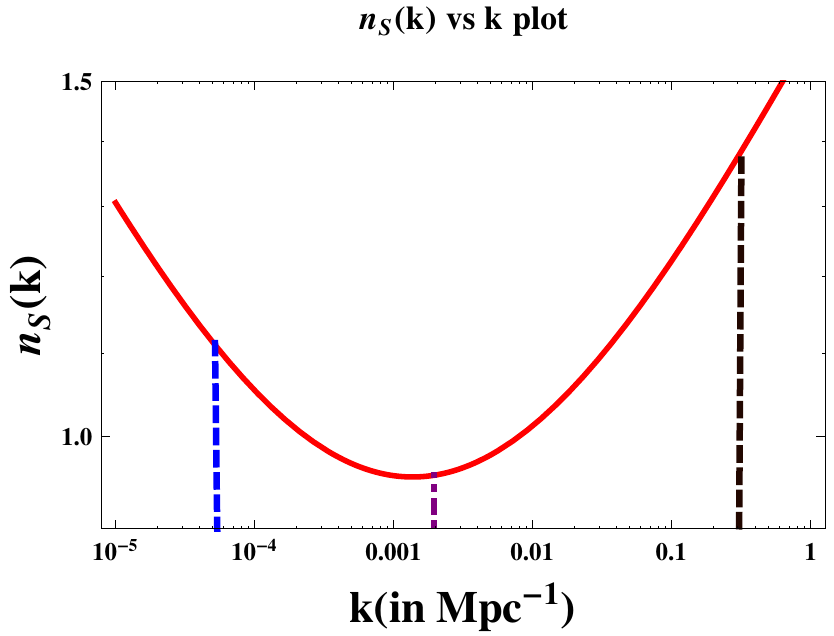}
    \label{fig:subfigv6b}
}
\subfigure[$\alpha_{S}$~vs~$k$]{
    \includegraphics[width=7cm, height=7cm] {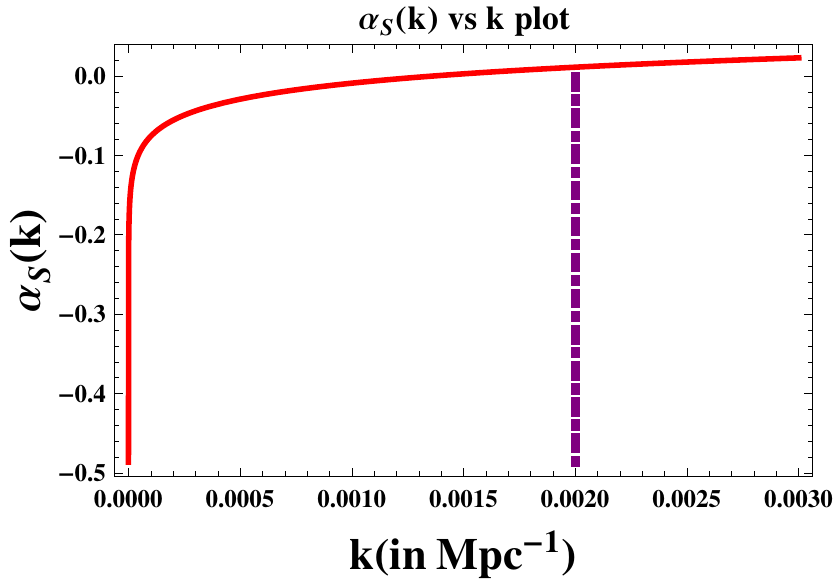}
    \label{fig:subfigv7b}
}
\caption[Optional caption for list of figures]{ In \ref{fig:subfigv5b}, I show the scalar Power spectrum $P_{s}(k)$ \ref{fig:subfigv6b}, I show the scalar spectral index $n_{s}(k)$,
 and in \ref{fig:subfigv7b}, I show the running of the scalar spectral index $\alpha_{s}(k)$, with respect to the momentum scale $k$.
The {\bf black} dotted line corresponds to $k_{max}=0.3~{\rm Mpc}^{-1}$ for $l_{max}=2500$, the \textcolor{blue}{\bf blue} dotted line corresponds to
$k_{min}=4.488\times 10^{-5}~{\rm Mpc}^{-1}$ for $l_{min}=2$,  and in all the plots \textcolor{violet}{\bf violet} dashed dotted line
 represents the pivot scale of momentum at $k_{\star}=0.002~{\rm Mpc}^{-1}$ for $l_{\star}\sim 80$ at which Planck (2015)+WMAP-9+high-L(TT) constraints
 $P_{S}(k_{\star})=2.2\times 10^{-9}$, $n_{S}=0.9569$, $\alpha_{S}=0.011$ and $N(k_{\star})=63.26$ are satisfied. Within $2<l<2500$ the
 value of the required momentum scale is determined by the relation,
$k_{reqd}\sim \frac{l_{reqd}}{\eta_{0}\pi}$ \cite{Choudhury:2013jya}, where the conformal time at the present epoch is $\eta_{0}\sim 14000~{\rm Mpc}$.
}
\label{figv2b}
\end{figure}

\begin{figure}[t]
\centering
\subfigure[$P_{S}$~vs~$k$]{
    \includegraphics[width=10.5cm, height=7.8cm] {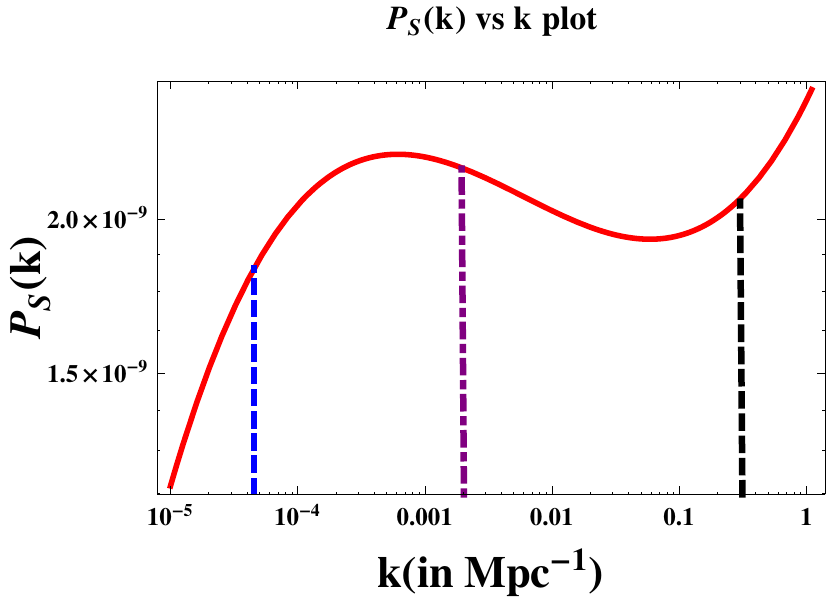}
    \label{fig:subfigv5c}
}
\subfigure[$n_{S}$~vs~$k$]{
    \includegraphics[width=7cm, height=7cm] {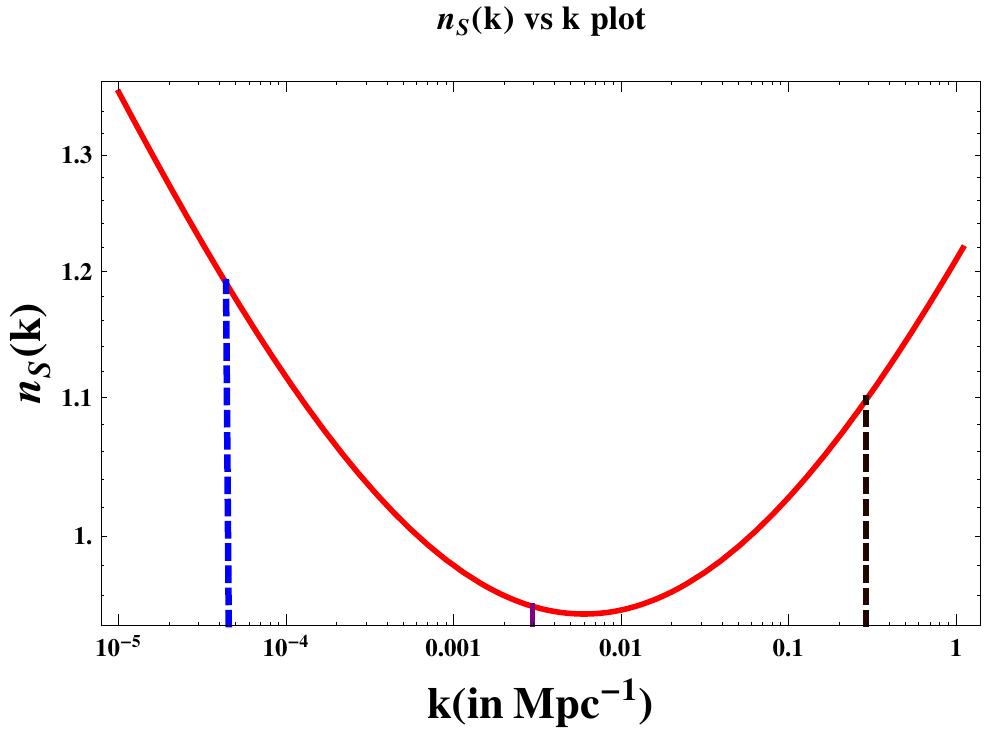}
    \label{fig:subfigv6c}
}
\subfigure[$\alpha_{S}$~vs~$k$]{
    \includegraphics[width=7cm, height=7cm] {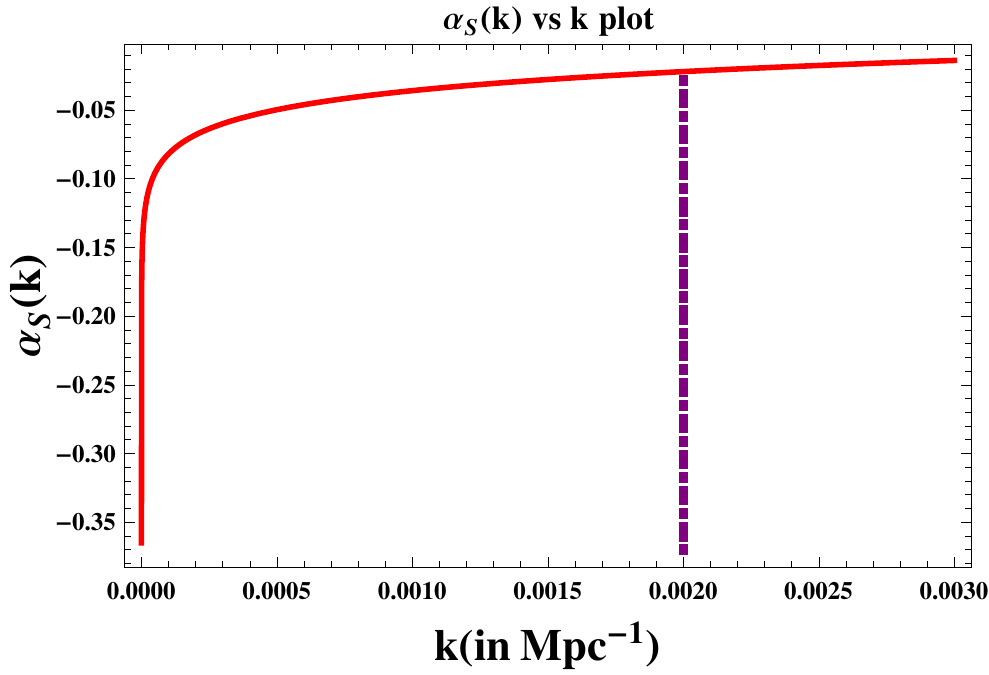}
    \label{fig:subfigv7c}
}
\caption[Optional caption for list of figures]{ In \ref{fig:subfigv5c}, I show the scalar Power spectrum $P_{s}(k)$ \ref{fig:subfigv6c}, I show the scalar spectral index $n_{s}(k)$,
 and in \ref{fig:subfigv7c}, I show the running of the scalar spectral index $\alpha_{s}(k)$, with respect to the momentum scale $k$.
The {\bf black} dotted line corresponds to $k_{max}=0.3~{\rm Mpc}^{-1}$ for $l_{max}=2500$, the \textcolor{blue}{\bf blue} dotted line corresponds to
$k_{min}=4.488\times 10^{-5}~{\rm Mpc}^{-1}$ for $l_{min}=2$,  and in all the plots \textcolor{violet}{\bf violet} dashed dotted line
 represents the pivot scale of momentum at $k_{\star}=0.002~{\rm Mpc}^{-1}$ for $l_{\star}\sim 80$ at which Planck (2015)+BICEP2/Keck Array joint constraint 
 $P_{S}(k_{\star})=2.2\times 10^{-9}$, $n_{S}=0.96$, $\alpha_{S}=-0.02$ and $N(k_{\star})=63.26$ are satisfied. Within $2<l<2500$ the
 value of the required momentum scale is determined by the relation,
$k_{reqd}\sim \frac{l_{reqd}}{\eta_{0}\pi}$ \cite{Choudhury:2013jya}, where the conformal time at the present epoch is $\eta_{0}\sim 14000~{\rm Mpc}$.
}
\label{figv2c}
\end{figure}

\begin{figure}[t]
\centering
\subfigure[$P_{T}$~vs~$k$]{
    \includegraphics[width=7cm, height=7cm] {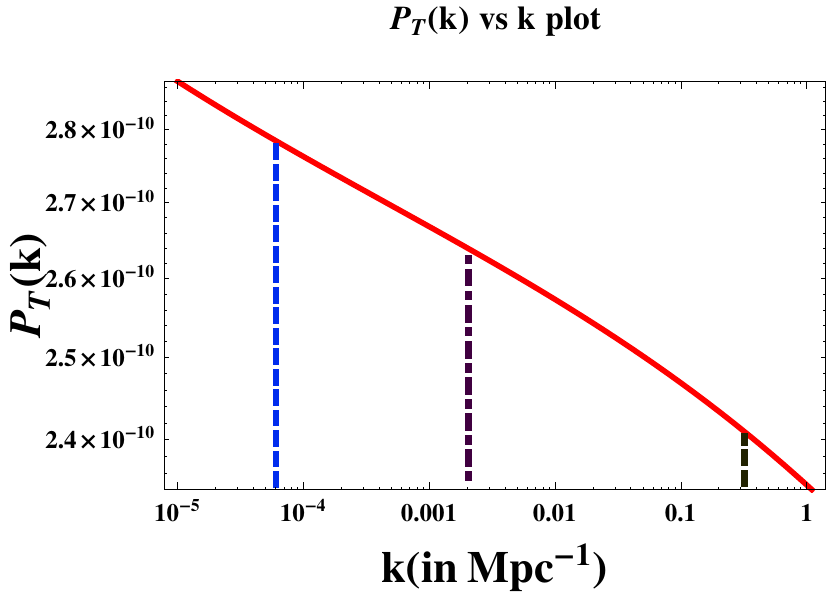}
    \label{fig:subfigz1aa}
}
\subfigure[$P_{T}$~vs~$k$]{
    \includegraphics[width=7cm, height=7cm] {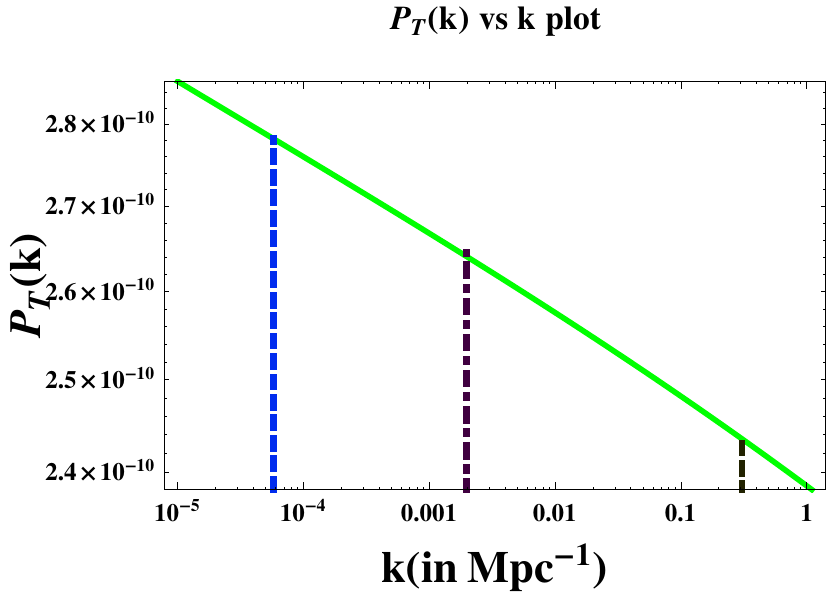}
    \label{fig:subfigz2aa}
}
\subfigure[$r(k)$~vs~$k$]{
    \includegraphics[width=7cm, height=7cm] {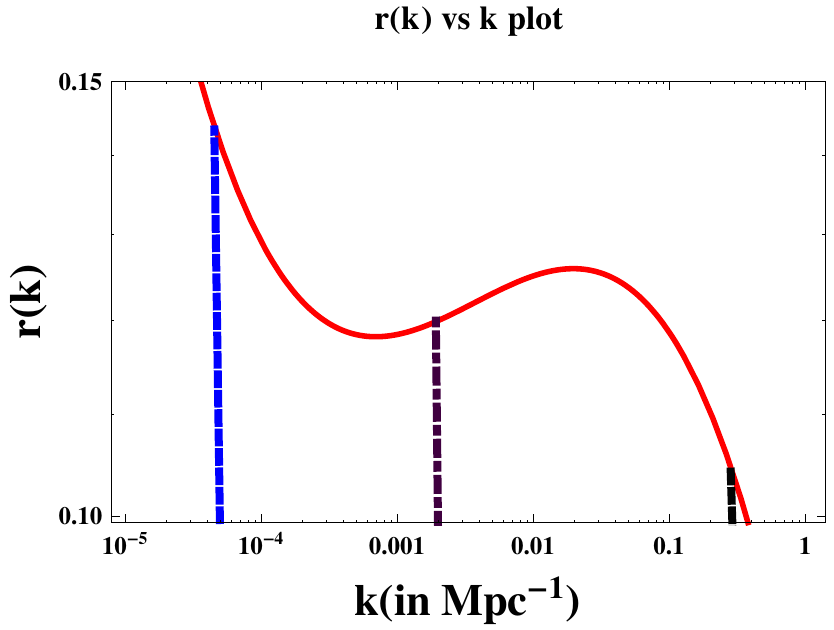}
    \label{fig:subfigz3aa}
}
\subfigure[$r(k)$~vs~$k$]{
    \includegraphics[width=7cm, height=7cm] {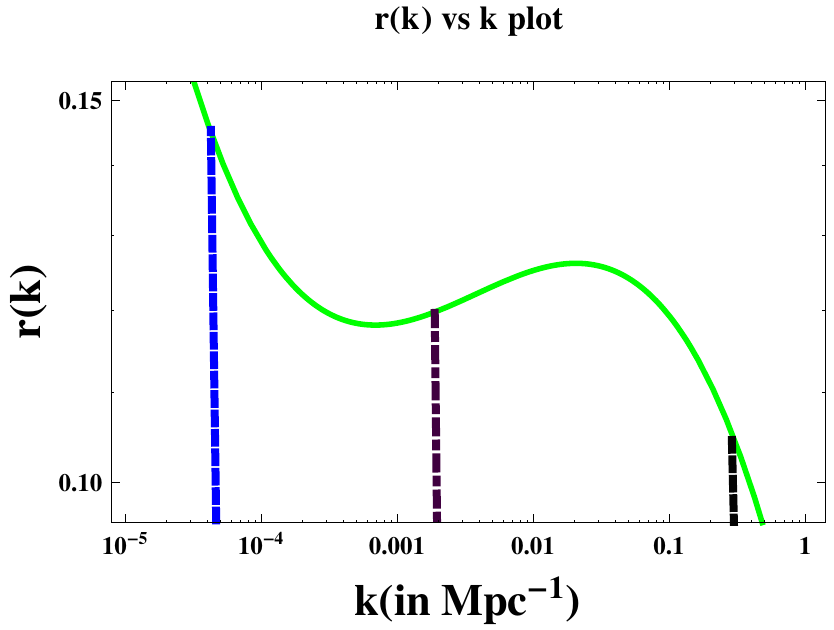}
    \label{fig:subfigz4aa}
}
\caption[Optional caption for list of figures]{ In \ref{fig:subfigz1aa}, I show the Tensor spectrum $P_{T}(k)$ by assuming $r_{0.002}\sim 0.12$
\ref{fig:subfigz2aa}, I show the Tensor spectrum $P_{T}(k)$ by assuming $r_{0.002}\sim 0.12$,
\ref{fig:subfigz3aa}, I show the scale dependence of tensor-to scalar ratio $r(k)$ with $k$ by assuming $r_{0.002}\sim 0.12$,
 and in \ref{fig:subfigz4aa}, I show the scale dependence of tensor-to scalar ratio $r(k)$ with $k$ by assuming $r_{0.002}\sim 0.12$.
The {\bf black} dotted line corresponds to $k_{max}=0.3~{\rm Mpc}^{-1}$ for $l_{max}=2500$, the \textcolor{blue}{\bf blue} dotted line corresponds to
$k_{min}=4.488\times 10^{-5}~{\rm Mpc}^{-1}$ for $l_{min}=2$,  and in all the plots \textcolor{violet}{\bf violet} dashed dotted line
 represents the pivot scale of momentum at $k_{\star}=0.002~{\rm Mpc}^{-1}$ for $l_{\star}\sim 80$ at which Planck (2013)+WMAP-9+high L constraints
 $P_{S}(k_{\star})=2.2\times 10^{-9}$, $n_{S}=0.9600$, $\alpha_{S}=-0.013$ and $N(k_{\star})=63.26$ are satisfied. Within $2<l<2500$ the
 value of the required momentum scale is determined by the relation,
$k_{reqd}\sim \frac{l_{reqd}}{\eta_{0}\pi}$ \cite{Choudhury:2013jya}, where the conformal time at the present epoch is $\eta_{0}\sim 14000~{\rm Mpc}$.
}
\label{figv22a}
\end{figure}

\begin{figure}[t]
\centering
\subfigure[$P_{T}$~vs~$k$]{
    \includegraphics[width=7cm, height=7cm] {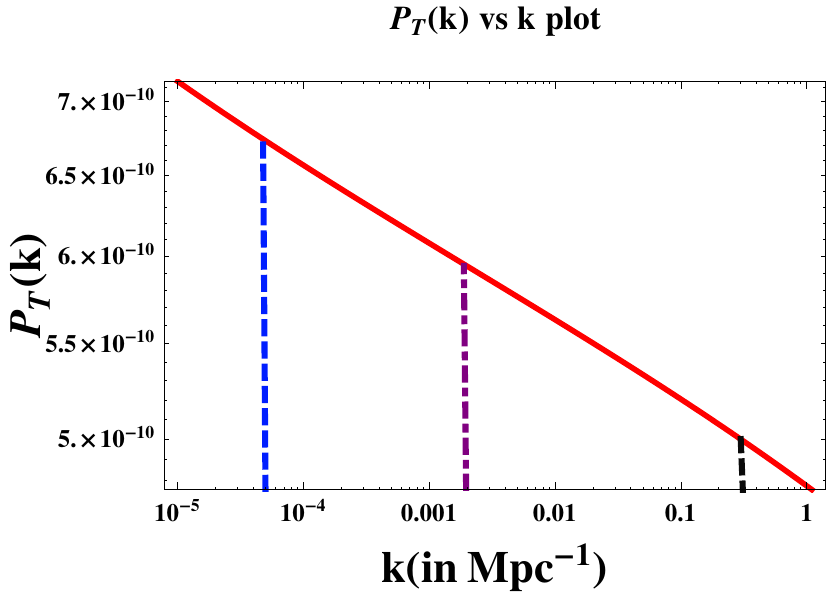}
    \label{fig:subfigz1x}
}
\subfigure[$P_{T}$~vs~$k$]{
    \includegraphics[width=7cm, height=7cm] {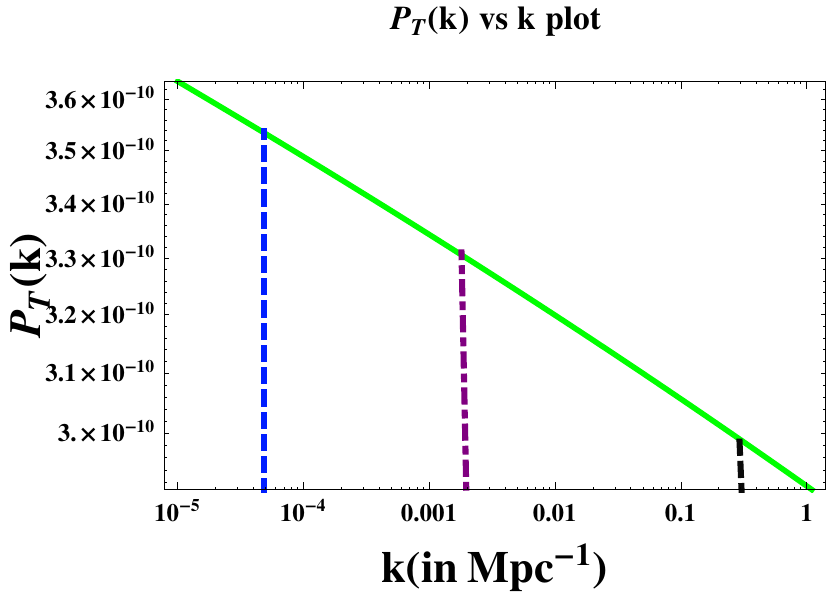}
    \label{fig:subfigz2x}
}
\subfigure[$r(k)$~vs~$k$]{
    \includegraphics[width=7cm, height=7cm] {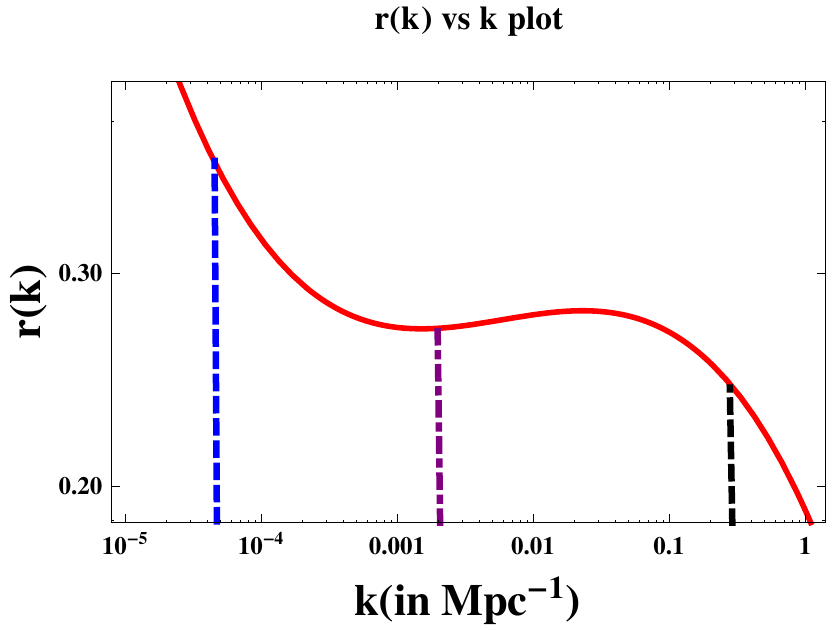}
    \label{fig:subfigz3x}
}
\subfigure[$r(k)$~vs~$k$]{
    \includegraphics[width=7cm, height=7cm] {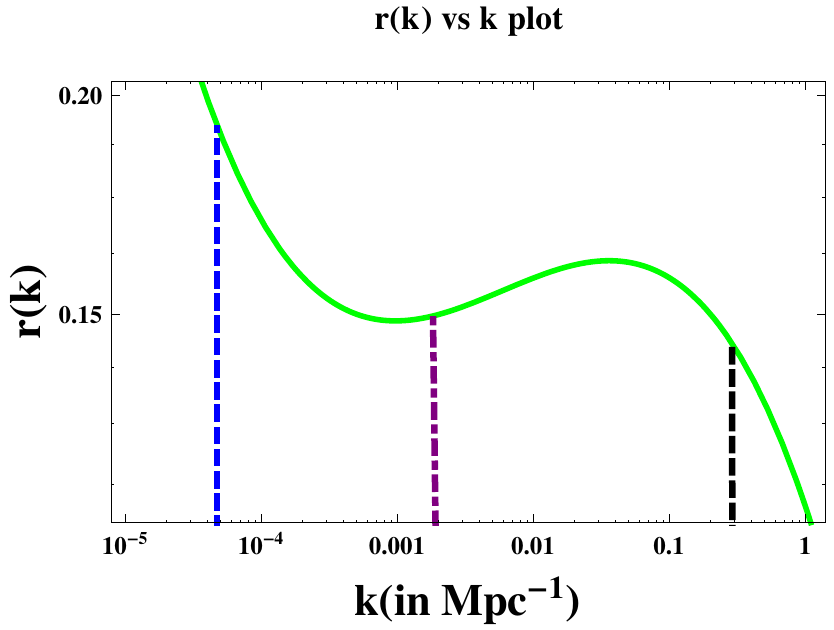}
    \label{fig:subfigz4x}
}
\caption[Optional caption for list of figures]{ In \ref{fig:subfigz1x}, I show the Tensor spectrum $P_{T}(k)$ by assuming $r_{0.002}\sim 0.27$
\ref{fig:subfigz2x}, I show the Tensor spectrum $P_{T}(k)$ by assuming $r_{0.002}\sim 0.15$,
\ref{fig:subfigz3x}, I show the scale dependence of tensor-to scalar ratio $r(k)$ with $k$ by assuming $r_{0.002}\sim 0.27$,
 and in \ref{fig:subfigz4x}, I show the scale dependence of tensor-to scalar ratio $r(k)$ with $k$ by assuming $r_{0.002}\sim 0.15$.
The {\bf black} dotted line corresponds to $k_{max}=0.3~{\rm Mpc}^{-1}$ for $l_{max}=2500$, the \textcolor{blue}{\bf blue} dotted line corresponds to
$k_{min}=4.488\times 10^{-5}~{\rm Mpc}^{-1}$ for $l_{min}=2$,  and in all the plots \textcolor{violet}{\bf violet} dashed dotted line
 represents the pivot scale of momentum at $k_{\star}=0.002~{\rm Mpc}^{-1}$ for $l_{\star}\sim 80$ at which Planck (2014)+WMAP-9+high L+BICEP2 (dust) constraints
 $P_{S}(k_{\star})=2.2\times 10^{-9}$, $n_{S}=0.96$, $\alpha_{S}=-0.022$ and $N(k_{\star})=63.26$ are satisfied. Within $2<l<2500$ the
 value of the required momentum scale is determined by the relation,
$k_{reqd}\sim \frac{l_{reqd}}{\eta_{0}\pi}$ \cite{Choudhury:2013jya}, where the conformal time at the present epoch is $\eta_{0}\sim 14000~{\rm Mpc}$.
}
\label{figv22}
\end{figure}

\begin{figure}[t]
\centering
\subfigure[$P_{T}$~vs~$k$]{
    \includegraphics[width=7cm, height=7cm] {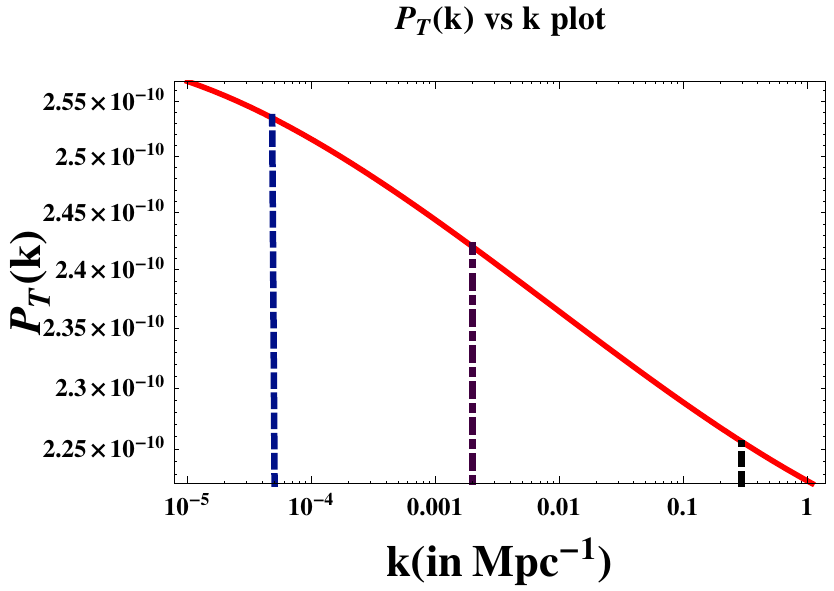}
    \label{fig:subfigz1bb}
}
\subfigure[$P_{T}$~vs~$k$]{
    \includegraphics[width=7cm, height=7cm] {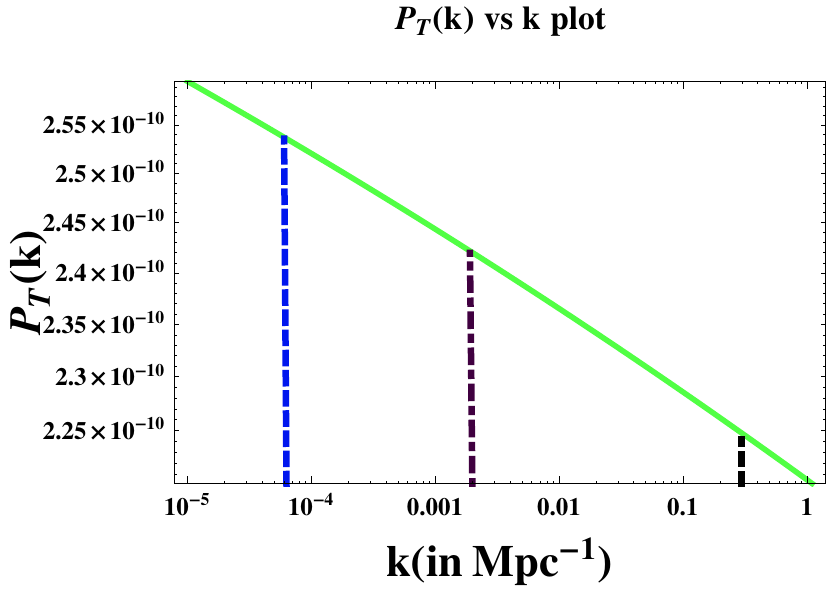}
    \label{fig:subfigz2bb}
}
\subfigure[$r(k)$~vs~$k$]{
    \includegraphics[width=7cm, height=7cm] {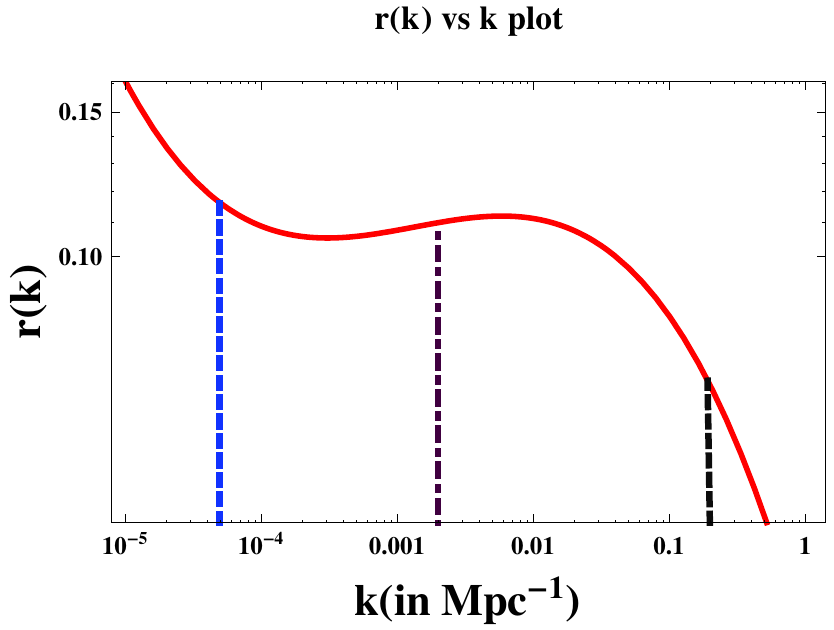}
    \label{fig:subfigz3bb}
}
\subfigure[$r(k)$~vs~$k$]{
    \includegraphics[width=7cm, height=7cm] {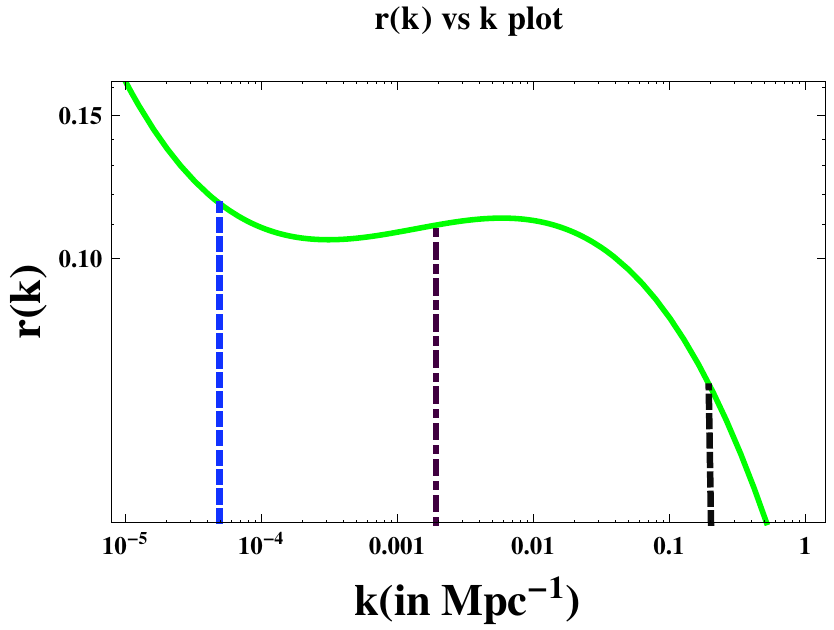}
    \label{fig:subfigz4bb}
}
\caption[Optional caption for list of figures]{ In \ref{fig:subfigz1bb}, I show the Tensor spectrum $P_{T}(k)$ by assuming $r_{0.002}\sim 0.11$
\ref{fig:subfigz2bb}, I show the Tensor spectrum $P_{T}(k)$ by assuming $r_{0.002}\sim 0.11$,
\ref{fig:subfigz3bb}, I show the scale dependence of tensor-to scalar ratio $r(k)$ with $k$ by assuming $r_{0.002}\sim 0.11$,
 and in \ref{fig:subfigz4bb}, I show the scale dependence of tensor-to scalar ratio $r(k)$ with $k$ by assuming $r_{0.002}\sim 0.11$.
The {\bf black} dotted line corresponds to $k_{max}=0.3~{\rm Mpc}^{-1}$ for $l_{max}=2500$, the \textcolor{blue}{\bf blue} dotted line corresponds to
$k_{min}=4.488\times 10^{-5}~{\rm Mpc}^{-1}$ for $l_{min}=2$,  and in all the plots \textcolor{violet}{\bf violet} dashed dotted line
 represents the pivot scale of momentum at $k_{\star}=0.002~{\rm Mpc}^{-1}$ for $l_{\star}\sim 80$ at which Planck (2015)+WMAP-9+high L constraints
 $P_{S}(k_{\star})=2.2\times 10^{-9}$, $n_{S}=0.96$, $\alpha_{S}=0.011$ and $N(k_{\star})=63.26$ are satisfied. Within $2<l<2500$ the
 value of the required momentum scale is determined by the relation,
$k_{reqd}\sim \frac{l_{reqd}}{\eta_{0}\pi}$ \cite{Choudhury:2013jya}, where the conformal time at the present epoch is $\eta_{0}\sim 14000~{\rm Mpc}$.
}
\label{figv22b}
\end{figure}

\begin{figure}[t]
\centering
\subfigure[$P_{T}$~vs~$k$]{
    \includegraphics[width=7cm, height=7cm] {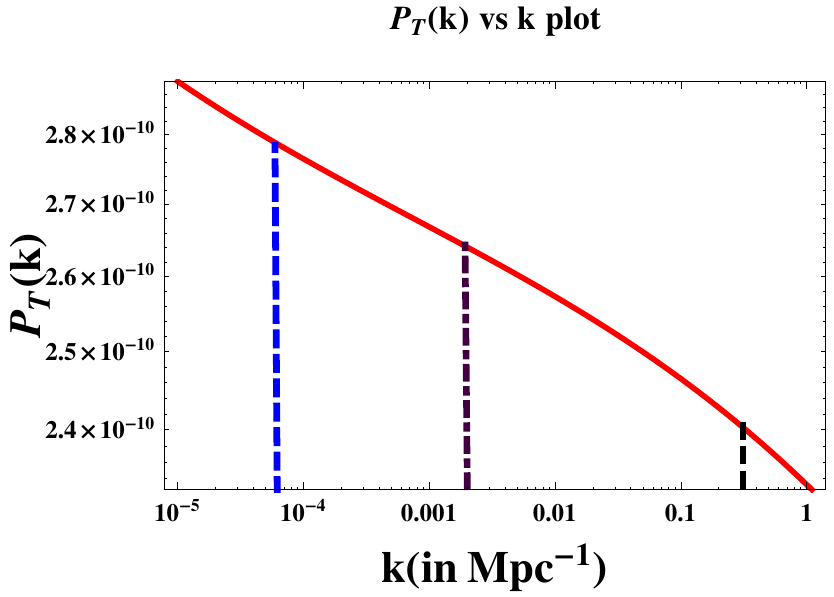}
    \label{fig:subfigz1cc}
}
\subfigure[$P_{T}$~vs~$k$]{
    \includegraphics[width=7cm, height=7cm] {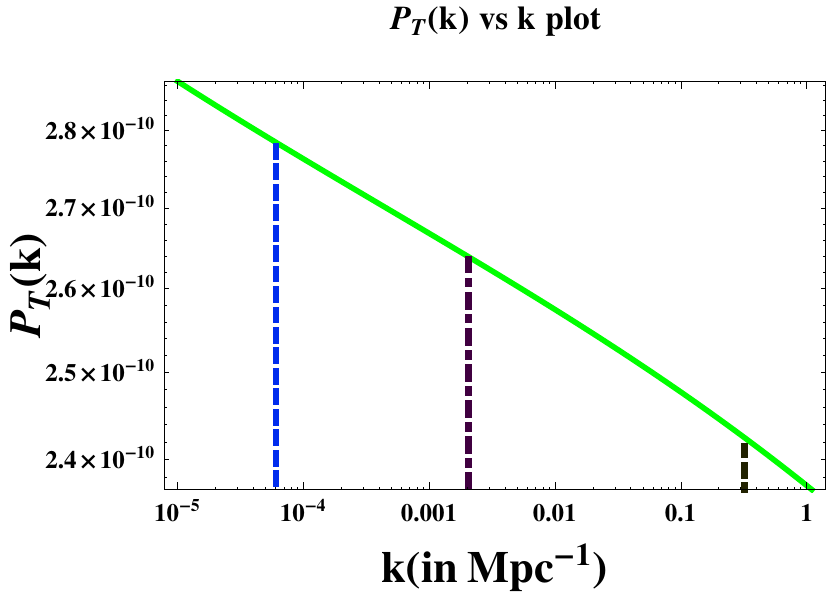}
    \label{fig:subfigz2cc}
}
\subfigure[$r(k)$~vs~$k$]{
    \includegraphics[width=7cm, height=7cm] {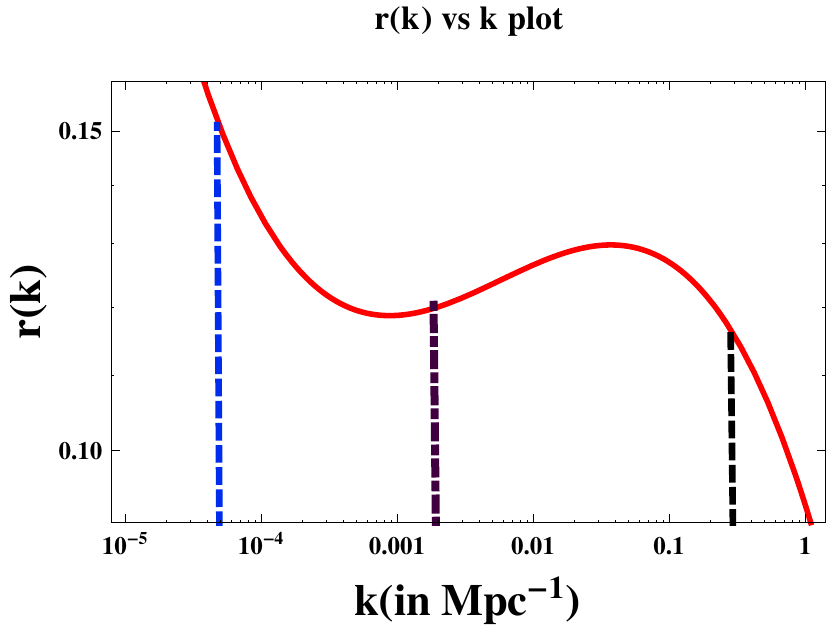}
    \label{fig:subfigz3cc}
}
\subfigure[$r(k)$~vs~$k$]{
    \includegraphics[width=7cm, height=7cm] {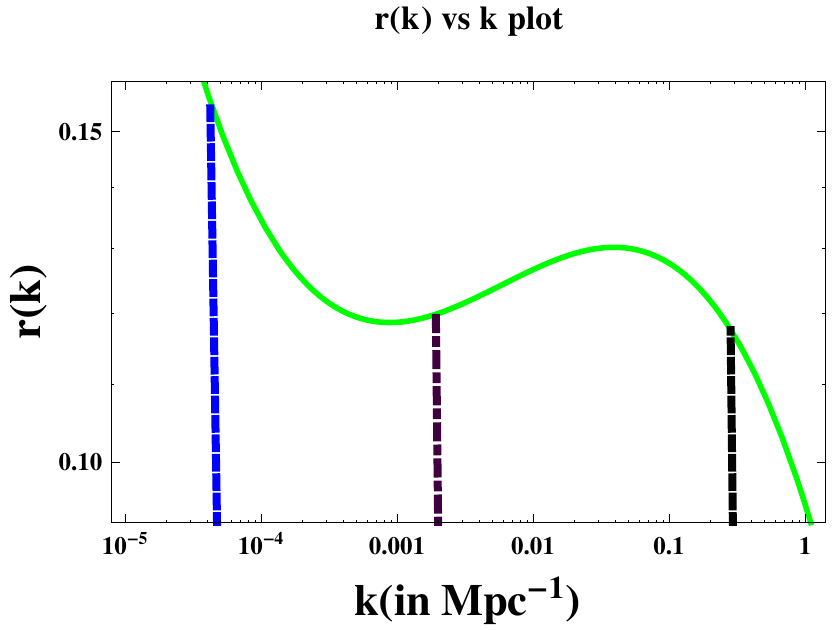}
    \label{fig:subfigz4cc}
}
\caption[Optional caption for list of figures]{ In \ref{fig:subfigz1cc}, I show the Tensor spectrum $P_{T}(k)$ by assuming $r_{0.002}\sim 0.12$
\ref{fig:subfigz2cc}, I show the Tensor spectrum $P_{T}(k)$ by assuming $r_{0.002}\sim 0.12$,
\ref{fig:subfigz3cc}, I show the scale dependence of tensor-to scalar ratio $r(k)$ with $k$ by assuming $r_{0.002}\sim 0.12$,
 and in \ref{fig:subfigz4cc}, I show the scale dependence of tensor-to scalar ratio $r(k)$ with $k$ by assuming $r_{0.002}\sim 0.12$.
The {\bf black} dotted line corresponds to $k_{max}=0.3~{\rm Mpc}^{-1}$ for $l_{max}=2500$, the \textcolor{blue}{\bf blue} dotted line corresponds to
$k_{min}=4.488\times 10^{-5}~{\rm Mpc}^{-1}$ for $l_{min}=2$,  and in all the plots \textcolor{violet}{\bf violet} dashed dotted line
 represents the pivot scale of momentum at $k_{\star}=0.002~{\rm Mpc}^{-1}$ for $l_{\star}\sim 80$ at which Planck (2015)+BICEP2/Keck Array joint constraints
 $P_{S}(k_{\star})=2.2\times 10^{-9}$, $n_{S}=0.96$, $\alpha_{S}=-0.022$ and $N(k_{\star})=63.26$ are satisfied. Within $2<l<2500$ the
 value of the required momentum scale is determined by the relation,
$k_{reqd}\sim \frac{l_{reqd}}{\eta_{0}\pi}$ \cite{Choudhury:2013jya}, where the conformal time at the present epoch is $\eta_{0}\sim 14000~{\rm Mpc}$.
}
\label{figv22c}
\end{figure}

In Fig.~(\ref{figv1}), Fig.~(\ref{fig:subfigv5a}-\ref{fig:subfigv7c}) and Fig.~(\ref{fig:subfigz1aa}-\ref{fig:subfigz4cc}), I have shown the variation of 
 $\ln N(k), P_{S}(k),~n_{S}(k)$, $\alpha_{S}(k)$, $P_{T}(k)$  and $r(k)$ (applying all the previously mentioned joint constraints), at any arbitrary momentum scale 
$k$. Here the {\bf black} dotted line corresponds to \be k_{max}=0.3~{\rm Mpc}^{-1} ~~~~{\bf  for}~~~~ l_{max}=2500,\ee the \textcolor{blue}{\bf blue} dotted line corresponds to
\be k_{min}=4.488\times 10^{-5}~{\rm Mpc}^{-1}~~~~ {\bf for}~~~~ l_{min}=2,\ee  and in all the plots \textcolor{violet}{\bf violet} dashed dotted line
 represents the pivot scale of momentum at \be k_{\star}=0.002~{\rm Mpc}^{-1}~~~~{\bf  for}~~~~ l_{\star}\sim 80\ee at which
 \bea P_{S}(k_{\star})&=&2.2\times 10^{-9},\\ n_{S}&=&0.96,\\ \alpha_{S}&=&-0.02,\\ N(k_{\star})&=&63.26.\eea Within $2<l<2500$ the
 value of the required momentum scale is determined by the relation \cite{Choudhury:2013jya},
\be k_{reqd}\sim \frac{l_{reqd}}{\eta_{0}\pi},\ee where the conformal time at the present epoch is given by: \be\eta_{0}\sim 14000~{\rm Mpc}.\ee

\begin{figure}[t]
\centering
\subfigure[$N$~vs~$(\phi-\phi_0)$]{
    \includegraphics[width=12.5cm, height=7.5cm] {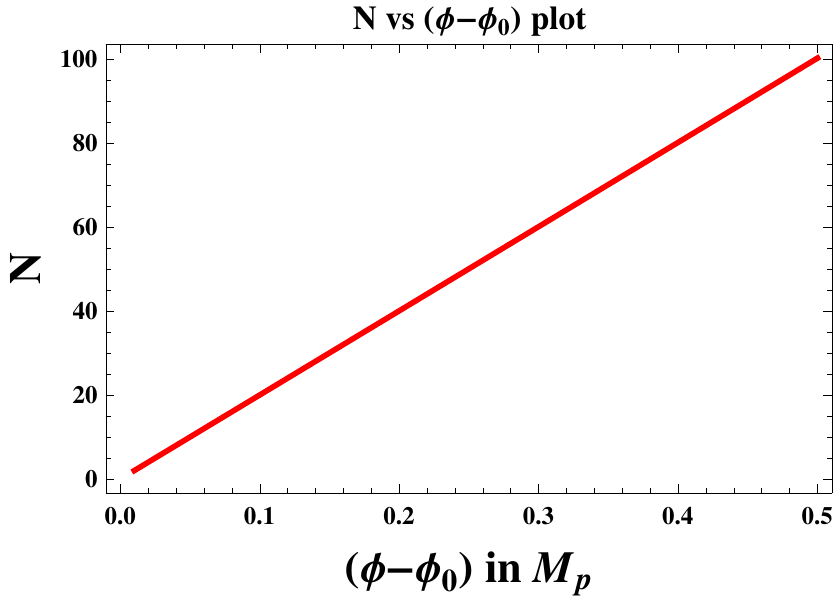}
    \label{xz1}
}
\subfigure[$\Delta\phi$~vs~$\Delta N$]{
    \includegraphics[width=12.5cm, height=7.5cm] {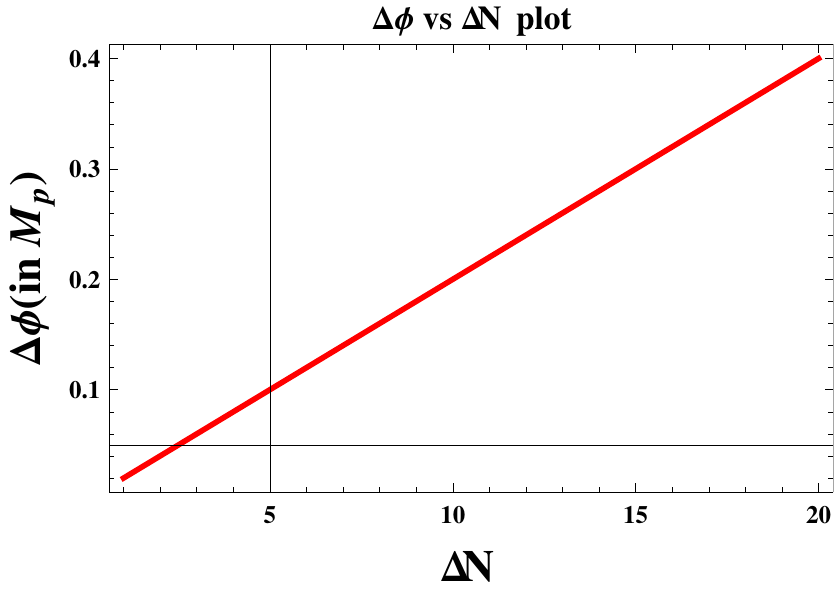}
    \label{xz2}
}
\caption{ We show in \ref{xz1} the total number of e-foldings $N$, with respect to the field $(\phi-\phi_0)$ and \ref{xz2} 
the field evolution $\Delta\phi$, with respect to $\Delta N$.
Within the depicted parameter space $2\sigma$ constraints are allowed as obtained from Planck (2013)+WMAP-9+high-L,\textcolor{red}{ Planck (2014)+WMAP-9+high~L+BICEP2 (dust)},
Planck (2015)+WMAP-9+high-L(TT) and Planck (2015)+BICEP2/Keck Array data.
This also shows that the observational scanning region, $\Delta N\sim 17$ \cite{Choudhury:2013iaa,Khatri:2013xwa,Clesse:2014pna} obtained
from CMB distortions is consistent with the sub-Planckian value of the field excursion. Here $|\Delta\phi|=
|(\phi_{cmb}-\phi_0)-(\phi_{e}-\phi_0)|=|\phi_{cmb}-\phi_e |\approx |\phi_{\star}-\phi_e |$. }\label{f1}
\end{figure}

 In Fig.~(\ref{xz1}) and Fig.~(\ref{xz2}), I have shown the total number of e-foldings $N$, with respect to the field $(\phi-\phi_0)$ and the field evolution $\Delta\phi$
with respect to $\Delta N$ respectively. In Fig.~(\ref{xz1}) the allowed $2\sigma$ region is shown as obtained from Planck (2013)+WMAP-9+high~L, Planck+WMAP-9+high~L+BICEP2 (dust),
Planck (2015)+WMAP-9+high~L(TT) and Planck (2015)  +BICEP2/Keck Array joint constraints.
This also shows that the observational scanning region, $\Delta N\sim 17$ is consistent with the bound on the sub-Planckian value of the field excursion
as obtained in earlier section of the paper. Fig.~(\ref{xz2}) depicts that at $\Delta N\sim 8$ and $\Delta N \sim 17$ the numerical value of the field evolution turn out to be 
$\Delta \phi\sim 0.14~{\rm M}_{p}$ and $\Delta\phi\sim 0.34~{\rm M}_{p}$ respectively.

\begin{figure}[t]
\centering
\subfigure[$P_{S}$~vs~$n_{S}$]{
    \includegraphics[width=12.5cm, height=8.2cm] {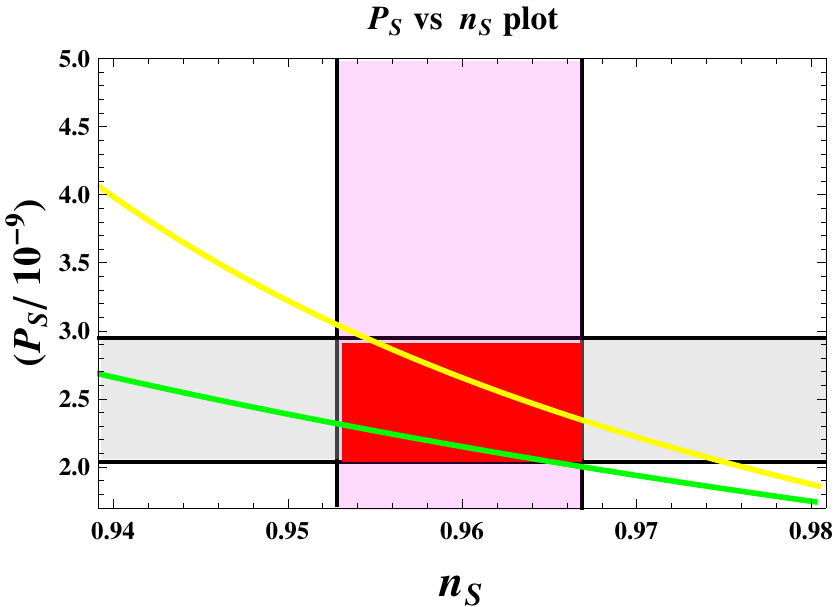}
    \label{fig:subfig5}
}
\subfigure[$P_{S}$~vs~$r_{0.002}$]{
    \includegraphics[width=12.5cm, height=8.2cm] {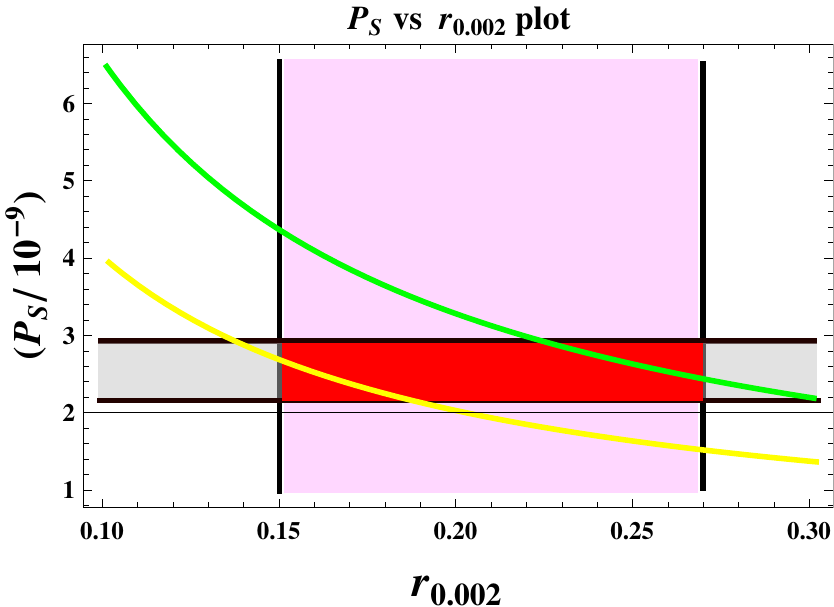}
    \label{fig:subfig6}
}
\caption[Optional caption for list of figures]{We have shown the variation of 
\subref{fig:subfig5} $P_{S}$~vs~$n_{S}$, and \subref{fig:subfig6} $P_{S}$~vs~$r_{0.002}$, for the pivot scale 
$k_{\star}=0.002~Mpc^{-1}$. The overlapping {\it red} patch shows the $2\sigma$ allowed region by the joint constraints obtained from 
Planck (2013)+WMAP-9+high-L,\textcolor{red}{ Planck (2014)+WMAP-9+high~L+BICEP2 (dust)},
Planck (2015)+WMAP-9+high-L(TT) and Planck (2015)+BICEP2/Keck Array data.  The upper ({\it \textcolor{green}{green}}) and lower ({\it \textcolor{yellow}{yellow}}) bounds are set by
Eqs.~(\ref{constraint1}-\ref{constraint20}). 
}
\label{fig5}
\end{figure}

In Fig.~(\ref{fig:subfig5}) and Fig.~(\ref{fig:subfig6}), I have shown the variation of 
 $P_{S},~n_{S}$ and   $r_{0.002}$, at the pivot scale 
$k_{\star}=0.002~Mpc^{-1}$. The overlapping {\it \textcolor{red}{red}} patch shows the allowed region by Planck (2013)+WMAP-9+high~L, Planck+WMAP-9+high~L+BICEP2 (dust),
Planck (2015)+WMAP-9+high~L(TT) and Planck (2015)+BICEP2/Keck Array joint constraints.
The upper {\it \textcolor{green}{green}} and lower {\it \textcolor{yellow}{yellow}} bounds are set by Eqs.~(\ref{constraint1}-\ref{constraint20}).

\begin{figure}[t]
{\centerline{\includegraphics[width=17cm, height=11.5cm] {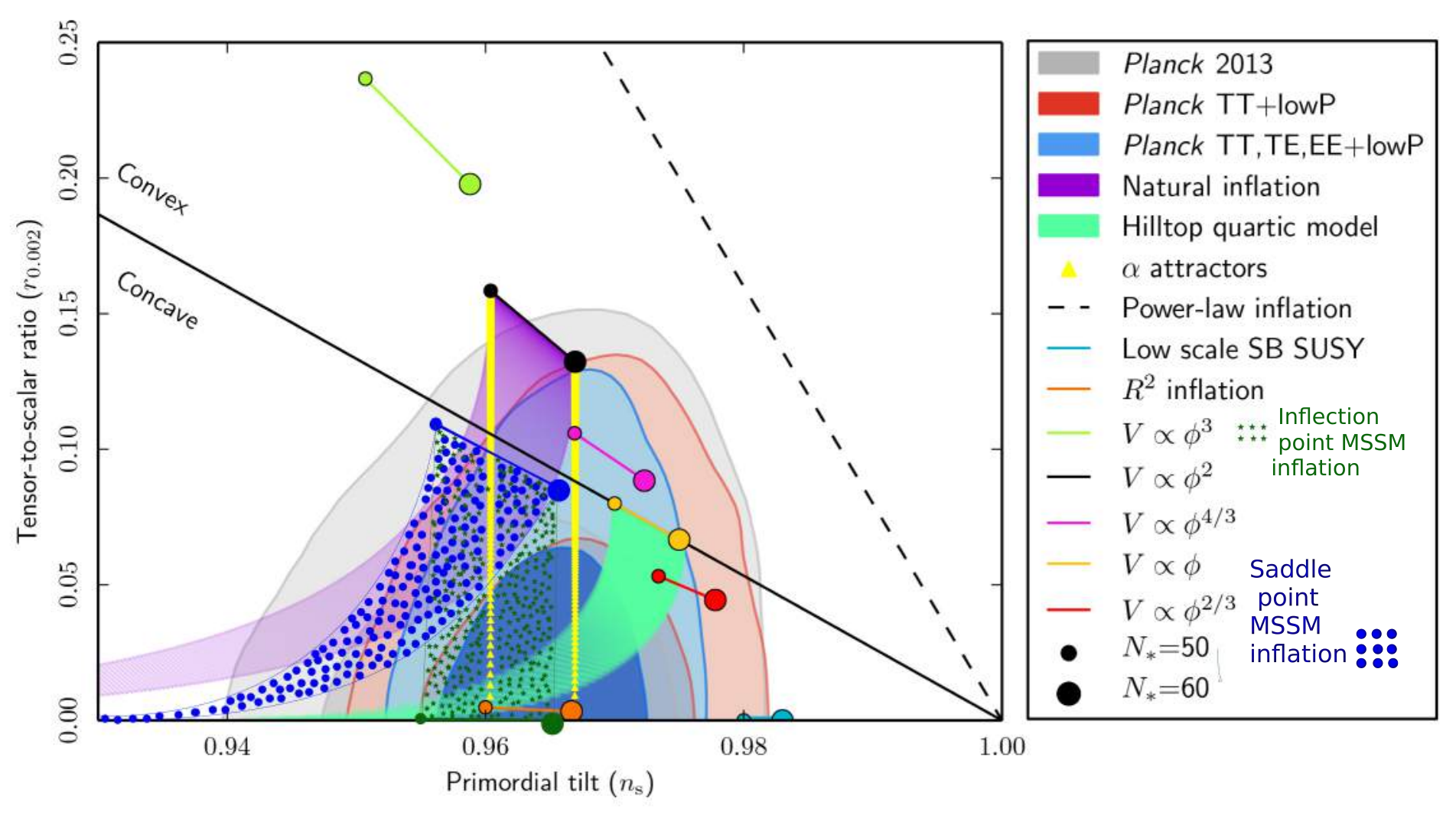}}}
\caption{We show the joint $1\sigma$ and $2\sigma$ CL. contours using
 Planck 2013, Planck 2015 +TT+low P and Planck 2015 +TT+TE+EE+low P data for $r$ vs $n_{S}$ plot at
 the momentum pivot $k_{\star}\sim 0.002~{\rm Mpc}^{-1}$. 
The small circle on the left corresponds to $N=50$, while the right big circle corresponds to $N=60$.
The allowed regions are shown by the shaded violet colour for $r_{\star}<0.12$ and  $0.952<n_{S}<0.967$.
The \textcolor{blue}{blue} and \textcolor{green}{green} bubbles and stars are drawn for saddle point MSSM inflation and inflection point MSSM inflation which are compatible with 
the constraints derived in Eq~(\ref{constraint21}-\ref{constraint24}) and Eq~(\ref{constraint29}-\ref{constraint36}). For Eq~(\ref{constraint25}-\ref{constraint29}) the range of $r_{\star}$ is 
outside the present upper bound from Planck 2015 and BICEP2/Keck Array joint data sets. The vertical black coloured lines are drawn to show the bounded regions of the sub-Planckian inflationary model, along which
the number of e-foldings are fixed. } \label{fig1}
\end{figure}

In Fig~(\ref{fig1}), I have shown  $r$ vs $n_{S}$  at
the  pivot scale: $k_{\star}\sim 0.002~{\rm Mpc}^{-1}$.
The allowed region is shown by the shaded violet colour, for $0.15<r_{\star}<0.27$ and  $0.952<n_{S}<0.967$.
The green and yellow lines are drawn for lower and upper bound of the constraints derived in Eq~(\ref{constraint21}-\ref{constraint36}).
We have used the relation between $n_{S}$ and $r_{\star}$ 
as mentioned in Eq~(\ref{para 21c},~\ref{para 21e}) in the appendix.

In Fig.~(\ref{fig:subfig7}-\ref{fig:subfig9}), I have depicted running of the tensor-to-scalar ratio: $n_{r}=dr/d\ln k$,
 running of the running of the tensor-to-scalar ratio: $\kappa_{r}=d^{2}r/d\ln k$,
running of the tensor spectral tilt: $\alpha_{T}=dn_{T}/d\ln k$,
 running of the running of tensor spectral tilt: $\kappa_{T}=d^{2}n_{T}/d\ln k$ vs scalar spectral tilt $n_{S}$.
Shaded violet colour region is the allowed overlapping region for Planck (2013)+WMAP-9+high~L, Planck+WMAP-9+high~L+BICEP2 (dust),
Planck (2015)+WMAP-9+high~L(TT) and Planck (2015)  +BICEP2/Keck Array joint constraints which further constrain $n_{r},~\kappa_{r},~\alpha_{T}$ and $\kappa_{T}$
within the specified range mentioned in Eq~(\ref{wq2}) and Eq~(\ref{wq5}).
The green and yellow lines are drawn for lower and upper bound on the constraints derived in Eq~(\ref{constraint21}-\ref{constraint26}), 
 We have used the relation between $\alpha_{T},~\kappa_{T}$ and $n_{S}$ 
as mentioned in Eq~(\ref{wq7}) and Eq~(\ref{wq9}).


\begin{figure}[t]
\centering
\subfigure[$\alpha_{T}$~vs~$n_{S}$]{
    \includegraphics[width=7cm, height=6.9cm] {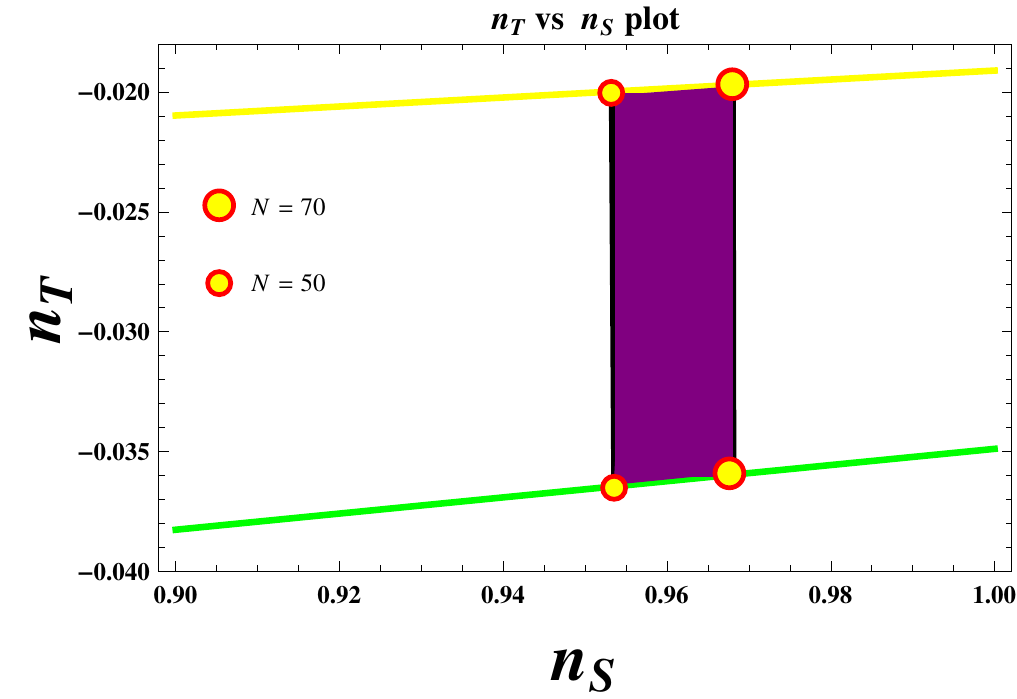}
    \label{fig:subfig7}
}
\subfigure[$\alpha_{T}$~vs~$n_{S}$]{
    \includegraphics[width=7cm, height=6.9cm] {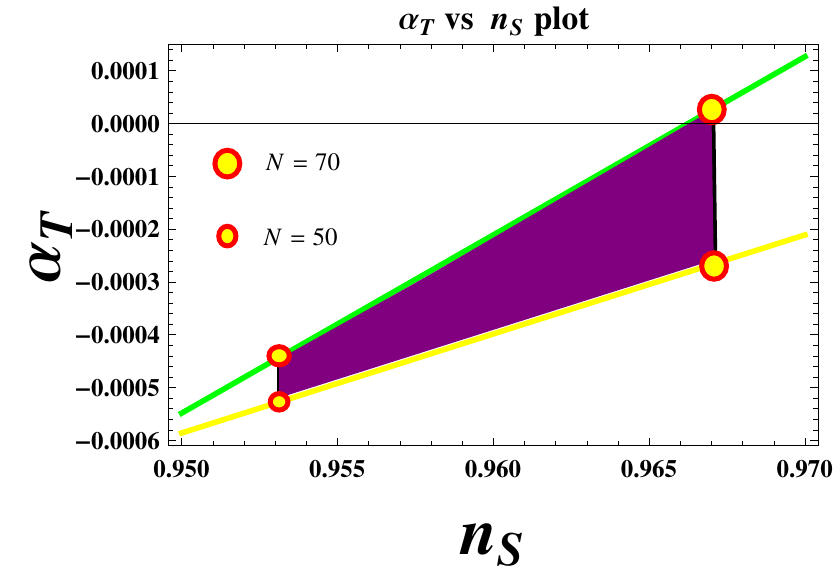}
    \label{fig:subfig8}
}
\subfigure[$\kappa_{T}$~vs~$n_{S}$]{
    \includegraphics[width=10cm, height=6.9cm] {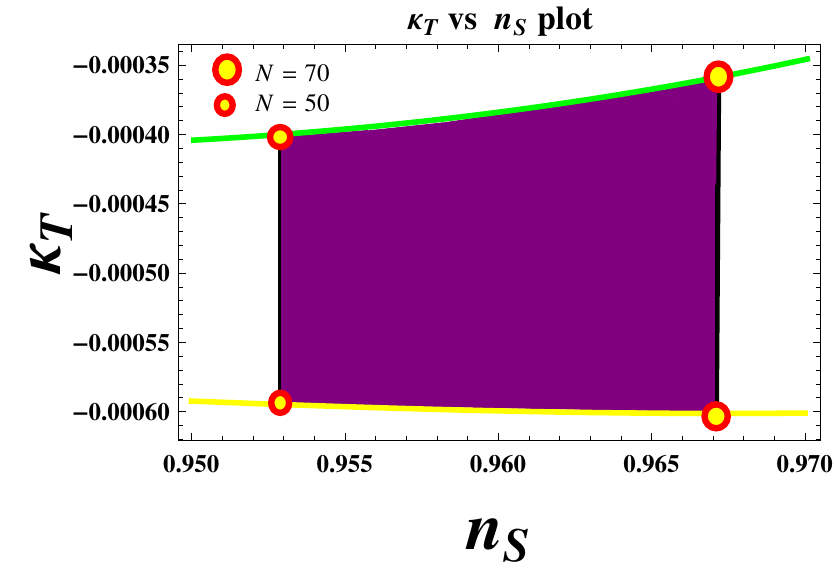}
    \label{fig:subfig9}
}
\caption[Optional caption for list of figures]{We show the \subref{fig:subfig7} tensor spectral tilt $n_{T}$, \subref{fig:subfig8} running of the tensor spectral tilt $\alpha_{T}=dn_{T}/d\ln k$,
\subref{fig:subfig9} running of the running of tensor spectral tilt $\kappa_{T}=d^{2}n_{T}/d\ln k$ vs scalar spectral tilt $n_{S}$ plot.
The small circle on the left corresponds to $N=50$, while the right big circle corresponds to $N=70$.
Shaded violet coloured regions are the allowed regions for Planck (2013)+WMAP-9+high~L, Planck+WMAP-9+high~L+BICEP2 (dust),
Planck (2015)+WMAP-9+high~L(TT) and Planck (2015)  +BICEP2/Keck Array joint data sets which will further constrain $\alpha_{T}$ and $\kappa_{T}$
within the specified ranges mentioned in Eq~(\ref{wq2}) and Eq~(\ref{wq5}).
The green and yellow lines are drawn for lower and upper bounds on the constraints derived in Eq~(\ref{constraint21}-\ref{constraint36}), 
The vertical black coloured lines are drawn to show the bounded regions of the sub-Planckian inflationary model, along which
the number of e-foldings are fixed. 
}
\label{fig4}
\end{figure}


\section{Example of Inflection point inflation within effective theory}
\label{w6}

Now I impose, \be V^{\prime\prime}(\phi_0)=0,\ee in order to study the {\it inflection point} scenario. The potential is given by~\footnote{The inflection point inflation 
has been studied in Refs.~\cite{Enqvist:2010vd}, with a constant potential energy density $V(\phi_0)$.}:
\be\label{rt1a}
V(\phi)=V(\phi_0)+V^{\prime}(\phi_0)(\phi-\phi_{0})+\frac{V^{\prime\prime\prime}(\phi_0)}{6}(\phi-\phi_{0})^{3}
+\frac{V^{\prime\prime\prime\prime}(\phi_0)}{24}(\phi-\phi_{0})^{4}+\cdots\,,
\ee
   %
We can express  $V(\phi_{\star}),V^{\prime}(\phi_{\star}),\cdots$ in terms of $\phi_0$ for a sub-Planckian regime as:
\bea\label{ws1}
V(\phi_{\star})&=&V(\phi_0)+\vartheta_{\star}V^{'}(\phi_0)+\frac{\vartheta^{3}_{\star}}{6}V^{'''}(\phi_0)+\frac{\vartheta^{4}_{\star}}{24}V^{''''}(\phi_0),\nonumber \\
V^{'}(\phi_{\star})&=&V^{'}(\phi_0)+\frac{\vartheta^{2}_{\star}}{2}V^{'''}(\phi_0)+\frac{\vartheta^{3}_{\star}}{6}V^{''''}(\phi_0),\nonumber \\
V^{''}(\phi_{\star})&=&\vartheta_{\star}V^{'''}(\phi_0)+\frac{\vartheta^{2}_{\star}}{2}V^{''''}(\phi_0),\nonumber \\
V^{'''}(\phi_{\star})&=&V^{'''}(\phi_0)+\vartheta_{\star}V^{''''}(\phi_0),\nonumber \\
V^{''''}(\phi_{\star})&=&V^{''''}(\phi_0).
\eea
Using Eq~(\ref{ws1}),  I obtain a {\it particular} solution for the coefficients: $V(\phi_0),~V^{\prime}(\phi_0),\cdots$, which can be written as
~\footnote{There will be in general $2$ solutions around an inflection point, here I will provide one of the two solutions which is the most interesting 
one for the general case of study.}: 
\bea\label{ws2}
V(\phi_0)&=&V(\phi_{\star})-\vartheta_{\star}V^{'}(\phi_{\star})+\frac{\vartheta^{3}_{\star}}{3}V^{'''}(\phi_{\star})-\frac{5\vartheta^{4}_{\star}}{24}V^{''''}(\phi_{\star}),\nonumber \\
V^{'}(\phi_0)&=&V^{'}(\phi_{\star})-\frac{\vartheta^{2}_{\star}}{2}V^{'''}(\phi_{\star})+\frac{\vartheta^{3}_{\star}}{3}V^{''''}(\phi_{\star}),\nonumber \\
V^{'''}(\phi_0)&=&V^{'''}(\phi_{\star})-\vartheta_{\star}V^{''''}(\phi_{\star}),\nonumber \\
V^{''''}(\phi_0)&=&V^{''''}(\phi_{\star}).
\eea
Now using the bound on $V(\phi_{\star}),~V^{\prime}(\phi_{\star}),\cdots$ as mentioned in Eqs.~(\ref{constraint1}-\ref{constraint20}), I can 
obtain the following constraints on the coefficients of $V(\phi_0),V^{'}(\phi_0),\cdots$:
 \\
\underline{\bf Planck (2013)+WMAP-9+high~L:}
 \bea\label{constraint1a}
 V(\phi_{0})\leq {\cal O}(3.79-3.94)\times 10^{-9}M^{4}_{p},\\  
   \label{constraint2a}
  V^{'}(\phi_{0})\leq {\cal O}(4.66-4.88)\times 10^{-10}M^{3}_{p}, \\
\label{constraint4a}
  V^{'''}(\phi_{0})\leq {\cal O}(8.13-(-1.25))\times 10^{-10}M_{p}, \\
\label{constraint5a}
    V^{''''}(\phi_{0})\leq {\cal O}(0.39-4.76)\times 10^{-9},
   \eea
 \underline{\bf\textcolor{red}{ Planck (2014)+WMAP-9+high~L+BICEP2 (dust)}:}
 \bea\label{constraint25a}
   5.26\times 10^{-9}M^{4}_{p}\leq V(\phi_0)\leq 9.50\times 10^{-9}M^{4}_{p},\\  
   \label{constraint26a}
   2.44\times 10^{-10}M^{3}_{p}\leq V^{'}(\phi_0)\leq 1.74\times 10^{-9}M^{3}_{p}, \\
\label{constraint28a}
   6.29\times 10^{-10}M_{p}\leq V^{'''}(\phi_0)\leq 7.08\times 10^{-10}M_{p}, \\
\label{constraint29a}
   5.56\times 10^{-10}\leq V^{''''}(\phi_0)\leq 4.82\times 10^{-9},
   \eea
\underline{\bf Planck (2015)+WMAP-9+high~L(TT):}
 \bea\label{constraint11a}
   V(\phi_{0})\leq {\cal O}(3.41-3.67)\times 10^{-9}M^{4}_{p},\\  
   \label{constraint12a}
   V^{'}(\phi_{0})\leq {\cal O}(4.06-4.31)\times 10^{-10}M^{3}_{p}, \\
\label{constraint14a}
   V^{'''}(\phi_{0})\leq {\cal O}((-6.00)-(-9.64))\times 10^{-10}M_{p}, \\
\label{constraint15a}
   V^{''''}(\phi_{0})\leq {\cal O}(5.52-5.76)\times 10^{-9},
   \eea
\underline{\bf Planck (2015)+BICEP2/Keck Array:}
 \bea\label{constraint16a}
  V(\phi_{0})\leq {\cal O}(3.75-3.95)\times 10^{-9}M^{4}_{p},\\  
   \label{constraint17a}
   V^{'}(\phi_{0})\leq {\cal O}(4.70-5.03)\times 10^{-10}M^{3}_{p}, \\
\label{constraint19a}
    V^{'''}(\phi_{0})\leq {\cal O}(8.13-30.34)\times 10^{-10}M_{p}, \\
\label{constraint20a}
    V^{''''}(\phi_{0})\leq {\cal O}(0.39-4.76)\times 10^{-9},
   \eea
We can compare these results with that of the constraints mentioned in Eq~(\ref{constraint21}-\ref{constraint36}) for a generic sub-Planckian inflationary 
setup for $\vartheta_\star =\phi_\star-\phi_0 \sim 10^{-1}M_p$ for a sub-Planckian VEV model of inflation. We find a very nice agreement which testifies 
the power of a model independent reconstruction of the potential.

Let us give an example of Hubble induced supergravity motivated MSSM inflation which is guided by inflection point prescription.
For the potential under consideration, I have \be V_{0}=3H^2M^{2}_{p}\sim M^{4}_{s}>>m_{\phi}^2|\phi|^2,\ee 
where $m_{\phi}\sim{\cal O}(\rm TeV)$ is the soft mass. In this case the 
contributions from the Hubble-induced terms are important compared to the soft SUSY breaking mass, $m_\phi$.
The potential, after stabilizing the angular
direction of the complex scalar field $\phi= |\phi|\exp[i\theta]$, see~\cite{Choudhury:2013jya,Choudhury:2014sxa,Choudhury:2014uxa}, reduces to a simple form along the real direction, which is dominated by a 
single scale, i.e. $H\sim H_{\star}$:
\begin{equation}\label{h1a}
 V(\phi)=V_{0}+c_{H}H^{2}|\phi|^{2}-
 \frac{a_{H}H\phi^n}{nM_{p}^{n-3}}+\frac{\lambda^2\lvert\phi\rvert^{2(n-1)}}{M_{p}^{2(n-3)}},
\end{equation}
where I take $\lambda=1$, and,the Hubble-induced mass parameter is $c_{H}$ and the trilinear A term is $a_{H}$.
Fortunately for this class of potential given by Eq~(\ref{h1a}), inflection point inflation can be accommodated, when  \be a_H^2\approx 8(n-1)c_H.\ee 
This can be characterized by a fine-tuning parameter, $\delta$, which is defined as:
\begin{equation}
\label{newbeta}
\frac{a_H^2}{8(n-1)c_H} = 1-\left(\frac{n-2}{2}\right)^2\delta^2\,.
\end{equation}
When $\vert\delta\vert$ is small~\footnote{We will consider a moderate tuning of order $\delta \sim 10^{-4}$ between $c_H$ and $a_H$.}, a point of inflection $\phi_0$ exists,
such that \be V^{\prime\prime}\left(\phi_0\right) =0,\ee with
\begin{equation}
\label{phi0}
\phi_0 = \left(\sqrt{\frac{c_H}{(n-1)}} H M_{p}^{n-3}\right)^{{1}/{n-2}}\, +{\cal O}(\delta^2).
\end{equation}
%
\begin{figure}[t]
{\centerline{\includegraphics[width=15.5cm, height=9cm] {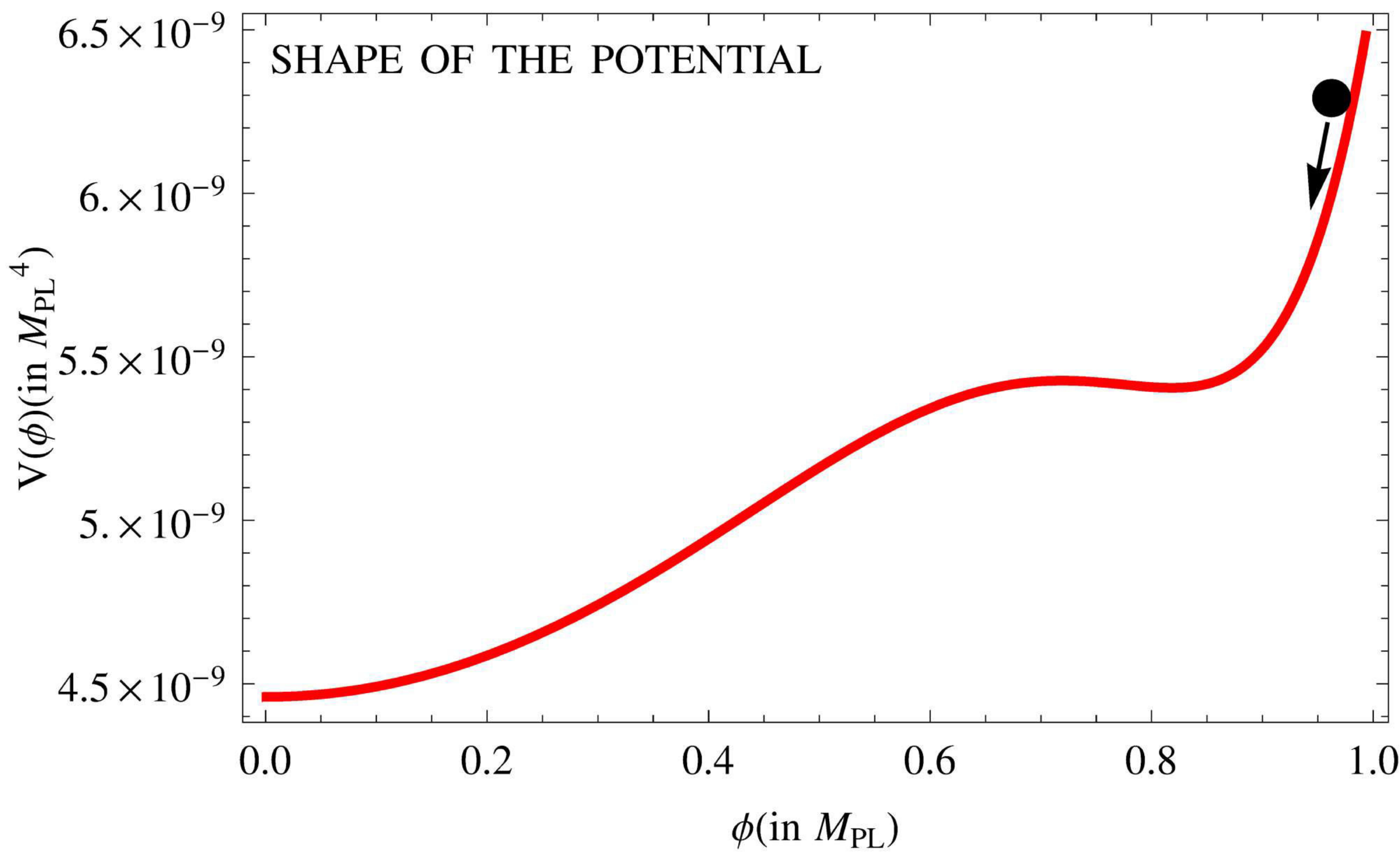}}}
\caption{We show the variation of the n=6 MSSM inflationary potential with respect to the field $\phi$, which satisfies all the obtained constraints from recent Planck 2015 data and Planck 2015 +BICEP2/Keck Array joint constraints.} \label{figbb}
\end{figure}


For $\delta <1$, I can Taylor-expand the inflaton potential around an inflection 
point, $\phi=\phi_{0}$, as~\cite{Choudhury:2013jya,Choudhury:2014sxa,Choudhury:2014uxa}:
\be\label{rt1a}
V(\phi)=\alpha+\beta(\phi-\phi_{0})+\gamma(\phi-\phi_{0})^{3}+\kappa(\phi-\phi_{0})^{4}+\cdots\,,
\ee 
where the expansion coefficients are now given by:
\begin{eqnarray}\label{p1}
     \alpha&=&V(\phi_{0})=V_{0}+\left(\frac{(n-2)^{2}}{n(n-1)}+\frac{(n-2)^2}{n}\delta^{2}\right)c_{H}H^{2}\phi^{2}_{0}+{\cal O}(\delta^{4}),\\
 \beta&=&V^{'}(\phi_{0})=2\left(\frac{n-2}{2}\right)^{2}\delta^{2}c_{H}H^{2}\phi_{0}+{\cal O}(\delta^{4}),\\
 \gamma&=&\frac{V^{'''}(\phi_{0})}{3!}=\frac{c_{H}H^{2}}{\phi_{0}}\left(4(n-2)^2-\frac{(n-1)(n-2)^3}{2}\delta^{2}\right)+{\cal O}(\delta^{4}),\\
 \small\kappa&=&\frac{V^{''''}(\phi_{0})}{4!}\\~~~~&=&\frac{c_{H}H^{2}}{\phi^{2}_{0}}\left(12(n-2)^3-\frac{(n-1)(n-2)(n-3)
(7n^2-27n+26)}{2}\delta^{2}\right)+{\cal O}(\delta^{4}).\nonumber 
   \end{eqnarray}
%
Note that once I specify $c_H$ and $H_{\star}$, all the terms in the potential are determined. In this regard the potential 
indeed simplifies a lot to study the cosmological observables.

As an concrete example, I considered $n=6$ case, where the flatness of the superfield $\Phi$ is lifted by the non-renormalizable operator.
This is appropriate for both $\widetilde u\widetilde d\widetilde d$ and $\widetilde L\widetilde L\widetilde e$ flat directions.

We fix $\lambda ={\cal O}(1)$ and $\delta \sim 10^{-4}$. 
In order to satisfy the Planck observational constraints on the amplitude of the power spectrum, $2.092\times 10^{-9}<P_S< 2.297\times 10^{-9}$
(within $2\sigma$), spectral tilt $0.958 < n_S<0.963$ (within $2\sigma$), sound speed $c_{S}=1$ (within $2\sigma$), and tensor-to-scalar ratio $r_\star \leq 0.12$, I obtain the following
constraints on our parameters for $H_{inf}\geq m_{\phi}\sim {\cal O}(\rm TeV)$, where successful inflation can occur via {\it inflection point}:
\begin{eqnarray}\label{P1-space}
c_{H} &\sim &{\cal O}(10-10^{-6})\,, ~~~~~~~~~~~~~~~{\bf for}~~~~~~~~~~10^{-22}<r_{\star}<0.12\nonumber \\
a_{H} & \sim &{\cal O}(30 - 10^{-3} )\,, ~~~~~~~~~~~~~~~{\bf for}~~~~~~~~~~10^{-22}<r_{\star}<0.12\nonumber \\
M_s &\sim & {\cal O}(9.50\times 10^{10}-1.77\times10^{16})~{\rm GeV}\,,~~~{\bf for}~~~~10^{-22}<r_{\star}<0.12\,.
\end{eqnarray}
This analysis justifies that the Hubble correction induced MSSM inflation is one of the examples of inflection point inflation where 
the present reconstruction technique holds good perfectly.

\section{Example of Saddle point inflation within effective theory}
\label{w6a}

Now I impose, \bea V^{'}(\phi_{0})=0,\\
V^{\prime\prime}(\phi_0)=0,\\ V^{'''}(\phi_{0})=0,\eea in order to study the {\it saddle point} scenario. The potential is given by~\footnote{The saddle point inflation 
has been studied in Refs.~\cite{Choudhury:2011jt}.}:
\be\label{rt1aa}
V(\phi)=V(\phi_0)
+\frac{V^{\prime\prime\prime\prime}(\phi_0)}{24}(\phi-\phi_{0})^{4}+\cdots\,,
\ee
   %
We can express  $V(\phi_{\star}),V^{\prime}(\phi_{\star}),\cdots$ in terms of $\phi_0$ for a sub-Planckian regime as:
\bea\label{ws1a}
V(\phi_{\star})&=&V(\phi_0)+\frac{\vartheta^{4}_{\star}}{24}V^{''''}(\phi_0),\nonumber \\
V^{'}(\phi_{\star})&=&\frac{\vartheta^{3}_{\star}}{6}V^{''''}(\phi_0),\nonumber \\
V^{''}(\phi_{\star})&=&\frac{\vartheta^{2}_{\star}}{2}V^{''''}(\phi_0),\nonumber \\
V^{'''}(\phi_{\star})&=&\vartheta_{\star}V^{''''}(\phi_0),\nonumber \\
V^{''''}(\phi_{\star})&=&V^{''''}(\phi_0).
\eea
Using Eq~(\ref{ws1a}),  I obtain a {\it particular} solution for the coefficients: $V(\phi_0),~V^{\prime}(\phi_0),\cdots$, which can be written as
~\footnote{There will be in general $2$ solutions around an inflection point, here I will provide one of the two solutions which is the most interesting 
one for the general case of study.}: 
\bea\label{ws2a}
V(\phi_0)&=&V(\phi_{\star})-\vartheta_{\star}V^{'}(\phi_{\star})+\frac{\vartheta^{3}_{\star}}{3}V^{'''}(\phi_{\star})-\frac{5\vartheta^{4}_{\star}}{24}V^{''''}(\phi_{\star}),\nonumber \\
V^{''''}(\phi_0)&=&V^{''''}(\phi_{\star}).
\eea
Now using the bound on $V(\phi_{\star}),~V^{\prime}(\phi_{\star}),\cdots$ as mentioned in Eqs.~(\ref{constraint1}-\ref{constraint20}), I can 
obtain the following constraints on the coefficients of $V(\phi_0),V^{'}(\phi_0),\cdots$:
 \\
\underline{\bf Planck (2013)+WMAP-9+high~L:}
 \bea\label{constraint1aa}
 V(\phi_{0})\leq {\cal O}(3.79-3.94)\times 10^{-9}M^{4}_{p},\\
\label{constraint5aa}
    V^{''''}(\phi_{0})\leq {\cal O}(0.39-4.76)\times 10^{-9},
   \eea
 \underline{\bf\textcolor{red}{ Planck (2014)+WMAP-9+high~L+BICEP2 (dust)}:}
 \bea\label{constraint25aa}
   5.26\times 10^{-9}M^{4}_{p}\leq V(\phi_0)\leq 9.50\times 10^{-9}M^{4}_{p},\\ 
\label{constraint29aa}
   5.56\times 10^{-10}\leq V^{''''}(\phi_0)\leq 4.82\times 10^{-9},
   \eea
\underline{\bf Planck (2015)+WMAP-9+high~L(TT):}
 \bea\label{constraint11aa}
   V(\phi_{0})\leq {\cal O}(3.41-3.67)\times 10^{-9}M^{4}_{p},\\ 
\label{constraint15aa}
   V^{''''}(\phi_{0})\leq {\cal O}(5.52-5.76)\times 10^{-9},
   \eea
\underline{\bf Planck (2015)+BICEP2/Keck Array:}
 \bea\label{constraint16aa}
  V(\phi_{0})\leq {\cal O}(3.75-3.95)\times 10^{-9}M^{4}_{p},\\ 
\label{constraint20aa}
    V^{''''}(\phi_{0})\leq {\cal O}(0.39-4.76)\times 10^{-9},
   \eea
We can compare these results with that of the constraints mentioned in Eq~(\ref{constraint21}-\ref{constraint36}) for a generic sub-Planckian inflationary 
setup for $\vartheta_\star =\phi_\star-\phi_0 \sim 10^{-1}M_p$ for a sub-Planckian VEV model of inflation. We find a very nice agreement which testifies 
the power of a model independent reconstruction of the potential.

Let us consider a concrete example of soft SUSY breaking induced MSSM inflation which is guided by the principle of saddle point inflation. Considering
the contribution from $\tilde{Q}\tilde{Q}\tilde{Q}\tilde{L}$, $\tilde{Q}\tilde{u}\tilde{Q}\tilde{d}$,
$\tilde{Q}\tilde{u}\tilde{L}\tilde{e}$ and $\tilde{u}\tilde{u}\tilde{d}\tilde{e}$, the $n=4$ indexed SUSY flat directions the effective potential 
within the framework of MSSM can be written as \cite{Choudhury:2011jt}:
\bea V(\phi,\theta)=V_{0}+\frac{1}{2}m^{2}_{\phi}|\phi|^{2}+\frac{\lambda A}{4M_{p}}|\phi|^{4}\cos\left(4\theta+\theta_{A}\right)
+\frac{\lambda^{2}|\phi|^{6}}{M^{2}_{p}},\eea
where $V_{0}$ is the vacuum energy dominated term which mimics the role of cosmological constant, $m_{\phi}$ represents the soft SUSY breaking mass term, the inflaton $|\phi|$ is the radial coordinate of the complex scalar 
field $\Phi=|\phi|e^{i\theta}$ and the second term is the trilinear A-term which has a periodicity of $2\pi$ in 2D
along with an extra phase $\theta_{A}$. The radiative correction slightly affects the soft term and the position of the saddle point in potential valley.
One can tune the vacuum energy term $V_{0}\approx 0$. But for the generality I keep this term. Once this term is switched on in the effective potential,
the scale of the potential goes up to the GUT scale. This will change the value of tensor-to-sclar ratio upto $r(k_{\star})\sim 0.12$. But the 
other inflationary observables computed from the model is insensitive to the addition of vacuum energy term $V_{0}$. After minimizing with respect to the 
angular coordinate $\theta$ the effective potential takes the following form \cite{Choudhury:2011jt}:
\bea V(\phi)=V_{0}+\frac{1}{2}m^{2}_{\phi}|\phi|^{2}-\frac{\lambda A}{4M_{p}}|\phi|^{4}
+\frac{\lambda^{2}|\phi|^{6}}{M^{2}_{p}}\eea
using which the position for the saddle point is computed from the model as:
\be \phi_{0}=\sqrt{\frac{M_{p}}{4\lambda(D_{3}+3)}\left[A\left(1+\frac{D_{2}}{2}\right)\pm \sqrt{A^2\left(1+\frac{D_{2}}{2}\right)^2-8m^{2}_{\phi}\left(D_{1}+1\right)
\left(D_{3}+3\right)}\right]}\ee
where $D_{1}, D_{2}, D_{3}$ are the contribution from one loop radiative corrections appearing as:
\bea 
\lambda&=& \lambda_{0}\left[1+2D_{3}\ln\left(\frac{\phi_{0}}{\mu_{0}}\right)\right],\\
A&=& A_{0}\left[1+2D_{2}\ln\left(\frac{\phi_{0}}{\mu_{0}}\right)\right]\left[1+2D_{3}\ln\left(\frac{\phi_{0}}{\mu_{0}}\right)\right]^{-1},\\
m^{2}_{\phi}&=& m^{2}_{0}\left[1+2D_{1}\ln\left(\frac{\phi_{0}}{\mu_{0}}\right)\right].
\eea
Also the trilenear $A$ term and the radiative correction term $D_{3}$ satisfy the following constraints:
\bea 
A&=&\sqrt{2\left(D_{3}+3\right)G_{1}G_{2}G_{3}}m_{0},\\
D_{3}&=&\frac{M_{p}A_{0}}{4\lambda_{0}\phi^2_{0}\left(37+60\ln\left(\frac{\phi_{0}}{\mu_{0}}\right)\right)}\left[D_{2}
\left(13+12\ln\left(\frac{\phi_{0}}{\mu_{0}}\right)\right)-\frac{2m^{2}_{0}D_{1}M_{p}}{\lambda_{0}m_{0}\phi^{2}_{0}}+6\left(1-
\frac{20\lambda_{0}\phi^{2}_{0}}{M_{p}A_{0}}\right)\right].~~~~~~~~~
\eea
Here $G_{1},G_{2}$ and $G_{3}$ is given by:
\bea 
G_{1}&=&\left[\frac{(D_{1}+1)}{(D_{3}+3)}(15+11D_{3})-(3D_{1}+1)\right]^{2},\\
G_{2}&=&\left[(D_{1}+1)\left(\frac{7}{2}D_{3}+3\right)-(3D_{1}+1)\left(1+\frac{D_{2}}{2}\right)\right]^{-1},\\
G_{3}&=&\left[\frac{\left(1+\frac{D_{2}}{2}\right)}{(D_{3}+3)}\left(11D_{3}+15\right)-\left(3+\frac{7D_{3}}{2}\right)\right]^{-1}.
\eea
After applying the saddle point technique aroun $\phi_{0}$ the effective potential can be recast as \cite{Choudhury:2011jt}:
\bea V(\phi)=\Delta_{1}+\Delta_{2}\left(\phi-\phi_{0}\right)^{4},\eea 
where $\Delta_{1}$ and $\Delta_{2}$ are the Taylor expansion co-efficients defined as:
\bea 
\Delta_{1}&=&V(\phi_{0})\nonumber\\
&=&V_{0}+\frac{m^{3}_{0}M_{p}}{6\sqrt{6}\lambda}\left[3\left(1+\frac{D_{1}}{2}-\frac{D_{3}}{6}\right)\left[1+2D_{1}\ln\left(\frac{\phi_{0}}{\mu_{0}}
\right)
\right]\right.\\ &&\left.~~~~~~~~~~~~~~~~~~~~~~~~~~~~~~~~~~~~~~~~~~~~~~~~~-2\left(1+\frac{D_{1}}{2}-\frac{D_{3}}{6}\right)^2\left[1+2D_{2}\ln\left(\frac{\phi_{0}}{\mu_{0}}
\right)
\right]\right],\nonumber\\
\Delta_{2}&=&\frac{V^{''''}(\phi_{0})}{4!}\nonumber\\
&=&\frac{m^{2}_{0}}{24\sqrt{6}\phi^{2}_{0}}\left(1+\frac{D_{1}}{2}-\frac{D_{3}}{6}\right)^2\left[\left(\frac{360}{\sqrt{6}}-12\sqrt{6}+684D_{3}
-50\sqrt{6}D_{2}-\frac{2\sqrt{6}D_{1}}{\left(1+\frac{D_{1}}{2}-\frac{D_{3}}{6}\right)^2}\right)
\right.\nonumber\\ &&\left.~~~~~~~~~~~~~~~~~~~~~~~~~~~~~~~~~~~~~~~~~~~~~~~~~+2\left(\frac{360D_{3}}
{\sqrt{6}}-12\sqrt{6}D_{2}\right)\ln\left(\frac{\phi_{0}}{\mu_{0}}\right)\right].
\eea
In order to satisfy the Planck observational constraints on the amplitude of the power spectrum, $2.092\times 10^{-9}<P_S< 2.297\times 10^{-9}$
(within $2\sigma$), spectral tilt $0.958 < n_S<0.963$ (within $2\sigma$), sound speed $c_{S}=1$ (within $2\sigma$), and tensor-to-scalar ratio $r_\star \leq 0.12$, I obtain the following
constraints on our parameters, where successful inflation can occur via {\it saddle point}:
\begin{eqnarray}\label{P2-space}
\Delta_{1} &\sim &{\cal O}(10^{-36}-10^{-9})M^{4}_{p}\,, ~~~~~~~~~~~~~~~{\bf for}~~~~~~~~~~10^{-29}<r_{\star}<0.12\nonumber \\
\Delta_{2} & \sim &{\cal O}(10^{-13}-10^{-9})\,, ~~~~~~~~~~~~~~~~~~~{\bf for}~~~~~~~~~~10^{-29}<r_{\star}<0.12.
\end{eqnarray}
This analysis justifies that the soft SUSY breaking induced MSSM inflation is one of the examples of inflection point inflation where 
the present reconstruction technique holds good perfectly.

\section{Multipole scanning of CMB spectra via reconstructed effective potential}
\label{w7}

In this section I study the CMB TT, TE, EE, BB-angular power spectrum
~\footnote{In this work I have not consider possibility of other cross correlators i.e. TB, EB as 
there is no observational evidence of such contributions in the CMB map. Also till date there is no observational evidence for 
inflationary origin of BB angular power spectrum except from CMB lensing. However, for the completeness 
in this paper I show the theoretical BB angular power 
spectra from the reconstructed potential. }. 
The angular power spectra are defined as:
\be
C_\ell^{XY} \equiv \frac{1}{2\ell + 1} \sum^{l}_{m=-l} \langle a_{X,\ell m}^* a_{Y,\ell m} \rangle \, , \qquad X, Y = T, E, B\, .
\ee
Further substituting the inflationary input spectra $P(k) \equiv \{ P_S (k), P_T (k) \}$ and the angular power spectra of CMB temperature fluctuations and polarization
\be
\label{equ:CXY}
\displaystyle
C_\ell^{XY} = \frac{2}{\pi} \int k^2 dk \underbrace{P(k)}_{\rm Inflation}\,  \underbrace{\Delta_{X \ell}(k) \Delta_{Y \ell}(k)}_{\rm Anisotropies} \, ,
\ee
where
\be
\label{equ:source1}
\Delta_{X \ell}(k) = \int_0^{\eta_0} d\eta\, \underbrace{S_X(k, \eta)}_{\rm Sources}\, \underbrace{P_{X\ell}(k[\eta_0-\eta])}_{\rm Projection}\, .
\ee
\begin{figure}[t]
{\centerline{\includegraphics[width=18cm, height=11.5cm] {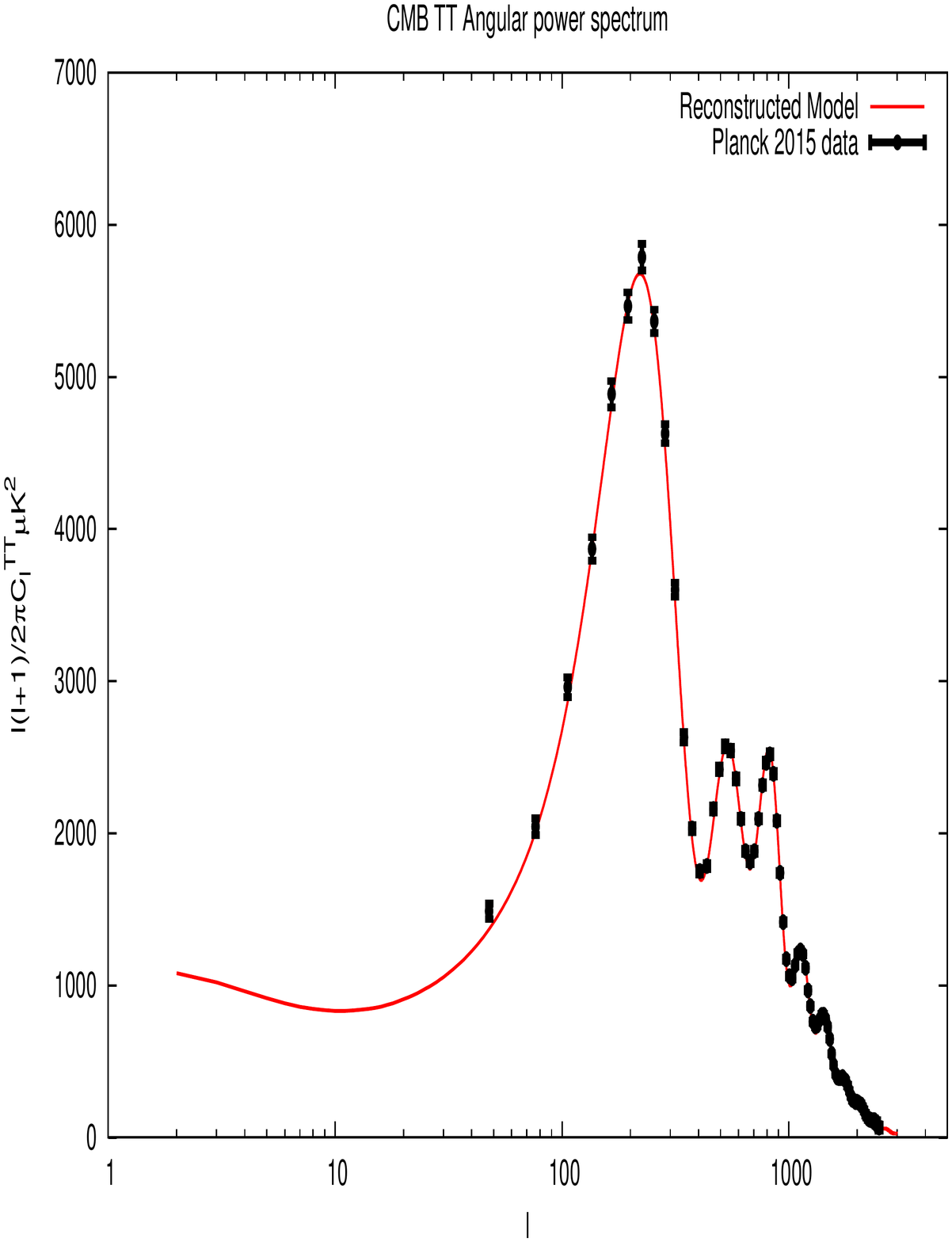}}}
\caption{We show the variation of CMB TT Angular power spectrum with respect to the multipole, $l$ for scalar modes with the choice of
 best fit reconstructed model parameters:
$V(\phi_0)\sim {\cal O}(10^{-9}~M^{4}_{p})$, $V^{'}(\phi_0)\sim {\cal O}(10^{-10}~M^{3}_{p})$, $V^{''}(\phi_0)\sim {\cal O}(10^{-11}~M^{2}_{p})$, $V^{'''}(\phi_0)\sim {\cal O}(10^{-10}~M_{p})$ and
 $V^{''''}(\phi_0)\sim {\cal O}(10^{-9})$ obtained from Planck 2015 data, which is consistent with the bound on field excursion value. From this analysis finally I get: $P_{S}(k_{\star})\sim 2.215\times 10^{-9}$, $n_{S}(k_{\star})\sim 0.962$, $r(k_{\star})\sim 0.2$,
 $\alpha_{S}(k_{\star})\sim-10^{-2}$
and $\kappa_{S}(k_{\star})\sim 5\times 10^{-3}$. } \label{figpol1}
\end{figure}

\begin{figure}[t]
{\centerline{\includegraphics[width=18cm, height=12.5cm] {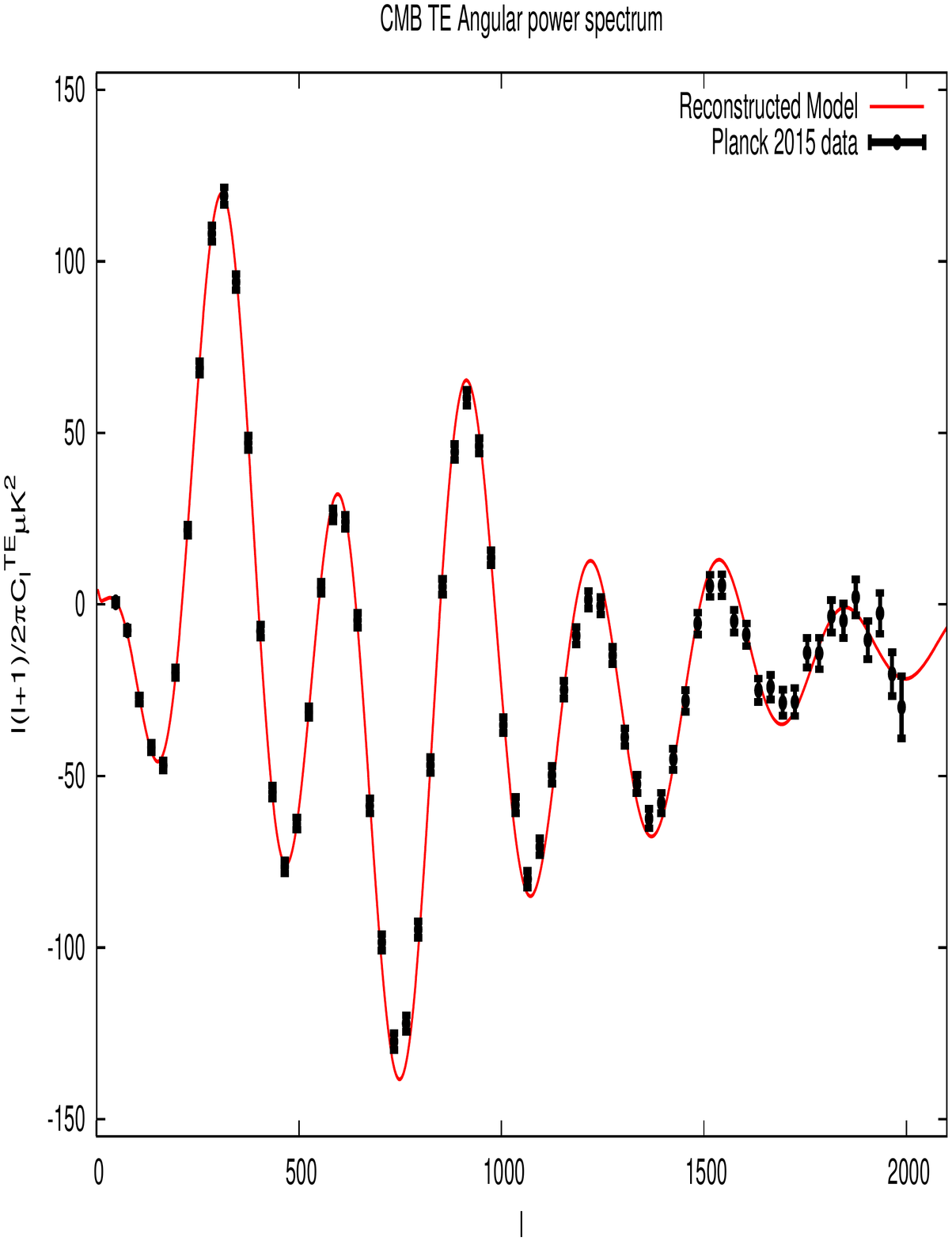}}}
\caption{We show the variation of CMB TE Angular power spectrum with respect to the multipole, $l$ for scalar modes with the choice of
 best fit reconstructed model parameters:
$V(\phi_0)\sim {\cal O}(10^{-9}~M^{4}_{p})$, $V^{'}(\phi_0)\sim {\cal O}(10^{-10}~M^{3}_{p})$, $V^{''}(\phi_0)\sim {\cal O}(10^{-11}~M^{2}_{p})$, $V^{'''}(\phi_0)\sim {\cal O}(10^{-10}~M_{p})$ and
 $V^{''''}(\phi_0)\sim {\cal O}(10^{-9})$ obtained from Planck 2015 data, which is consistent with the bound on field excursion value. From this analysis finally I get: $P_{S}(k_{\star})\sim 2.215\times 10^{-9}$, $n_{S}(k_{\star})\sim 0.962$, $r(k_{\star})\sim 0.2$,
 $\alpha_{S}(k_{\star})\sim-10^{-2}$
and $\kappa_{S}(k_{\star})\sim 5\times 10^{-3}$. } \label{figpol2}
\end{figure}

\begin{figure}[t]
{\centerline{\includegraphics[width=18cm, height=12.5cm] {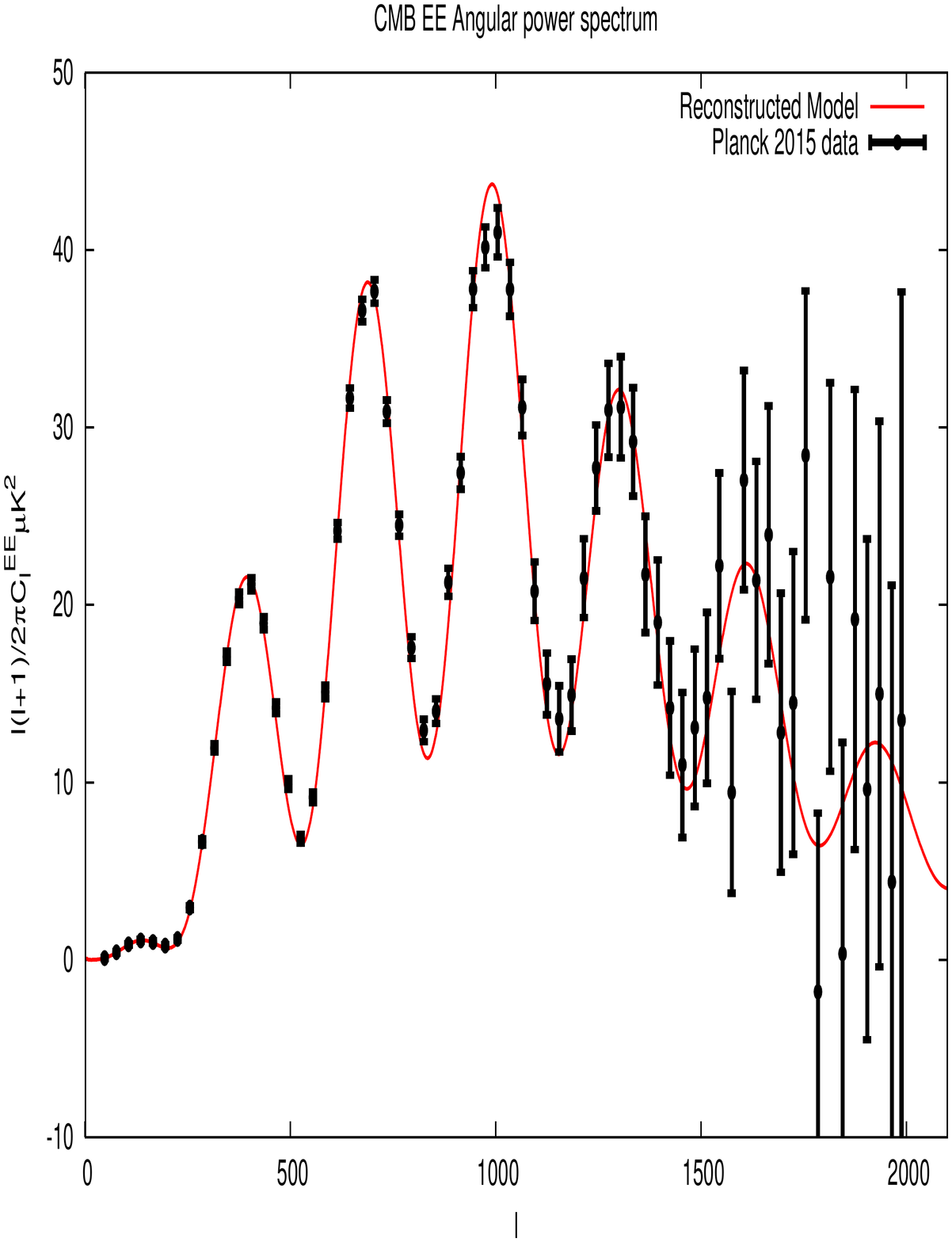}}}
\caption{We show the variation of CMB EE Angular power spectrum with respect to the multipole, $l$ for scalar modes with the choice of
 best fit reconstructed model parameters:
$V(\phi_0)\sim {\cal O}(10^{-9}~M^{4}_{p})$, $V^{'}(\phi_0)\sim {\cal O}(10^{-10}~M^{3}_{p})$, $V^{''}(\phi_0)\sim {\cal O}(10^{-11}~M^{2}_{p})$, $V^{'''}(\phi_0)\sim {\cal O}(10^{-10}~M_{p})$ and
 $V^{''''}(\phi_0)\sim {\cal O}(10^{-9})$ obtained from Planck 2015 data, which is consistent with the bound on field excursion value. From this analysis finally I get: $P_{S}(k_{\star})\sim 2.215\times 10^{-9}$, $n_{S}(k_{\star})\sim 0.962$, $r(k_{\star})\sim 0.2$,
 $\alpha_{S}(k_{\star})\sim-10^{-2}$
and $\kappa_{S}(k_{\star})\sim 5\times 10^{-3}$. } \label{figpol3}
\end{figure}


\begin{figure}[t]
\centering
\subfigure[$l(l+1)C^{BB}_{l}/2\pi$~vs~$l$~(tensor)]{
    \includegraphics[width=7.2cm, height=7.7cm] {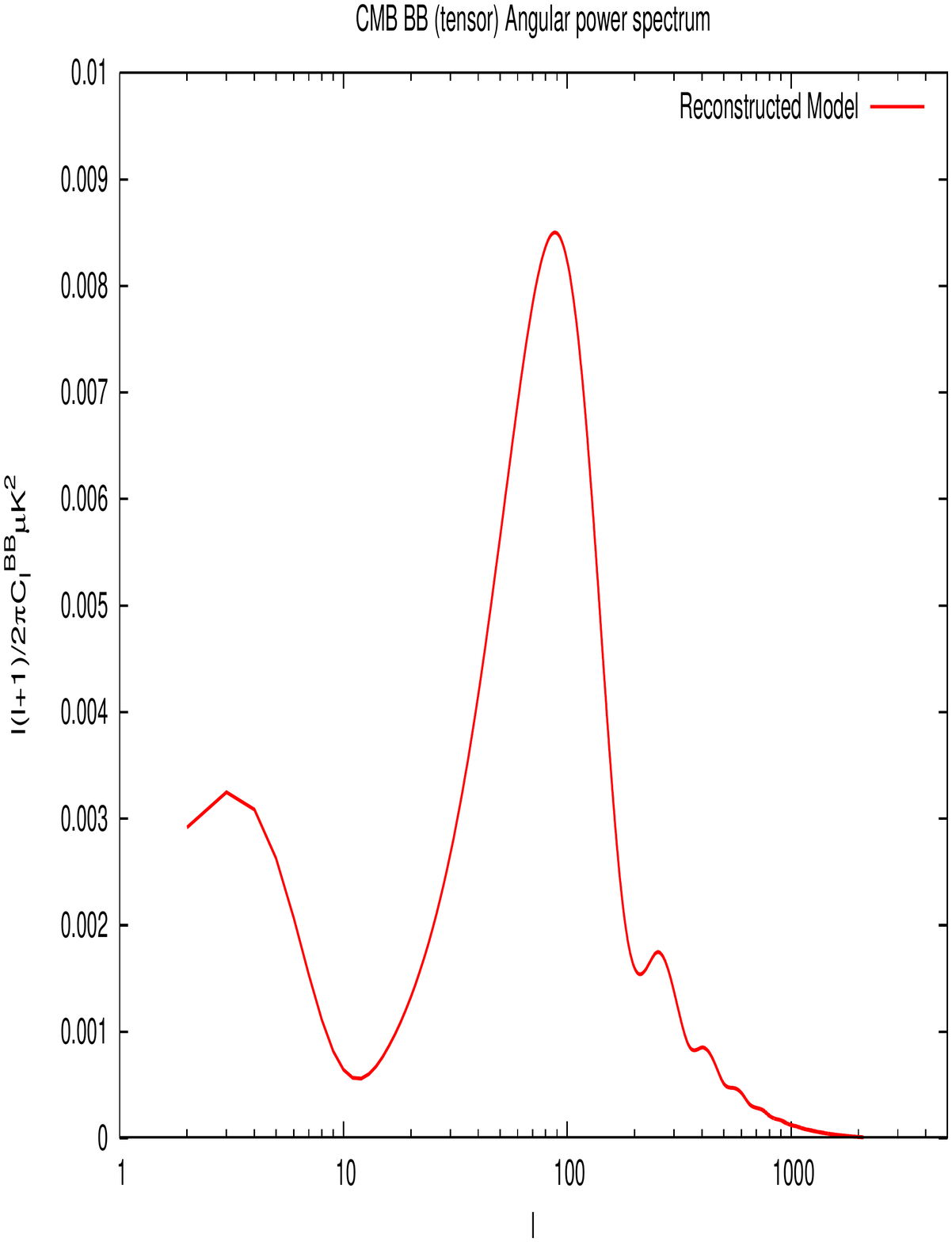}
    \label{figbb1}
}
\subfigure[$l(l+1)C^{TT}_{l}/2\pi$~vs~$l$~(tensor)]{
    \includegraphics[width=7.2cm, height=7.7cm] {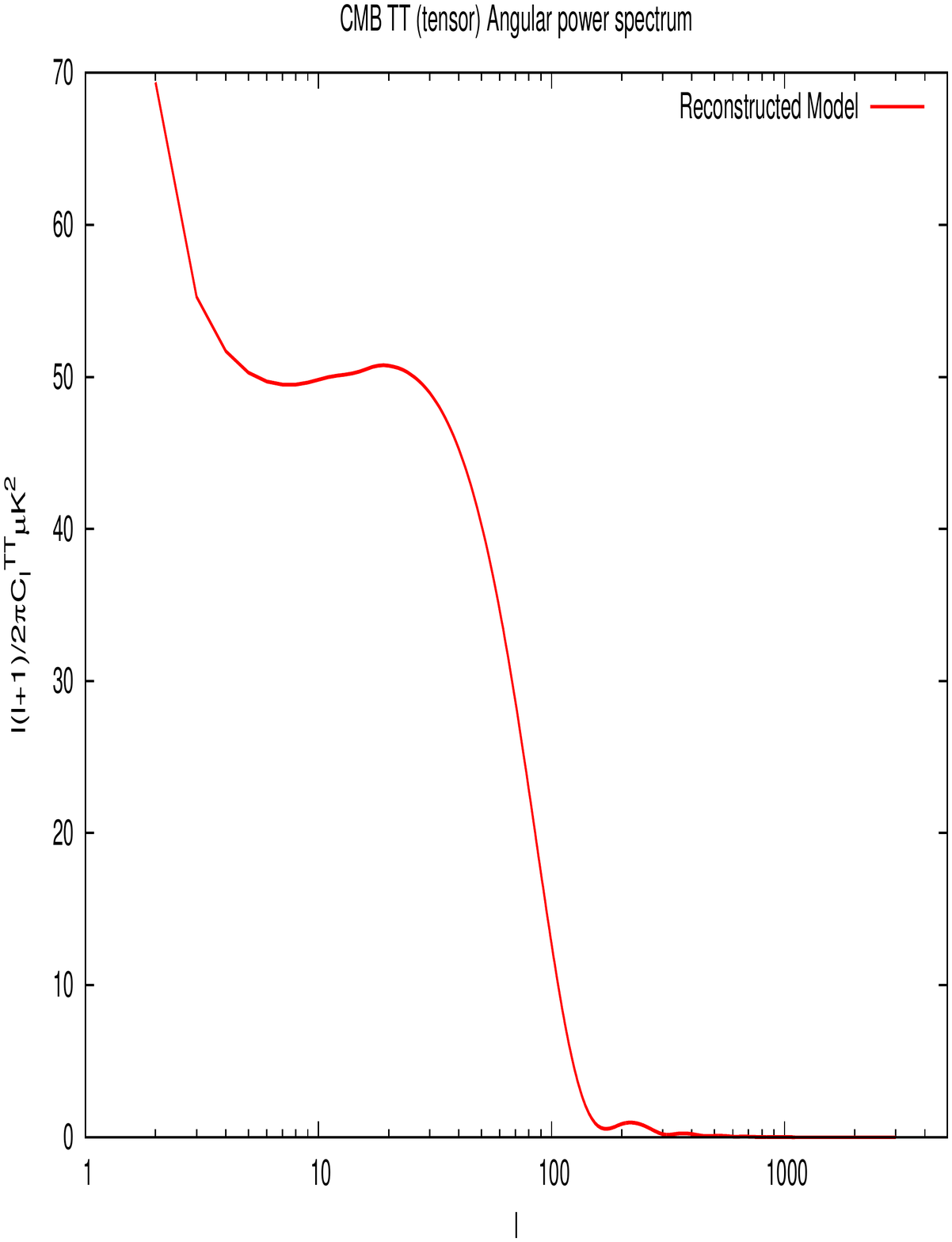}
    \label{figtt2}
}
\subfigure[$l(l+1)C^{TE}_{l}/2\pi$~vs~$l$~(tensor)]{
    \includegraphics[width=7.2cm, height=7.7cm] {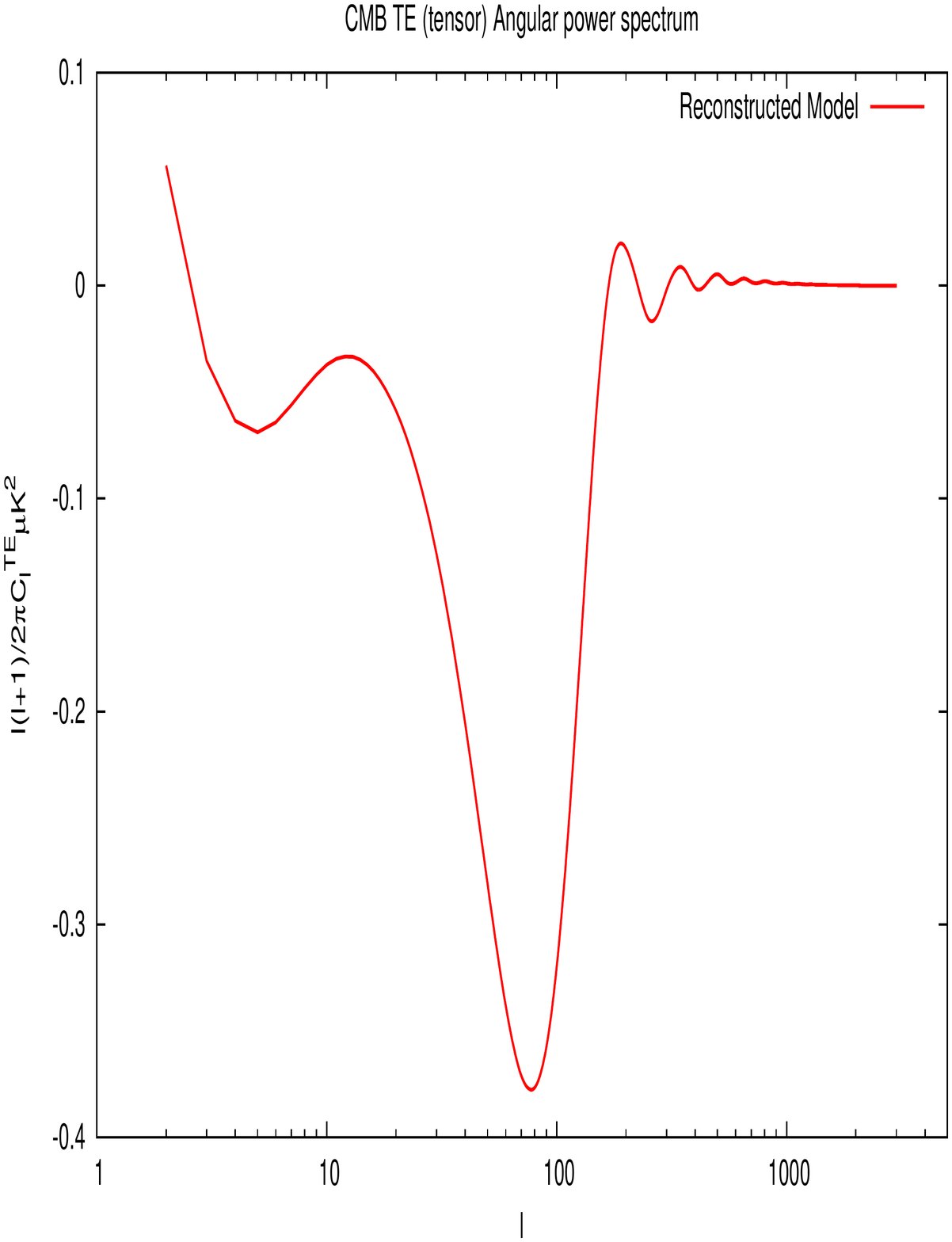}
    \label{figte3}
}
\subfigure[$l(l+1)C^{EE}_{l}/2\pi$~vs~$l$~(tensor)]{
    \includegraphics[width=7.2cm, height=7.7cm] {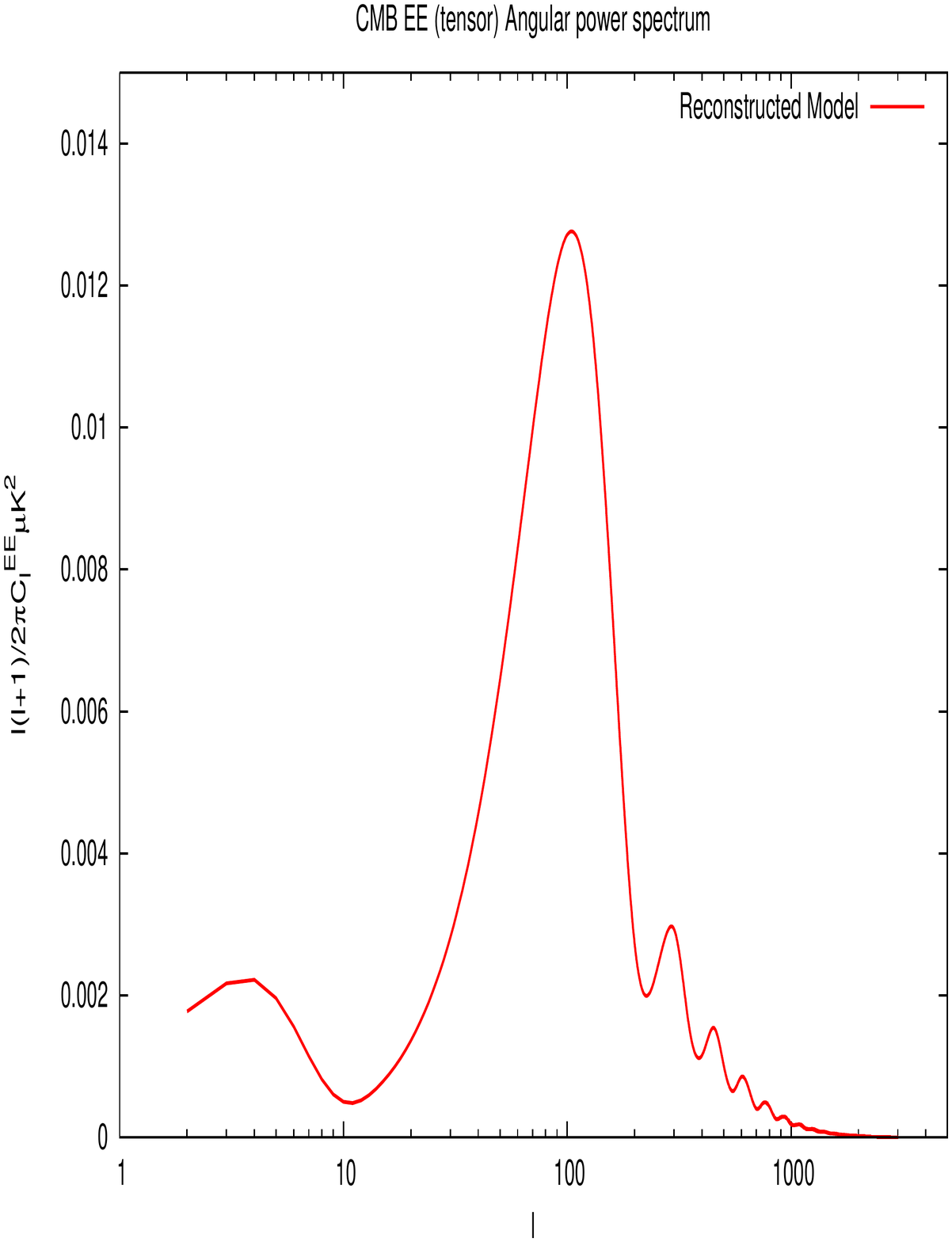}
    \label{figee4}
}
\caption[Optional caption for list of figures]{We show the variation of \subref{figbb1} CMB BB Angular power spectrum, \subref{figtt2} CMB TT Angular power spectrum, \subref{figte3} CMB TE Angular power spectrum
and \subref{figee4} CMB EE Angular power spectrum with respect to the multipole, $l$ for scalar modes with the choice of
 best fit reconstructed model parameters:
$V(\phi_0)\sim {\cal O}(10^{-9}~M^{4}_{p})$, $V^{'}(\phi_0)\sim {\cal O}(10^{-10}~M^{3}_{p})$, $V^{''}(\phi_0)\sim {\cal O}(10^{-11}~M^{2}_{p})$, $V^{'''}(\phi_0)\sim {\cal O}(10^{-10}~M_{p})$ and
 $V^{''''}(\phi_0)\sim {\cal O}(10^{-9})$ obtained from Planck 2015 data, which is consistent with the bound on the field excursion value.
 From this analysis finally I get: $P_{S}(k_{\star})\sim 2.215\times 10^{-9}$, $n_{S}(k_{\star})\sim 0.962$, $r(k_{\star})\sim 0.1$,
 $\alpha_{S}(k_{\star})\sim-10^{-2}$
and $\kappa_{S}(k_{\star})\sim 5\times 10^{-3}$.
}
\label{figbb}
\end{figure}


The integral (\ref{equ:CXY}) relates the inhomogeneities predicted by inflation, $P(k)$, to the anisotropies observed in the CMB, $C_\ell^{XY}$. The correlations between the different $X$ and $Y$ modes are related by the transfer functions $\Delta_{X\ell}(k)$ and $\Delta_{Y\ell}(k)$.
The transfer functions may be written as the line-of-sight integral Eq~(\ref{equ:source1}) which factorizes into physical source terms $S_X(k, \eta)$ and
 geometric projection factors $P_{X\ell}(k[\eta_0-\eta])$ through combinations of Bessel functions.

Further expressing Eq~(\ref{equ:CXY}) in terms of TT, TE, EE, BB correlation I get:
\begin{eqnarray}
 \underline{\bf For~ scalar}:~~~~C_\ell^{TT} &=& \frac{2}{\pi} \int k^2 dk ~{P_{S}(k)}\,  {\Delta_{T \ell}(k) \Delta_{T \ell}(k)} \, ,\\
C_\ell^{TE} &=& (4\pi)^2 \int k^2 dk ~{P_{S}(k)}\,  {\Delta_{T \ell}(k) \Delta_{E \ell}(k)} \, ,\\
C_\ell^{EE} &=& (4\pi)^2 \int k^2 dk ~{P_{S}(k)}\,  {\Delta_{E \ell}(k) \Delta_{T \ell}(k)} \, ,\\
\underline{\bf For~ tensor}:~~~~C_\ell^{BB} &=& (4\pi)^2 \int k^2 dk ~{P_{T}(k)}\,  {\Delta_{E \ell}(k) \Delta_{T \ell}(k)} \,\\
C_\ell^{TT} &=& \frac{2}{\pi} \int k^2 dk ~{P_{T}(k)}\,  {\Delta_{T \ell}(k) \Delta_{T \ell}(k)} \, ,\\
C_\ell^{TE} &=& (4\pi)^2 \int k^2 dk ~{P_{T}(k)}\,  {\Delta_{T \ell}(k) \Delta_{E \ell}(k)} \, ,\\
C_\ell^{EE} &=& (4\pi)^2 \int k^2 dk ~{P_{T}(k)}\,  {\Delta_{E \ell}(k) \Delta_{T \ell}(k)} \,.
\end{eqnarray}

where the inflationary power spectra $\{ P_S (k), P_T (k) \}$ are parametrized at any arbitrary momentum scale as:
\begin{eqnarray}
 P_{S}(k)&=& P_{S}(k_{\star})\left(\frac{k}{k_{\star}}\right)^{n_{S}-1+\frac{\alpha_{S}}{2}\ln\left(\frac{k}{k_{\star}}\right)
+\frac{\kappa_{S}}{6}\ln^{2}\left(\frac{k}{k_{\star}}\right)+....}\,,\\
 P_{T}(k)&=& r(k)P_{S}(k)=P_{T}(k_{\star})\left(\frac{k}{k_{\star}}\right)^{n_{T}+\frac{\alpha_{T}}{2}\ln\left(\frac{k}{k_{\star}}\right)
+\frac{\kappa_{T}}{6}\ln^{2}\left(\frac{k}{k_{\star}}\right)+....}\,.
\end{eqnarray}

It is important to note that the cosmological significance of the $E$ and $B$ decomposition of CMB polarization carries the following significant features:
\begin{itemize}
\item Scalar (density) perturbations create only $E$-modes.
\item Vector (vorticity) perturbations create mainly $B$-modes. However, the contributions of vectors decay
 with the expansion of the universe and are therefore sub-dominant at the epoch of recombination. For this reason I have neglected such sub-dominant 
effects from our analysis.
\item Tensor (gravitational wave) perturbations create both $E$-modes and $B$-modes.
\end{itemize}

To compute the momentum integrals numerically and to analyze the various features of CMB angular power spectra from the prescribed reconstruction
 algorithm I use a numerical code ``CAMB'' \cite{camb}. 
For the numerical analysis I use here the best fit reconstructed model parameters:
\bea V(\phi_0)&\sim& {\cal O}(10^{-9}~M^{4}_{p}),\\ 
V^{'}(\phi_0)&\sim& {\cal O}(10^{-10}~M^{3}_{p}),\\
V^{''}(\phi_0)&\sim& {\cal O}(10^{-11}~M^{2}_{p}),\\
V^{'''}(\phi_0)&\sim& {\cal O}(10^{-10}~M_{p}),\\ 
 V^{''''}(\phi_0)&\sim& {\cal O}(10^{-9})\eea which is compatible with Planck 2015 data. For this specific choice of multi parameter
 space finally the inflationary observables are estimated as:
 \bea P_{S}(k_{\star})&\sim& 2.215\times 10^{-9},\\
 n_{S}(k_{\star})&\sim& 0.962,\\
 r(k_{\star})&\sim& 0.1,\\
 \alpha_{S}(k_{\star})&\sim&-10^{-2},\\
\kappa_{S}(k_{\star})&\sim& 5\times 10^{-3}.\eea
Additionally I take $\Lambda$CDM background in which I fix: 
\bea \Omega_{b}h^2&=&0.022220,\\
\Omega_{c}h^{2}&=&0.119700,\\
\Omega_{\nu}h^{2}&=&0.000640,\\
\Omega_{K}&\approx& 0,\\
100~\theta&=&1.040832,\\
N_{eff}&=&3.046.\eea 
Further I put all these inputs to ``CAMB'' and modify the
 inbuilt parameterization of power spectrum for scalar and tensor 
modes accordingly. Finally from the analysis I get: 
\bea \Omega_{\Lambda}&=&0.685342,\\
\Omega_{m}(=1-\Omega_{K}-\Omega_{\Lambda})&=&0.314658,\\
z_{Reion}&=&9.959,\\
z_{eq}&=&3391.45,\\
t_{0}&=&13.814~{\rm Gyr},\\
z_{\star}&=&1089.87,\\
r_{S}(z_{\star})&=&144.64~{\rm Mpc},\\
D_{A}(z_{\star})&=&13.89228~{\rm Mpc},\\
z_{drag}&=&1059.40,\\
r_{S}(z_{drag})&=&147.35~{\rm Mpc},\\
k_{D}(z_{\star})&=&0.1407~{\rm Mpc}^{-1},\\
100~\theta_{D}&=&0.160737,\\
100~\theta_{eq}&=&0.814740,\\
\eta_{recomb}&=&280.75~{\rm Mpc},\\
\eta_{now}&=&14164.5~{\rm Mpc},\\
\sigma_{8}({\rm all~ matter})&=&0.8234.\eea 
After performing all the numerical computations via ``CAMB'' finally from our analysis I have
generated all the theoretical CMB angular power spectra from which I observed the following significant features:-
\begin{itemize}
\item In Fig.~(\ref{figpol1}),
at low $\ell$ region $(2<l<49)$ the contributions from the running ($\alpha_{S},\alpha_{T}$), and running of running ($\kappa_{S},\kappa_{T}$) are very small.
Their additional contribution to the CMB power spectrum for scalar and tensor modes becomes unity $({\cal O}(1))$ within low-$l$ region
 and the original
 power spectrum becomes unchanged. As a result the reconstructed model will be well fitted with the CMB TT spectrum at low-$l$ region within 
 high cosmic variance as observed by Planck except for a few outliers according to the Planck 2013 data release. But to update our analysis with the latest Planck 2015 data set I have not shown such high cosmic variance explicitly. In case of WMAP9 data for CMB TT spectrum the cosmic variance in the low-$l$ region is not very large compared to the 
Planck low-$l$ data. But our reconstructed model is also well fitted with WMAP9 low-$l$ data also which I have not shown explicitly in the plot to update the analysis using Planck 2015 data.
It is also important to mention that if I incorporate the uncertainties 
in the scanning multipole $l$ for the measurement of CMB TT spectrum, then also our prescribed analysis is pretty consistent with the 
Planck 2015 data. Further if I move
 towards high $\ell$ regime ($47<l<2500$) the contribution of running and running of running become stronger and
 this will enhance the power spectrum to a permissible value such that it will accurately fit high-$l$ 
data within very small cosmic variance as observed by Planck. In this way one can easily scan over all the multipoles starting from low-$l$ to high-$l$ using the
same momentum dependent parameterization of tensor-to-scalar ratio as prescribed in Eq~(\ref{con5}).

\item From Fig.~(\ref{figpol1}), we see that the Sachs-Wolfe plateau obtained
from our proposed reconstructed model is non flat, confirming the appearance of running, and running of the running in the spectrum 
observed for low $l$ region ($l<47$).
 For larger value of the multipole ($47<l<2500$),
CMB anisotropy spectrum is dominated by the Baryon
Acoustic Oscillations (BAO), giving rise to several ups and
downs in the CMB TT spectrum. In the low $l$ region due to the presence of very 
large cosmic variance there may be other pre-inflationary scenarios which might be able to describe the TT-power spectrum better. 
In my study I have considered only the possibility for which the behaviour of reconstructed model is analyzed for both 
low and high $l$ regions. 

\item From Fig.~(\ref{figpol1}), Fig.~(\ref{figpol2}), Fig.~(\ref{figpol3}), we observe that if I include the uncertainties in multipole $l$ as well in the observed CMB angular power spectra 
then the proposed reconstructed model is pretty consistent with the CMB TT, TE, EE for scalar mode
from Planck 2015 data. 

\item In Fig.~(\ref{figbb1}), Fig.~(\ref{figtt2}), Fig.~(\ref{figte3}) and Fig.~(\ref{figee4}) I have explicitly shown the 
theoretical CMB BB, TT, TE and EE angular power spectrum from tensor mode. Most importantly, if inflationary paradigm is responsible
for the nearly de-Sitter expansion of the early universe then the CMB BB spectra for tensor modes is one of the prime components trough which one can detect the 
contribution for primordial gravitational waves via tensor-to-scalar ratio. But till date only the contribution from the lesing B-modes are detected via South Pole Telescope \cite{Hanson:2013hsb} and Planck 2015
+BICEP2/Keck Array joint mission \cite{Ade:2015tva}.
But confirming the sole inflationary origin from the detection of the de-lensed version of the signal is not sufficient enough to draw any final conclusion~\footnote{In this respect one may consider the alternative frameworks of 
inflationary paradigm as well \cite{Choudhury:2015baa}.}. There are other possibilities as
well through which it is possible to generate 
CMB B-modes, those components are:
\begin{enumerate}
 \item Primordial Magnetic Field \cite{Bonvin:2014xia,Choudhury:2014hua,Choudhury:2015jaa},
 \item Gravitational Lensing \cite{Hanson:2013hsb,Zaldarriaga:1998ar},
 \item CP asymmetry in the lepton sector of particle theory \cite{Choudhury:2014hua,Choudhury:2015jaa} etc. 
\end{enumerate}
  \end{itemize}

 \section{Conclusion}
\label{w8}
In this paper I have obtained the following results:
\begin{itemize}
 \item  We derive the most generalized analytical expressions for the field excursion $|\Delta \phi|/M_p$ in terms of $r(k_\star)$, $V(\phi_\star)$, $H(k_\star)$ for a generic sub-Plackian model
of inflation by taking into account of higher order slow roll corrections and Buch-Davies initial condition. We derive the results for various parameterization of the primordial power spectrum 
by allowing:\begin{enumerate}
             \item scale invariant feature,
             \item further modification in presence of spectral tilt,
             \item the modification in presence of spectral tilt and running of the tilt,
             \item the effect of spectral tilt, running of the tilt and running of the running of tilt,
            \end{enumerate}
 through which it is possible to evade the well known {\it Lyth bound}. Using these relations one can easily rule out all classes of super Planckian or 
 trans-Planckian inflationary models available in inflationary literature.
 
 \item For the completeness, I also extend the idea by deriving the most generalized analytical expressions for the field excursion $|\Delta \phi|/M_p$ in terms of $r(k_\star)$, $V(\phi_\star)$, $H(k_\star)$ for a generic sub-Plackian model
 for the various limiting situations of non Buch-Davies vacuum. 
 
 \item In order to satisfy the observational constraints from Planck 2015 and Planck 2015+BICEP2/Keck Array joint constraints,
 one requires a non-monotonic evolution of the first slow-roll parameter, $\epsilon_V$, if one wants to build a model of inflation with a sub-Planckian field excursion and VEV, 
as pointed out in earlier Refs.~\cite{BenDayan:2009kv,Hotchkiss:2011gz,Choudhury:2013iaa,Choudhury:2014kma}.
 
 \item Hence in this paper I have reconstructed the inflationary potential 
around the VEV $\phi_0$ by computing the Taylor expansion co-efficients $V(\phi_0)$,$V^{\prime}(\phi_0)$,$V^{\prime\prime}(\phi_0)$, $V^{\prime\prime\prime}(\phi_0)$ and $V^{\prime\prime\prime\prime}(\phi_0)$
in terms of the Taylor expansion co-efficients $V(\phi_{\star})$,$V^{\prime}(\phi_{\star})$, $V^{\prime\prime}(\phi_{\star})$, $V^{\prime\prime\prime}(\phi_{\star})$ and $V^{\prime\prime\prime\prime}(\phi_{\star})$ at the CMB pivot scale $\phi_{\star}$
using matrix inversion technique using Planck 2015 and Planck 2015+BICEP2/Keck Array joint data.
 
 \item In order to satisfy the current observational constraints from Planck 2015 and Planck 2015+BICEP2/Keck Array joint data,  I have found that the upper bound of 
 the scale of inflation to be within: $\sqrt[4]{V_{\star}}\leq {\cal O}(1.856 - 1.926)\times 10^{16}~{\rm GeV}$, for the upper bound on field excursion
 of inflaton varying within: $\frac{\left |\Delta\phi\right|}{M_p}\,\leq {\cal O}(0.223-0.242)$. 
 
 \item We have also derived a new set of second order {\it consistency relationships} for any generic sub-Planckian model of inflation within slow-roll prescription. In particular,
 this can be treated as a discriminator to break the degenracy between various cosmological models of inflation. Most importantly, using these relations one can break also various class of models of inflation as well.
 
 \item Furthermore, if the
data could be refined to constrain tensor spectral tilt $n_T$, running $\alpha_T$ and running of the running 
$\kappa_T$, then this would play a very crucial role to break the degeneracy between various cosmological 
parameters esmated from different class of inflationary models and also rule out various models available in 
inflationary literature.
 
 \item Within our prescribed methodology I have also
 found that the choice of the field interval $\vartheta$ is sensitive to the number of e-foldings $\Delta N\sim {\cal O}(8-17)$, which is 
 necessarily required to solve the horizon problem associated with standard big bang cosmology and 
 to produce sufficient amount of inflation by constraining the Taylor expansion co-efficients of the generic potential 
 using Planck 2015 and Planck 2015+BICEP2/Keck Array joint data.
 
 \item Finally, I have checked the validity of our prescribed 
reconstruction technique by fitting the theoretical CMB angular power spectra from TT, TE, EE for scalar mode 
and BB, TT, TE, EE correlation for tensor mode with the observed Planck 2015 data within multipole scanning region 
$2<l<2500$, in which I cover the momentum scale within the window, $4.488\times 10^{-5}~{\rm Mpc}^{-1}<k<0.3~{\rm Mpc}^{-1}$.

\item Typically, within the framework of particle physics, the nature and shape of the inflationary potential
will not {\it just} be a single monomial. In principle the inflationary potential could
contain quadratic, cubic and quartic renormalizable interactions for effective field theory,
or even higher order non-renormalizable UV cut-off scale suppressed effective operators arising from integrating out the
heavy degrees of freedom originated from string theory inspired hidden sector. In this respect our
derived expressions and numerical results are important for reconstructing a particle physics motivated effective potential
for inflation in a successful fashion. One can extend the prescribed methodology to various string theory motivated 
effective potentials, where the VEV of the inflaton field and
field excursion both are sub-Planckian. Consequently, the reconstructed version of the inflationary 
potential allow only the {\it concave} feature, 
which is compatible with both particle physics and string theory framework.

The future prospects of the present work are appended below:
\begin{enumerate}
 \item First of all, using the non Bunch-Davies initial conditions one can repeat the prescribed methology presented in this paper.
       Then applying the observational constraints from Planck 2015 and Planck 2015+BICEP2/Keck Array data one can constrain the 
       exact version of the non Bunch-Davies initial condition applicable for effective field theory description of inflationary paradigm.
       
 \item Hence applying the observational constraints one can derive the first order and second order inflationary consistency relations
  from the exact version of the non Bunch-Davies initial condition.
 
 \item Further, one can compare the results obtained from non Bunch-Davies and Bunch-Davies vacuum and check which initial condition is more
 compatible with the observed data. This will in turn fix the ambiguity of choosing the proper initial condition for inflationary paradigm.
 Also by doing this analysis one can further constrain the nature and shape of the inflationary potential.
 
 \item Using various other parameterization of primordial power spectra one can also derive the most generic relationship 
 between tensor-to-scalar ratio, scale of inflation, Hubble parameter and the field excursion and also repeat the
rest of the reconstruction methodology presented in this paper. 
 \item As I have already mentioned at the end of last section that one of the prime source of generating CMB B-mode polarization 
 is primordial magnetic field, by applying the observational constraint 
 one can reconstruct an effective field theory of inflationary magnetogenesis for sub-Planckian 
 inflation.
 
 \item Finally using the results for non Bunch-Davies initial condition one can also check the compatibility of
 the theoretical CMB angular power spectra with the temperature anisotropy and polarization data obtained from Planck 2015.
\end{enumerate}

\end{itemize}


\section*{Acknowledgments}
SC would like to thank Department of Theoretical Physics, Tata Institute of Fundamental
Research, Mumbai for providing me Visiting (Post-Doctoral) Research Fellowship.  SC take this opportunity to thank sincerely
to Prof. Sandip P. Trivedi, Prof. Shiraz Minwalla, Prof. Soumitra SenGupta, Prof. Subhabrata Majumdar, Prof. Sudhakar Panda,
Prof. Sayan Kar and Dr. Supratik Pal
for their constant support and inspiration. SC take this opportunity to thank all the active members and the
regular participants of weekly student discussion meet ``COSMOMEET'' from Department of Theoretical Physics and Department of Astronomy and Astrophysics, Tata Institute of Fundamental
Research for their strong support. SC additionally Indian Association for the Cultivation of Science (IACS), Kolkata and Physics and Applied Mathematics Unit (PAMU), Indian Statistical Institute (ISI), Kolkata for
extending hospitality during the work. Additionally SC take this opportunity to thank the organizers of STRINGS, 2015, International Centre for Theoretical Science, Tata Institute of Fundamental Research (ICTS,TIFR)
and Indian Institute of Science (IISC)
for providing the local hospitality during the work and give a chance to discuss with Prof. Nima Arkani-Hamed on related issues, which finally helped us to improve the qualitative and quantitative discussion in the paper.
Last but not the least, I would all like to acknowledge our debt to the people of
India for their generous and steady support for research in natural sciences, especially for various areas in
theoretical high energy physics i.e. cosmology, string theory and particle physics.

\section*{Appendix}

\subsection*{\Large A. Slow-roll Integration from the reconstructed potential:}
The expressions for the slow-roll parameters ($\epsilon_{V},\eta_{V},\xi^{2}_{V},\sigma^{3}_{V}$) can be expressed as:
\be\begin{array}{llll}\label{sl1s}
    \displaystyle \epsilon_{V}(k)=\epsilon_{V}-\frac{\alpha_{T}}{2}\ln\left(\frac{k}{k_{\star}}\right)+\frac{\kappa_{T}}{4}\ln^{2}\left(\frac{k}{k_{\star}}\right)+.......,\\
 \displaystyle \eta_{V}(k)=\eta_{V}-\frac{\left(\alpha_{S}-3\alpha_{T}\right)}{2}\ln\left(\frac{k}{k_{\star}}\right)+\frac{1}{2}\left(
\kappa_{S}-3\kappa_{T}+\left\{n^{2}_{T}+\alpha_{T}\right\}\left[\frac{n_{S}-3n_{T}-1}{2}\right]\right.\\ \left.~~~~~~~~~~~~~~~~~~~~~~~~~~~~~~~~~~~
~~~~~~~~~~~~~~~~~~~~~~~~~~~~~~~\displaystyle +\frac{\xi^{2}_{V}}{2}\left\{1-n_{S}-3n_{T}\right\}
\right)\ln^{2}\left(\frac{k}{k_{\star}}\right)+.......,\\
 \displaystyle \xi^{2}_{V}(k)=\xi^{2}_{V}-\frac{1}{2}\left(\kappa_{S}-4\kappa_{T}+4n^{2}_{T}\left\{n_{S}-n_{T}-1
\right\}\right)\ln\left(\frac{k}{k_{\star}}\right)\\ ~~~~~~~
\displaystyle +\frac{1}{4}\left(\xi^{2}_{V}\left\{16n^{2}_{T}+9n_{T}\left[n_{S}-3n_{T}-1\right]+\left[n_{S}-3n_{T}-1\right]^{2}+2\xi^{2}_{V}
\right\}\right)\ln^{2}\left(\frac{k}{k_{\star}}\right)+.......,\\
 \displaystyle \sigma^{3}_{V}(k)=\sigma^{3}_{V}+\sigma^{3}_{V}\left(1-n_{S}\right)
\ln\left(\frac{k}{k_{\star}}\right)\\~~~~~~~~~~~~~~~\displaystyle +\frac{\sigma^{3}_{V}}{4}\left(30n^{2}_{T}+20n_{T}\left[n_{S}-3n_{T}-1\right]+\xi^{2}_{V}+2\left[n_{S}-3n_{T}-1\right]^{2}
\right)\ln^{2}\left(\frac{k}{k_{\star}}\right)+...... 
   \end{array}\ee

where at the pivot scale $k=k_{\star}$ the slow roll parameters are defined as follows:
    \begin{eqnarray}\label{para 2}
     \epsilon_{V}=\frac{M^{2}_{P}}{2}\left(\frac{V^{\prime}}{V}\right)^{2}, 
     \eta_{V}={M^{2}_{P}}\left(\frac{V^{\prime\prime}}{V}\right),
     \xi^{2}_{V}=M^{4}_{P}\left(\frac{V^{\prime}V^{\prime\prime\prime}}{V^{2}}\right),
     \sigma^{3}_{V}=M^{6}_{P}\left(\frac{V^{\prime 2}V^{\prime\prime\prime\prime}}{V^{3}}\right).
    \end{eqnarray}
Further in terms of the reconstructed potential I get:
\be\begin{array}{llll}\label{slpara}
\epsilon_{V}\approx\frac{M^{2}_{p}}{2V(\phi_0)^{2}}\left[V^{'}(\phi_0)+V^{''}(\phi_0)\left(\phi-\phi_0\right)+\frac{V^{'''}(\phi_0)}{2}\left(\phi-\phi_0\right)^{2}
+\frac{V^{''''}(\phi_0)}{6}\left(\phi-\phi_0\right)^3+\cdots\right]^{2},\\
\eta_{V}\approx\frac{M^{2}_{p}}{V(\phi_0)}\left[V^{''}(\phi_0)+V^{'''}(\phi_0)\left(\phi-\phi_0\right)+\frac{V^{''''}(\phi_0)}{2}\left(\phi-\phi_0\right)^2+\cdots\right],\\
\xi^{2}_{V}\approx\frac{M^{4}_{p}}{V(\phi_0)^{2}}\left[V^{'''}(\phi_0)V^{'}(\phi_0)+\left(V^{''''}(\phi_0) V^{'}(\phi_0)+V^{'''}(\phi_0)V^{''}(\phi_0)\right)
\left(\phi-\phi_0\right)\right. \\ \left. ~~~~~~~~~~~~~~~~~~~~~~~~~~~~~~~~~~~~~~~~+\left(V^{''''}(\phi_0)V^{''}(\phi_0)+\frac{V^{'''}(\phi_0)^{2}}{2}\right)
\left(\phi-\phi_0\right)^{2}+\cdots\right],\\
\sigma^{3}_{V}\approx\frac{V^{''''}(\phi_0) M^{6}_{p}}{V(\phi_0)^{3}}\left[V^{'}(\phi_0)+V^{''}(\phi_0)\left(\phi-\phi_0\right)+\frac{V^{'''}(\phi_0)}{2}\left(\phi-\phi_0\right)^{2}
+\frac{V^{''''}(\phi_0)}{6}\left(\phi-\phi_0\right)^3+\cdots\right]^{2}.
\end{array}\ee
Furthermore, the inflationary observables, i.e. the amplitude of scalar and tensor power spectrum $(P_{S},P_{T})$, spectral tilt $(n_{S},n_{T})$, and tensor-to-scalar 
ratio $(r_{\star})$ at the pivot scale $k_{\star}$ can be expressed as:  
\be\begin{array}{lll}\label{ps}
\displaystyle P_{S}(k_{\star}) 
=\left[1-(2{\cal C}_{E}+1)\epsilon_{V}+{\cal C}_{E}\eta_{V}\right]^{2}\frac{V}{24\pi^{2}M^{4}_{p}\epsilon_{V}}\\
\displaystyle ~~~~~~~~~\approx\frac{V(\phi_0)^{2}}{12\pi^{2}M^{6}_{p}V^{'}(\phi_0)^{2}}\left[V(\phi_0)+V^{'}(\phi_0)(\phi_{\star}-\phi_{0})+\frac{V^{''}(\phi_0)}{2}(\phi_{\star}-\phi_{0})^{2}
+\frac{V^{'''}(\phi_0)}{6}(\phi_{\star}-\phi_{0})^{3}\right.\\ \left. \displaystyle ~~~~~~~~~~~~~~~~~~~~~~~~~~~~~+\frac{V^{''''}(\phi_0)}{24}(\phi_{\star}-\phi_{0})^{4}+\cdots\,\right]
\left[1-(2{\cal C}_{E}+1)\frac{V^{'}(\phi_0)^{2}M^{2}_{p}}{2V(\phi_0)^{2}}+{\cal C}_{E}\frac{M^{2}_{p} V^{''}(\phi_0)}{V(\phi_0)}\right]^{2},\\
\end{array}\ee
\be\begin{array}{lll}\label{pT}
\displaystyle  P_{T}(k_{\star})=\left[1-({\cal C}_{E}+1)\epsilon_{V}\right]^{2}\frac{2V}{3\pi^{2}M^{4}_{p}}\\
\displaystyle ~~~~~~~~~\approx\frac{2}{3\pi^{2}M^{4}_{p}}\left[V(\phi_0)+V^{'}(\phi_0)(\phi_{\star}-\phi_{0})+\frac{V^{''}(\phi_0)}{2}(\phi_{\star}-\phi_{0})^{2}
+\frac{V^{'''}(\phi_0)}{6}(\phi_{\star}-\phi_{0})^{3}\right.\\ \left. \displaystyle ~~~~~~~~~~~~~~~~~~~~~~~~~~~~~+\frac{V^{''''}(\phi_0)}{24}(\phi_{\star}-\phi_{0})^{4}+\cdots\,\right]
\left[1-({\cal C}_{E}+1)\frac{V^{'}(\phi_0)^{2}M^{2}_{p}}{2V(\phi_0)^{2}}\right]^{2},\\
\end{array}\ee
\be\begin{array}{lllll}\label{para 21c} \displaystyle  n_{S}-1
 \approx (2\eta_{V}
-6\epsilon_{V})+\cdots\\
~~~~~~~~~\displaystyle =M^{2}_{p}\left[\left(\frac{2V^{''}(\phi_0)}{V(\phi_0)}-\frac{V^{'}(\phi_0)^{2}}{2V(\phi_0)^{2}}\right)
+\left(\frac{2V^{'''}(\phi_0)}{V(\phi_0)}-\frac{2V^{'}(\phi_0)V^{''}(\phi_0)}{V(\phi_0)^{2}}\right)\left(\phi_{\star}-\phi_0\right)+\cdots\right],\end{array}\ee
\be\begin{array}{lllll}\label{para 21cc} \displaystyle  n_{T}
 \approx -2\epsilon_{V}+\cdots=-\frac{M^{2}_{p}}{V(\phi_0)^{2}}\left[V^{'}(\phi_0)+V^{''}(\phi_0)\left(\phi_{\star}-\phi_0\right)+\cdots\right]^{2},\end{array}\ee
\be\begin{array}{lllll}\label{para 21e} \displaystyle  r(k_{\star})\approx16\epsilon_{V}\left[1+2{\cal C}_{E}(\epsilon_{V}-\eta_{V})\right]+\cdots\\
~~~~~~~\displaystyle=\frac{8M^{2}_{p}}{V(\phi_0)^{2}}\left[V^{'}(\phi_0)+V^{''}(\phi_0)\left(\phi_{\star}-\phi_0\right)+\cdots\right]^{2}
\left[1+2{\cal C}_{E}M^{2}_{p}\left(\frac{V^{'}(\phi_0)^{2}}{2V(\phi_0)^{2}}-\frac{V^{''}(\phi_0)}{V(\phi_0)}\right)\right]
\end{array}\ee

\be\begin{array}{lllll}\label{para 21f}  \displaystyle \alpha_{S}(k_{\star})=\left(\frac{dn_{S}}{d\ln k}\right)_{\star}
\approx\left(16\eta_{V}\epsilon_{V}-24\epsilon^{2}_{V}-2\xi^{2}_{V}\right)+\cdots,\\
~~~~~~~~~\displaystyle =\frac{8M^{4}_{p}}{V(\phi_0)^3}\left[V^{'}(\phi_0)+V^{''}(\phi_0)\left(\phi_{\star}-\phi_0\right)+\cdots\right]^{2}
\left[V^{''}(\phi_0)+V^{'''}(\phi_0)\left(\phi_{\star}-\phi_0\right)+\cdots\right]\\
~~~~~~~~~~~~~~~~~~~~~~~~~~~~~~~~~~~~~~~~~~~~~~~\displaystyle -\frac{6M^{4}_{p}}{V(\phi_0)^{4}}\left[V^{'}(\phi_0)+V^{''}(\phi_0)\left(\phi_{\star}-\phi_0\right)
\cdots\right]^{2}\\
\displaystyle ~~~~~~~~~~~~~~~~~~~-\frac{2M^{4}_{p}}{V(\phi_0)^{2}}\left[V^{'''}(\phi_0)V^{'}(\phi_0)+\left(V^{''''}(\phi_0) V^{'}(\phi_0)+V^{'''}(\phi_0)V^{''}(\phi_0)\right)
\left(\phi_{\star}-\phi_0\right)\right. \\ \left. ~~~~~~~~~~~~~~~~~~~~~~~~~~~~~~~~~~~~~~~~+\left(V^{''''}(\phi_0)V^{''}(\phi_0)+\frac{V^{'''}(\phi_0)^{2}}{2}\right)
\left(\phi_{\star}-\phi_0\right)^{2}+\cdots\right]
\end{array}\ee

\be\begin{array}{lllll}\label{para 21h} \displaystyle   \kappa_{S}(k_{\star})=\left(\frac{d^{2}n_{S}}{d\ln k^{2}}\right)_{\star}\approx192\epsilon^{2}_{V}\eta_{V}-192\epsilon^{3}_{V}+2\sigma^{3}_{V}
-24\epsilon_{V}\xi^{2}_{V}+2\eta_{V}\xi^{2}_{V}-32\eta^{2}_{V}\epsilon_{V}+\cdots\\
~~~~~~~~~\displaystyle =\frac{48M^{6}_{p}V^{'}(\phi_0)^4V^{''}(\phi_0)}{V(\phi_0)^5}-\frac{24M^{6}_{p}V^{'}(\phi_0)^6}{V(\phi_0)^6}
-\frac{12M^{6}_{p}V^{'}(\phi_0)^3V^{'''}(\phi_0)}{V(\phi_0)^4}\\
~~~~~~~~~~~~~~~~~~\displaystyle +\frac{2M^{6}_{p}V^{'}(\phi_0)V^{''}(\phi_0)V^{'''}(\phi_0)}{V(\phi_0)^3}
-\frac{16M^{6}_{p}V^{'}(\phi_0)^2V^{''}(\phi_0)^2}{V(\phi_0)^4}\\
~~~~~~~~~~~~~~~~~~~~~~~~~~~~~~~~~~~~~~~~~~~~~~~~~\displaystyle +\frac{2M^{6}_{p}V^{'}(\phi_0)^2V^{''''}(\phi_0)}{V(\phi_0)^3}+\cdots\end{array}\ee

\be\begin{array}{lllll}\label{para 21red} \displaystyle   n_{r}(k_{\star})=\left(\frac{dr}{d\ln k}\right)_{\star}\approx
32\epsilon_{V}\eta_{V}-64\epsilon^{2}_{V}+\cdots\\
~~~~~~~~~\displaystyle =16M^{4}_{p}\left(\frac{V^{'}(\phi_0)^2V^{"}(\phi_0)}{V(\phi_0)^3}-\frac{V^{'}(\phi_0)^4}{V(\phi_0)^4}\right)+\cdots\end{array}\ee
The crucial integrals of the first and second slow-roll parameters ($\epsilon_{V},~\eta_{V}$)
appearing in the right hand side of Eq.~(\ref{con4}),
 which can be written up to the leading order as: 
%
\be\begin{array}{llll}\label{hj1}
    \displaystyle \int^{{\phi}_{\star}}_{{\phi}_{e}}d {\phi}~\epsilon_{V}
\approx\frac{1}{2}\sum^{\infty}_{p=0} \frac{M^{p+2}_{p}~{\bf C}_{p}}{(p+1)}\left(\frac{\phi_{e}-\phi_{0}}{M_p}\right)^{p+1}
\left\{\left(1+\frac{\Delta\phi}{M_p}\left(\frac{\phi_{e}-\phi_{0}}{M_p}\right)^{-1}\right)^{p+1}-
1\right \}+\cdots\,
   \end{array}\ee
 \be\begin{array}{llll}\label{hj2}
    \displaystyle \int^{{\phi}_{\star}}_{{\phi}_{e}}d {\phi}~\eta_{V}
\approx \sum^{\infty}_{q=0} \frac{M^{q+2}_{p}~{\bf D}_{q}}{(q+1)}\left(\frac{\phi_{e}-\phi_{0}}{M_p}\right)^{q+1}
\left\{\left(1+\frac{\Delta\phi}{M_p}\left(\frac{\phi_{e}-\phi_{0}}{M_p}\right)^{-1}\right)^{q+1}-
1\right \}+\cdots\,
   \end{array}\ee
%
where I have used the  $(\phi-\phi_{0})<M_p$  (including at $\phi=\phi_{\star}$ and $\phi=\phi_{e}$) around $\phi_0$.
The leading order dimensionful Planck scale suppressed expansion co-efficients (${\bf C}_{p}$) and (${\bf D}_{q}$)
 are given in terms of the model parameters $(V(\phi_0),V^{'}(\phi_0),\cdots)$, which determine the
hight and shape of the potential in terms of the model parameters as:
%
\be\begin{array}{llll}\label{ceff}
    \displaystyle {\bf C}_{0}=\frac{V^{'}(\phi_0)^{2}}{V(\phi_0)^{2}},~~{\bf C}_{1}=\frac{2V^{''}(\phi_0)V^{'}
(\phi_0)}{V(\phi_0)^{2}}-\frac{2V^{'}(\phi_0)^{3}}{V(\phi_0)^{3}},\\
\displaystyle {\bf C}_{2}=\frac{V^{''}(\phi_0)^{2}}{V(\phi_0)^{2}}-\frac{5V^{'}(\phi_0)^{2}V^{''}(\phi_0)}{V(\phi_0)^{3}}+\frac{V^{'}(\phi_0)V^{'''}(\phi_0)}{V(\phi_0)^{2}},\\
\displaystyle {\bf C}_{3}=\frac{V^{'}(\phi_0)V^{''''}(\phi_0)}{3V(\phi_0)^{2}}-\frac{7V^{'}(\phi_0)^{2}V^{'''}(\phi_0)}{3V(\phi_0)^{3}}+\frac{V^{''}(\phi_0)V^{'''}(\phi_0)}{V(\phi_0)^{2}}
\\~~~~~~~~~~~~~~~~~~~~~~~~~~~~~~~~~~~~~~\displaystyle-\frac{4V^{'}(\phi_0)V^{''}(\phi_0)^{2}}{V(\phi_0)^{3}}+\frac{9V^{'}(\phi_0)^{3}V^{''}(\phi_0)}{V(\phi_0)^{4}},\\
\displaystyle \cdots \cdots \cdots \cdots \cdots \cdots\cdots \cdots \cdots \cdots \cdots \cdots\cdots \cdots \cdots\\
    \displaystyle {\bf D}_{0}=\frac{V^{''}(\phi_0)}{V(\phi_0)},~~{\bf D}_{1}=\frac{V^{'''}(\phi_0)}{V(\phi_0)}-\frac{V^{'}(\phi_0)V^{''}(\phi_0)}{V(\phi_0)^{2}},\\
\displaystyle {\bf D}_{2}=\frac{V^{''''}(\phi_0)}{2V(\phi_0)}-\frac{V^{'}(\phi_0)V^{'''}(\phi_0)}{V(\phi_0)^{2}}-\frac{V^{''}(\phi_0)}{V(\phi_0)^{2}}+\frac{V^{''}(\phi_0) V^{'}(\phi_0)^{2}}{V(\phi_0)^{3}},\\
\displaystyle {\bf D}_{3}=\frac{4V^{'}(\phi_0)\delta^{2}}{V(\phi_0)^{3}}-\frac{2V^{'}(\phi_0)^{3}\delta}{V(\phi_0)^{4}}+\frac{V^{'}(\phi_0)^{2}V^{'''}(\phi_0)}{V(\phi_0)^{3}}
-\frac{2V^{''}(\phi_0)V^{'''}(\phi_0)}{3V(\phi_0)^{2}}-\frac{V^{''''}(\phi_0) V^{'}(\phi_0)}{2V(\phi_0)^{2}},\\
\displaystyle \cdots \cdots \cdots \cdots \cdots \cdots\cdots \cdots \cdots \cdots \cdots \cdots\cdots \cdots \cdots
   \end{array}\ee
Here $V(\phi_0),V^{'}(\phi_0),V^{'''}(\phi_0),V^{''''}(\phi_0)\neq 0,V^{''}(\phi_0)=0$ and $V(\phi_0),,V^{'''}(\phi_0),V^{''''}(\phi_0)\neq 0,V^{'}(\phi_0),V^{''}(\phi_0)=0$ 
are two limiting situations which signifies the {\it inflection point} and {\it saddle point} inflationary setup. For details, see Ref.~\cite{Choudhury:2013iaa}.
\subsection*{\Large B. Momentum Integration from various parameterization of tensor-to-scalar ratio:}

In general, the tensor-to-scalar ratio can be parametrized at any arbitrary momentum scale as:
\be\begin{array}{lll}\label{rk1}
  \displaystyle r(k)\displaystyle =\left\{\begin{array}{ll}
                    \displaystyle  r(k_{\star}) &
 \mbox{ {\bf for \underline{Case I}}}  \\ 
         \displaystyle  r(k_{\star})\left(\frac{k}{k_{\star}}\right)^{n_{T}(k_{\star})-n_{S}(k_{\star})+1} & \mbox{ {\bf for \underline{Case II}}}\\ 
\displaystyle  r(k_{\star})\left(\frac{k}{k_{\star}}\right)^{n_{T}(k_{\star})-n_{S}(k_{\star})+1+\frac{\alpha_{T}(k_{\star})-\alpha_{S}(k_{\star})}{2!}\ln\left(\frac{k}{k_{\star}}\right)} & \mbox{ {\bf for \underline{Case III}}} \\
\displaystyle  r(k_{\star})\left(\frac{k}{k_{\star}}\right)^{n_{T}(k_{\star})-n_{S}(k_{\star})+1+\frac{\alpha_{T}(k_{\star})-\alpha_{S}(k_{\star})}{2!}\ln\left(\frac{k}{k_{\star}}\right)
+\frac{\kappa_{T}(k_{\star})-\kappa_{S}(k_{\star})}{3!}\ln^2\left(\frac{k}{k_{\star}}\right)} & \mbox{ {\bf for \underline{Case IV}}}.
          \end{array}
\right.
\end{array}\ee

where $k_{\star}$ be the pivot scale of momentum. Here these four possibilities are:-
\begin{itemize}
                        \item \underline{\bf Case I} stands for a situation where the spectrum is scale invariant. This is the similar situation 
                                            as considered in case of Lyth bound \cite{Lyth}. This possibility also surmounts to the Harrison \& Zeldovich
spectrum, which is completely ruled out by Planck+WMAP9 data within 5$\sigma$ C.L.

\item  \underline{\bf Case II} stands for a situation where spectrum follows power law feature 
through the spectral tilt $(n_{S},n_{T})$. This possibility is also tightly constrained by
the WMAP9 and Planck+WMAP9 data within 2$\sigma$ C.L. Recently in Ref.~\cite{Hazra:2014aea} the authors have explicitly shown that power law feature in the 
primordial power spectrum is ruled out at more than 3$\sigma$ C.L.,
\item \underline{\bf Case III} signifies a situation where the specrum shows deviation from power low in presence of running of the
 spectral tilt $(\alpha_{S},\alpha_{T})$ along with logarithmic correction in the momentum scale as appearing in the exponent.
 This possibility is favoured by WMAP9 data
and tightly constrained within 2$\sigma$ window by Planck+WMAP9 data,

\item \underline{\bf Case IV} characterizes
 a physical situation in which the spectrum is further modified compared to the \underline{\bf Case III},
by allowing running of the running of spectral tilt $(\kappa_{S},\kappa_{T})$ along with square of the momentum dependent logarithmic correction.
This case is satisfied by both WMAP9 and
Planck+WMAP9 data within 2$\sigma$ C.L. This is the only criteria which is
always satisfied by a general class of inflationary potentials. In this article, I have
only focused on this possibility, from which I have derived all the
constraint conditions for a generic model of sub-Planckian inflationary potentials.

                       \end{itemize}

Let us start with the computation of momentum integration where I investigate the possibility of four physical 
situations as mentioned in Eq~(\ref{rk1}) finally leading to: 

\be\begin{array}{lll}\label{rk2}
 \displaystyle\int^{k_{cmb}}_{k_{e}}d\ln k ~\sqrt{r(k)}\displaystyle \\ =\left\{\begin{array}{ll}
                    \displaystyle  \sqrt{r(k_\star)}\ln\left(\frac{k_{cmb}}{k_{e}}\right) &
 \mbox{\small {\bf for \underline{Case I}}}  \\  \\
         \displaystyle  \frac{2\sqrt{r(k_\star)}}{n_{T}(k_{\star})-n_{S}(k_{\star})+1}\left[\left(\frac{k_{cmb}}{k_{\star}}\right)^{\frac{n_{T}(k_{\star})-n_{S}(k_{\star})+1}{2}}-\left(\frac{k_{e}}{k_{\star}}\right)^{\frac{n_{T}(k_{\star})-n_{S}(k_{\star})+1}{2}}\right] & \mbox{\small {\bf for \underline{Case II}}} \\ \\
\displaystyle  \sqrt{r(k_\star)}e^{-\frac{(n_{T}(k_{\star})-n_{S}(k_{\star})+1)^2}{2(\alpha_{T}(k_{\star})-\alpha_{S}(k_{\star}))}}\sqrt{\frac{2\pi}{(\alpha_{T}(k_{\star})-\alpha_{S}(k_{\star}))}}
\\ \displaystyle \left[{\rm erfi}\left(\frac{n_{T}(k_{\star})-n_{S}(k_{\star})+1}{\sqrt{
2(\alpha_{T}(k_{\star})-\alpha_{S}(k_{\star}))}}+\sqrt{\frac{(\alpha_{T}(k_{\star})-\alpha_{S}(k_{\star}))}{8}}\ln\left(\frac{k_{cmb}}{k_{\star}}\right)\right)
\right.\\ \left. \displaystyle ~~~~~~
-{\rm erfi}\left(\frac{n_{T}(k_{\star})-n_{S}(k_{\star})+1}{\sqrt{2(\alpha_{T}(k_{\star})-\alpha_{S}(k_{\star}))}}+\sqrt{\frac{(\alpha_{T}(k_{\star})-\alpha_{S}(k_{\star}))}{8}}\ln\left(\frac{k_{e}}{k_{\star}}\right)\right)\right] & \mbox{\small {\bf for \underline{Case III}}}\\  \\ 
\displaystyle  \sqrt{r(k_{\star})}\left[\left(\frac{3}{2}-\frac{n_{T}(k_{\star})-n_{S}(k_{\star})}{2}+\frac{\alpha_{T}(k_{\star})-\alpha_{S}(k_{\star})}{8}\right.\right.\\ \left.\left. 
\displaystyle ~~~~~\displaystyle -\frac{\kappa_{T}(k_{\star})-\kappa_{S}(k_{\star})}{24}\right)\displaystyle \left\{\frac{k_{cmb}}{k_{\star}}-\frac{k_{e}}{k_{\star}}\right\}-
\left(\frac{1}{2}-\frac{n_{T}(k_{\star})-n_{S}(k_{\star})}{2}\right.\right.\\ \left.\left.\displaystyle \displaystyle+\frac{\alpha_{T}(k_{\star})-\alpha_{S}(k_{\star})}{8} 
 -\frac{\kappa_{T}(k_{\star})-\kappa_{S}(k_{\star})}{24}\right)\displaystyle \left\{\frac{k_{cmb}}{k_{\star}}\ln\left(\frac{k_{cmb}}{k_{\star}}\right)
-\frac{k_{e}}{k_{\star}}\ln\left(\frac{k_{e}}{k_{\star}}\right)\right\}\right.\\ \left.\displaystyle +\left(\frac{\kappa_{T}(k_{\star})
-\kappa_{S}(k_{\star})}{48}-\frac{\alpha_{T}(k_{\star})-\alpha_{S}(k_{\star})}{16}\right)\displaystyle\left\{\frac{k_{cmb}}{k_{\star}}\ln^2\left(\frac{k_{cmb}}{k_{\star}}\right)
-\frac{k_{e}}{k_{\star}}\ln^2\left(\frac{k_{e}}{k_{\star}}\right)\right\}\right.\\ \left.\displaystyle 
-\frac{\kappa_{T}(k_{\star})
-\kappa_{S}(k_{\star})}{144}\displaystyle\left\{\frac{k_{cmb}}{k_{\star}}\ln^3\left(\frac{k_{cmb}}{k_{\star}}\right)
-\frac{k_{e}}{k_{\star}}\ln^3\left(\frac{k_{e}}{k_{\star}}\right)\right\}\right] & \mbox{\small {\bf for \underline{Case IV}}}.
          \end{array}
\right.
\end{array}\ee

where in a realistic physical situation one assumes the pivot scale of momentum $k_{\star}\approx k_{cmb}$.
Now further substituting Eq~(\ref{intnk1}) on Eq~(\ref{rk2}) I get:
  
\be\begin{array}{lll}\label{rk3}
 \displaystyle\int^{k_{cmb}}_{k_{e}}d\ln k ~\sqrt{r(k)}\displaystyle \\=\left\{\begin{array}{ll}
                    \displaystyle  \sqrt{r(k_\star)}\Delta {N} &
 \mbox{\small {\bf for \underline{Case I}}}  \\  \\
         \displaystyle  \frac{2\sqrt{r(k_\star)}}{n_{T}(k_{\star})-n_{S}(k_{\star})+1}\left[1-e^{-\Delta {N}\left(\frac{n_{T}(k_{\star})-n_{S}(k_{\star})+1}{2}\right)}\right] & \mbox{\small {\bf for \underline{Case II}}}\\ \\
\displaystyle  \sqrt{r(k_\star)}e^{-\frac{(n_{T}(k_{\star})-n_{S}(k_{\star})+1)^2}{2(\alpha_{T}(k_{\star})-\alpha_{S}(k_{\star}))}}\sqrt{\frac{2\pi}{(\alpha_{T}(k_{\star})-\alpha_{S}(k_{\star}))}}
\\ \displaystyle \left[{\rm erfi}\left(\frac{n_{T}(k_{\star})-n_{S}(k_{\star})+1}{\sqrt{
2(\alpha_{T}(k_{\star})-\alpha_{S}(k_{\star}))}}\right)
\right.\\ \left. \displaystyle ~~~~~~
-{\rm erfi}\left(\frac{n_{T}(k_{\star})-n_{S}(k_{\star})+1}{\sqrt{2(\alpha_{T}(k_{\star})-\alpha_{S}(k_{\star}))}}-\sqrt{\frac{(\alpha_{T}(k_{\star})-\alpha_{S}(k_{\star}))}{8}}
\Delta {N}\right)\right] & \mbox{\small {\bf for \underline{Case III}}}\\ \\
\displaystyle  \sqrt{r(k_{\star})}\left[\left(\frac{3}{2}-\frac{n_{T}(k_{\star})-n_{S}(k_{\star})}{2}+\frac{\alpha_{T}(k_{\star})-\alpha_{S}(k_{\star})}{8}\right.\right.\\ \left.\left. 
\displaystyle ~~~~~\displaystyle -\frac{\kappa_{T}(k_{\star})-\kappa_{S}(k_{\star})}{24}\right)\displaystyle \left\{1-e^{-\Delta {N}}\right\}-
\left(\frac{1}{2}-\frac{n_{T}(k_{\star})-n_{S}(k_{\star})}{2}\right.\right.\\ \left.\left.\displaystyle \displaystyle+\frac{\alpha_{T}(k_{\star})-\alpha_{S}(k_{\star})}{8} 
 -\frac{\kappa_{T}(k_{\star})-\kappa_{S}(k_{\star})}{24}\right)\displaystyle 
\Delta {N}e^{-\Delta {N}}\right.\\ \left.\displaystyle -\left(\frac{\kappa_{T}(k_{\star})
-\kappa_{S}(k_{\star})}{48}-\frac{\alpha_{T}(k_{\star})-\alpha_{S}(k_{\star})}{16}\right)\displaystyle
(\Delta {N})^{2}e^{-\Delta {N}}\right.\\ \left.\displaystyle 
-\frac{\kappa_{T}(k_{\star})
-\kappa_{S}(k_{\star})}{144}\displaystyle
(\Delta {N})^{3}e^{-\Delta {N}}\right] & \mbox{\small {\bf for \underline{Case IV}}}.
          \end{array}
\right.
\end{array}\ee
\subsection*{\Large C. Taylor expansion co-efficients of inflationary potential:}
To write down all the Taylor expansion co-efficients in terms of the inflationary observables at the pivot scale $k_\star$ I start with the slow-roll parameters
$\epsilon_V, \eta_V, \xi^{2}_{V}, \sigma^{3}_{V}$ which can be expressed as:
\bea 
\epsilon_V(k_\star)&\approx&\frac{M^{2}_{p}}{2}\left(\frac{V^{'}(\phi_\star)}{V(\phi_\star)}\right)^{2}=\frac{r(k_\star)}{16}=\frac{V(\phi_\star)}{24\pi^{2}M^{4}_{p}P_{S}(k_\star)} ,\\
\eta_V (k_\star)&=& M^{2}_{p}\left(\frac{V^{''}(\phi_\star)}{V(\phi_\star)}\right)=\left(n_{S}(k_{\star})-1+\frac{3r(k_{\star})}{8}\right),\\
\xi^{2}_{V}(k_\star)&=& M^{4}_{p}\left(\frac{V^{'}(\phi_\star)V^{'''}(\phi_\star)}{(V(\phi_\star))^2}\right),\nonumber \\
&=&\frac{1}{2}\left[\eta_V(k_\star) r(k_\star)-\left(\frac{r(k_\star)}{8}\right)^{2}-\alpha_{S}(k_\star)\right],\\
\sigma^{3}_{V}(k_\star)&=&M^{6}_{p}\left(\frac{(V^{'}(\phi_\star))^2 V^{''''}(\phi_\star)}{(V(\phi_\star))^3}\right),\\
&=&\frac{1}{2}\left[\kappa_{S}(k_\star)-\left(\frac{r(k_\star)}{8}\right)^{2}\eta_{V}+6\left(\frac{r(k_\star)}{8}\right)^{3}
\right.\\ &&\left.~~~~~~~~~~~~ +\left(\frac{3r(k_\star)}{8}-\eta_{V}(k_\star)\right)\sqrt{\frac{r(k_\star)}{8}}X_\star+\eta^{2}_{V}r(k_\star)\right]~~~~~~
\eea
where
\bea 
V(\phi_\star)&=&V(k_\star)=V_\star=\frac{3}{2}P_{S}(k_\star)\pi^2 M^{4}_{p},\\
X_\star &=& \left[\eta_{V}\sqrt{2r(k_\star)}-\frac{1}{2}\left(\frac{r(k_\star)}{8}\right)^{\frac{3}{2}}-\alpha_{S}(k_\star)\sqrt{\frac{2}{r(k_\star)}}~~\right].
\eea
Finally we are left with the following Taylor expansion co-effiecients (derivatives of the potential) at the pivot scale in terms of the inflationary observables:
\bea\label{aq1d1}
\displaystyle V^{'}(\phi_{\star})&=& \frac{3}{2}P_{S}(k_{\star})r(k_{\star})\pi^{2}\sqrt{\frac{r(k_{\star})}{8}}M^{3}_{p},\\
\displaystyle V^{''}(\phi_{\star})&=& \frac{3}{4}P_{S}(k_{\star})r(k_{\star})\pi^{2}\left(n_{S}(k_{\star})-1+\frac{3r(k_{\star})}{8}\right)M^{2}_{p},\\
\displaystyle V^{'''}(\phi_{\star})&=& \frac{3}{2}P_{S}(k_{\star})r(k_{\star})\pi^{2}X_\star M_{p},\\
\displaystyle V^{''''}(\phi_{\star})&=& 12P_{S}(k_{\star})\pi^{2}\left\{\frac{\kappa_{S}(k_{\star})}{2}-
\frac{1}{2}\left(\frac{r(k_{\star})}{8}\right)^{2}\left(n_{S}(k_{\star})-1+\frac{3r(k_{\star})}{8}\right)\nonumber\right.\\ &&\left.\displaystyle 
~~~~~~~~~~~~~~~~~~~~~~~~+12\left(\frac{r(k_{\star})}{8}\right)^{3}+r(k_{\star})\left(n_{S}(k_{\star})-1+\frac{3r(k_{\star})}{8}\right)^{2}
\nonumber\right.\\ &&\left.\displaystyle 
~~~~~~~~~~~~~~~~~+\sqrt{\frac{r(k_{\star})}{8}}X_\star \left(n_{S}(k_{\star})-1+\frac{3r(k_{\star})}{8}\right)
-6X_\star\left(\frac{r(k_{\star})}{8}\right)^{\frac{3}{2}}\right\}.~~~~~~~~~
\eea

\subsection*{\Large D. Field excursion from various parameterization of tensor-to-scalar ratio:}

Using the result of Eq~(\ref{hj1}), Eq~(\ref{hj2}), Eq~(\ref{rk3}) I get the following expression for the field excursion in terms of the tensor-to-scalar ratio and other inflationary observables as: 
\be\begin{array}{lll}\label{rk4}
 \displaystyle\left|\frac{\Delta\phi}{M_{p}}\right|=\left\{\begin{array}{ll}
                    \displaystyle  \sqrt{\frac{r(k_\star)}{8}}\Delta {N} &
 \mbox{\small {\bf for \underline{Case I}}}  \\  \\ \\
         \displaystyle  \frac{2\sqrt{\frac{r(k_\star)}{8}}}{n_{T}(k_{\star})-n_{S}(k_{\star})+1}\left[1-e^{-\Delta {N}\left(\frac{n_{T}(k_{\star})-n_{S}(k_{\star})+1}{2}\right)}\right] & \mbox{\small {\bf for \underline{Case II}}}\\ \\ \\
\displaystyle  \sqrt{\frac{r(k_\star)}{8}}e^{-\frac{(n_{T}(k_{\star})-n_{S}(k_{\star})+1)^2}{2(\alpha_{T}(k_{\star})-\alpha_{S}(k_{\star}))}}\sqrt{\frac{2\pi}{(\alpha_{T}(k_{\star})-\alpha_{S}(k_{\star}))}}
\\ \displaystyle \left[{\rm erfi}\left(\frac{n_{T}(k_{\star})-n_{S}(k_{\star})+1}{\sqrt{
2(\alpha_{T}(k_{\star})-\alpha_{S}(k_{\star}))}}\right)
\right.\\ \left. \displaystyle ~~~~~~
-{\rm erfi}\left(\frac{n_{T}(k_{\star})-n_{S}(k_{\star})+1}{\sqrt{2(\alpha_{T}(k_{\star})-\alpha_{S}(k_{\star}))}}-\sqrt{\frac{(\alpha_{T}(k_{\star})-\alpha_{S}(k_{\star}))}{8}}
\Delta {N}\right)\right] & \mbox{\small {\bf for \underline{Case III}}}\\ \\ \\
\displaystyle  \sqrt{\frac{r(k_\star)}{8}}\left[\left(\frac{3}{2}-\frac{n_{T}(k_{\star})-n_{S}(k_{\star})}{2}+\frac{\alpha_{T}(k_{\star})-\alpha_{S}(k_{\star})}{8}\right.\right.\\ \left.\left. 
\displaystyle ~~~~~\displaystyle -\frac{\kappa_{T}(k_{\star})-\kappa_{S}(k_{\star})}{24}\right)\displaystyle \left\{1-e^{-\Delta {N}}\right\}-
\left(\frac{1}{2}-\frac{n_{T}(k_{\star})-n_{S}(k_{\star})}{2}\right.\right.\\ \left.\left.\displaystyle \displaystyle+\frac{\alpha_{T}(k_{\star})-\alpha_{S}(k_{\star})}{8} 
 -\frac{\kappa_{T}(k_{\star})-\kappa_{S}(k_{\star})}{24}\right)\displaystyle 
\Delta {N}e^{-\Delta {N}}\right.\\ \left.\displaystyle -\left(\frac{\kappa_{T}(k_{\star})
-\kappa_{S}(k_{\star})}{48}-\frac{\alpha_{T}(k_{\star})-\alpha_{S}(k_{\star})}{16}\right)\displaystyle
(\Delta {N})^{2}e^{-\Delta {N}}\right.\\ \left.\displaystyle 
-\frac{\kappa_{T}(k_{\star})
-\kappa_{S}(k_{\star})}{144}\displaystyle
(\Delta {N})^{3}e^{-\Delta {N}}\right] & \mbox{\small {\bf for \underline{Case IV}}}.
          \end{array}
\right.
\end{array}\ee
Substituting $\Delta N\sim {\cal O}(8-17)$ in Eq~(\ref{rk3}) for the previously mentioned four physical
situations and further using Eq~(\ref{con4}) the field excursion can be constrained 
 as:
\\
\underline{\bf Planck (2013)+WMAP-9+high~L:}
\be\begin{array}{lll}\label{rk9fin}
 \displaystyle\left|\frac{\Delta\phi}{M_p}\right|\displaystyle \leq\left\{\begin{array}{ll}
                    \displaystyle  {\cal O}(0.98-2.08) &
 \mbox{\small {\bf for \underline{Case I}}}  
\\ 
         \displaystyle  {\cal O}(0.87-1.88) & \mbox{\small {\bf for \underline{Case II}}}
\\ 
\displaystyle  {\cal O}(0.51-0.96) & \mbox{\small {\bf for \underline{Case III}}}
\\ 
\displaystyle  {\cal O}(0.239-0.241) & \mbox{\small {\bf for \underline{Case IV}}}.
          \end{array}
\right.
\end{array}\ee
\underline{\bf\textcolor{red}{ Planck (2014)+WMAP-9+high~L+BICEP2 (dust)}:}
\be\begin{array}{lll}\label{rk9fin}
 \displaystyle\left|\frac{\Delta\phi}{M_p}\right|\displaystyle =\left\{\begin{array}{ll}
                    \displaystyle  {\cal O}(2.32-3.12) &
 \mbox{\small {\bf for \underline{Case I}}}  
\\ 
         \displaystyle  {\cal O}(1.73-2.63) & \mbox{\small {\bf for \underline{Case II}}}
\\ 
\displaystyle  {\cal O}(0.62-0.94) & \mbox{\small {\bf for \underline{Case III}}}
\\ 
\displaystyle  {\cal O}(0.242-0.354) & \mbox{\small {\bf for \underline{Case IV}}}.
          \end{array}
\right.
\end{array}\ee
\underline{\bf Planck (2015)+WMAP-9+high~L(TT):}
\be\begin{array}{lll}\label{rk9fin}
 \displaystyle\left|\frac{\Delta\phi}{M_p}\right|\displaystyle \leq \left\{\begin{array}{ll}
                    \displaystyle  {\cal O}(0.94-2) &
 \mbox{\small {\bf for \underline{Case I}}}  
\\ 
         \displaystyle  {\cal O}(0.82-1.42) & \mbox{\small {\bf for \underline{Case II}}}
\\ 
\displaystyle  {\cal O}(0.56-0.96) & \mbox{\small {\bf for \underline{Case III}}}
\\ 
\displaystyle  {\cal O}(0.230-0.231) & \mbox{\small {\bf for \underline{Case IV}}}.
          \end{array}
\right.
\end{array}\ee
\underline{\bf Planck (2015)+BICEP2/Keck Array:}
\be\begin{array}{lll}\label{rk9fin}
 \displaystyle\left|\frac{\Delta\phi}{M_p}\right|\displaystyle \leq \left\{\begin{array}{ll}
                    \displaystyle  {\cal O}(0.98-2.08) &
 \mbox{\small {\bf for \underline{Case I}}}  
\\ 
         \displaystyle  {\cal O}(0.84-1.49) & \mbox{\small {\bf for \underline{Case II}}}
\\ 
\displaystyle  {\cal O}(0.51-0.97) & \mbox{\small {\bf for \underline{Case III}}}
\\ 
\displaystyle  {\cal O}(0.223-0.242) & \mbox{\small {\bf for \underline{Case IV}}}.
          \end{array}
\right.
\end{array}\ee
In this paper I have
only focused on the last possibility, from which I have derived all the
constraint conditions for a generic model of sub-Planckian inflationary potentials. Also the last possibility is important because within this it is possible to 
generate large value of tensor-to-scalar ratio along with field excursion $\Delta\phi\lesssim M_p$. This also validates the effective field theory prescription  
within the regime of inflationary paradigm. There are other possibilities as well through which one can address this crucial issue in the context of inflation. Those 
possibilities are:-
\begin{itemize}
 \item Multi-field inflationary prescription,
 \item Randall-Sundrum braneworld \cite{Choudhury:2014sua},
 \item Higher curvature gravity \cite{Choudhury:2014hna},
 \item Other ghost-free modifications in GR \cite{Biswas:2011ar},
 \item Inflation from torsion \cite{Choudhury:2014hja} etc. 
\end{itemize}



\begin{thebibliography}{}






\bibitem{Guth:1980zm} 
  A.~H.~Guth,
  \textcolor{blue}{\it``The Inflationary Universe: A Possible Solution to the Horizon and Flatness Problems,''}
   \textcolor{purple}{Phys.\ Rev.\ D {\bf 23}, 347 (1981)}.\\
A.~A.~Starobinsky,
 \textcolor{blue}{\it``A New Type of Isotropic Cosmological Models Without Singularity,''}
  \textcolor{purple}{Phys.\ Lett.\ B {\bf 91}, 99 (1980)}.
  
  \bibitem{Linde}
A.~D.~Linde,
  \textcolor{blue}{\it``A New Inflationary Universe Scenario: A Possible Solution of the Horizon, Flatness, Homogeneity, Isotropy and Primordial Monopole Problems,''}
  \textcolor{purple}{Phys.\ Lett.\ B {\bf 108}, 389 (1982)}.
  
  \bibitem{Albrecht}
   A.~Albrecht and P.~J.~Steinhardt,
  \textcolor{blue}{\it``Cosmology for Grand Unified Theories with Radiatively Induced Symmetry Breaking,''}
   \textcolor{purple}{Phys.\ Rev.\ Lett.\  {\bf 48}, 1220 (1982)}.
  
  \bibitem{Mukhanov:1981xt} 
  V.~F.~Mukhanov and G.~V.~Chibisov,
  \textcolor{blue}{\it``Quantum Fluctuation and Nonsingular Universe. (In Russian),''}
  \textcolor{purple}{JETP Lett.\  {\bf 33}, 532 (1981)
  [Pisma Zh.\ Eksp.\ Teor.\ Fiz.\  {\bf 33}, 549 (1981)]}.\\
   A.~A.~Starobinsky,
  \textcolor{blue}{\it``Relict Gravitation Radiation Spectrum and Initial State of the Universe. (In Russian),''}
  \textcolor{purple}{JETP Lett.\  {\bf 30}, 682 (1979)
  [Pisma Zh.\ Eksp.\ Teor.\ Fiz.\  {\bf 30}, 719 (1979)]}.
  
  
  \bibitem{Mukhanov:1990me} 
  V.~F.~Mukhanov, H.~A.~Feldman and R.~H.~Brandenberger,
  \textcolor{blue}{\it``Theory of cosmological perturbations. Part 1. Classical perturbations. Part 2. Quantum theory of perturbations. Part 3. Extensions,''}
  \textcolor{purple}{Phys.\ Rept.\  {\bf 215}, 203 (1992)}.


  

\bibitem{Hinshaw:2012aka} 
  G.~Hinshaw {\it et al.}  [WMAP Collaboration],
  \textcolor{blue}{\it``Nine-Year Wilkinson Microwave Anisotropy Probe (WMAP) Observations: Cosmological Parameter Results,''}
  \textcolor{purple}{Astrophys.\ J.\ Suppl.\  {\bf 208}, 19 (2013)
  [arXiv:1212.5226 [astro-ph.CO]]}.
  
  \bibitem{Planck-1}
  P.~A.~R.~Ade {\it et al.}  [Planck Collaboration],
  \textcolor{blue}{\it``Planck 2013 results. XVI. Cosmological parameters,''}
  \textcolor{purple}{arXiv:1303.5076 [astro-ph.CO]}.
  
  \bibitem{Planck-infl}
   P.~A.~R.~Ade {\it et al.}  [Planck Collaboration],
  \textcolor{blue}{\it``Planck 2013 results. XXII. Constraints on inflation,''}
  \textcolor{purple}{arXiv:1303.5082 [astro-ph.CO]}.
  
 \bibitem{Ade:2014xna}
  P.~A.~R.~Ade {\it et al.}  [BICEP2 Collaboration],
  \textcolor{blue}{\it``Detection of $B$-Mode Polarization at Degree Angular Scales by BICEP2,''}
  \textcolor{purple}{Phys.\ Rev.\ Lett.\  {\bf 112} (2014) 24,  241101
  [arXiv:1403.3985 [astro-ph.CO]]}.
  
  
 

\bibitem{Liu:2014mpa}
  H.~Liu, P.~Mertsch and S.~Sarkar,
  \textcolor{blue}{\it``Fingerprints of Galactic Loop I on the Cosmic Microwave Background,''}
  \textcolor{purple}{Astrophys.\ J.\  {\bf 789} (2014) L29
  [arXiv:1404.1899 [astro-ph.CO]]}.

\bibitem{Mortonson:2014bja}
  M.~J.~Mortonson and U.~Seljak,
  \textcolor{blue}{\it``A joint analysis of Planck and BICEP2 B modes including dust polarization uncertainty,''}
  \textcolor{purple}{arXiv:1405.5857 [astro-ph.CO]}.

\bibitem{Flauger:2014qra}
  R.~Flauger, J.~C.~Hill and D.~N.~Spergel,
  \textcolor{blue}{\it``Toward an Understanding of Foreground Emission in the BICEP2 Region,''}
  \textcolor{purple}{JCAP {\bf 1408} (2014) 039
  [arXiv:1405.7351 [astro-ph.CO]]}. 

\bibitem{Adam:2014gaa}
  R.~Adam {\it et al.}  [ Planck Collaboration],
  \textcolor{blue}{\it``Planck intermediate results. XXXII. The relative orientation between the magnetic field and structures traced by interstellar dust,''}
  \textcolor{purple}{arXiv:1409.6728 [astro-ph.GA]}.

 \bibitem{Ade:2015tva}
  P.~A.~R.~Ade {\it et al.}  [BICEP2 and Planck Collaborations],
  \textcolor{blue}{\it``A Joint Analysis of BICEP2/Keck Array and Planck Data,''}
  \textcolor{purple}{arXiv:1502.00612 [astro-ph.CO]}.
  
  
  
  \bibitem{Ade:2015lrj}
  P.~A.~R.~Ade {\it et al.}  [Planck Collaboration],
  \textcolor{blue}{\it``Planck 2015. XX. Constraints on inflation,''}
  \textcolor{purple}{arXiv:1502.02114 [astro-ph.CO]}.


  
  \bibitem{BD}
  N.~D.~Birrell and P.~C.~W.~Davies,
  \textcolor{blue}{\it``Quantum Fields in Curved Space,''}
  \textcolor{purple}{Cambridge Monogr. Math. Phys}..
  


\bibitem{Allahverdi:2006iq} 
  R.~Allahverdi, K.~Enqvist, J.~Garcia-Bellido and A.~Mazumdar,
  \textcolor{blue}{\it``Gauge invariant MSSM inflaton,''}
  \textcolor{purple}{Phys.\ Rev.\ Lett.\  {\bf 97}, 191304 (2006)
  [hep-ph/0605035]}.\\
   R.~Allahverdi, K.~Enqvist, J.~Garcia-Bellido, A.~Jokinen and A.~Mazumdar,
  \textcolor{blue}{\it``MSSM flat direction inflation: Slow roll, stability, fine tunning and reheating,''}
  \textcolor{purple}{JCAP {\bf 0706}, 019 (2007)
  [hep-ph/0610134]}.
  
\bibitem{assisted}
 A.~R.~Liddle, A.~Mazumdar and F.~E.~Schunck,
  \textcolor{blue}{\it``Assisted inflation,''}
  \textcolor{purple}{Phys.\ Rev.\ D {\bf 58}, 061301 (1998)
  [astro-ph/9804177]}.
  
  
  \bibitem{kanti}
   P.~Kanti and K.~A.~Olive,
  \textcolor{blue}{\it``Assisted chaotic inflation in higher dimensional theories,''}
   \textcolor{purple}{Phys.\ Lett.\ B {\bf 464}, 192 (1999)
  [hep-ph/9906331]}.

\bibitem{Choudhury:2014sua}
  S.~Choudhury,
  \textcolor{blue}{\it``Can Effective Field Theory of inflation generate large tensor-to-scalar ratio within Randall Sundrum single braneworld?,''}
  \textcolor{purple}{ Nucl.\ Phys.\ B {\bf 894} (2015) 29 [arXiv:1406.7618 [hep-th]]}.

\bibitem{BenDayan:2009kv}
  I.~Ben-Dayan and R.~Brustein,
  \textcolor{blue}{\it``Cosmic Microwave Background Observables of Small Field Models of Inflation,''}
  \textcolor{purple}{JCAP {\bf 1009} (2010) 007
  [arXiv:0907.2384 [astro-ph.CO]]}.
  
  \bibitem{Shafi}
 M.~U.~Rehman, Q.~Shafi and J.~R.~Wickman,
 \textcolor{blue}{\it``Observable Gravity Waves from Supersymmetric Hybrid Inflation II,''}
  \textcolor{purple}{Phys.\ Rev.\ D {\bf 83} (2011) 067304
  [arXiv:1012.0309 [astro-ph.CO]]}.\\
 Q.~Shafi and J.~R.~Wickman,
  \textcolor{blue}{\it``Observable Gravity Waves From Supersymmetric Hybrid Inflation,''}
  \textcolor{purple}{Phys.\ Lett.\ B {\bf 696} (2011) 438
  [arXiv:1009.5340 [hep-ph]]}.\\
  N.~Okada, M.~U.~Rehman and Q.~Shafi,
  \textcolor{blue}{\it``Non-Minimal B-L Inflation with Observable Gravity Waves,''}
  \textcolor{purple}{Phys.\ Lett.\ B {\bf 701} (2011) 520
  [arXiv:1102.4747 [hep-ph]]}.\\
  M.~Civiletti, M.~U.~Rehman, Q.~Shafi and J.~R.~Wickman,
   \textcolor{blue}{\it``Red Spectral Tilt and Observable Gravity Waves in Shifted Hybrid Inflation,''}
  \textcolor{purple}{Phys.\ Rev.\ D {\bf 84} (2011) 103505
  [arXiv:1104.4143 [astro-ph.CO]]}.

 
  
  
  \bibitem{Choudhury:2013iaa} 
  S.~Choudhury and A.~Mazumdar,
  \textcolor{blue}{\it``An accurate bound on tensor-to-scalar ratio and the scale of inflation,''}
  \textcolor{purple}{Nucl.\ Phys.\ B {\bf 882} (2014) 386
  [arXiv:1306.4496 [hep-ph]]}.
  
  \bibitem{Hotchkiss:2011gz}
  S.~Hotchkiss, A.~Mazumdar and S.~Nadathur,
  \textcolor{blue}{\it``Observable gravitational waves from inflation with small field excursions,''}
  \textcolor{purple}{JCAP {\bf 1202} (2012) 008
  [arXiv:1110.5389 [astro-ph.CO]]}.
  
   
\bibitem{Lyth:1996im}
 D.~H.~Lyth,
  \textcolor{blue}{\it``What would we learn by detecting a gravitational wave signal in the cosmic microwave background anisotropy?,''}
  \textcolor{purple}{Phys.\ Rev.\ Lett.\  {\bf 78} (1997) 1861
  [hep-ph/9606387]}.\\
  A.~Kehagias and A.~Riotto,
  \textcolor{blue}{\it``Remarks about the Tensor Mode Detection by the BICEP2 Collaboration and the Super-Planckian Excursions of the Inflaton Field,''}
  \textcolor{purple}{arXiv:1403.4811 [astro-ph.CO]}.
  
   
  \bibitem{Choudhury:2014kma}
  S.~Choudhury and A.~Mazumdar,
  \textcolor{blue}{\it``Reconstructing inflationary potential from BICEP2 and running of tensor modes,''}
  \textcolor{purple}{arXiv:1403.5549 [hep-th]}.

\bibitem{Choudhury:2014wsa}
  S.~Choudhury and A.~Mazumdar,
   \textcolor{blue}{\it``Sub-Planckian inflation \& large tensor to scalar ratio with $r\geq 0.1$,''}
  \textcolor{purple}{arXiv:1404.3398 [hep-th]}.
  
  

\bibitem{Khatri:2013xwa}
  R.~Khatri,
  \textcolor{blue}{\it``Mixing of blackbodies: Increasing our view of inflation to 17 e-folds with spectral distortions from Silk damping,''}
  \textcolor{purple}{arXiv:1302.5633 [astro-ph.CO]}.

 \bibitem{Clesse:2014pna}
  S.~Clesse, B.~Garbrecht and Y.~Zhu,
  \textcolor{blue}{\it``Testing Inflation and Curvaton Scenarios with CMB Distortions,''}
  \textcolor{purple}{arXiv:1402.2257 [astro-ph.CO]}.
  
  \bibitem{Easther:2006tv}
  R.~Easther and H.~Peiris,
  \textcolor{blue}{\it``Implications of a Running Spectral Index for Slow Roll Inflation,''}
  \textcolor{purple}{JCAP {\bf 0609} (2006) 010
  [astro-ph/0604214]}.

  \bibitem{Choudhury:2013jya}
 S.~Choudhury, A.~Mazumdar and S.~Pal,
  \textcolor{blue}{\it``Low \& High scale MSSM inflation, gravitational waves and constraints from Planck,''}
  \textcolor{purple}{JCAP {\bf 1307} (2013) 041
  [arXiv:1305.6398 [hep-ph]]}.
  
  
  \bibitem{Burgess:2005sb} 
  C.~P.~Burgess, et.al.
  \textcolor{blue}{\it``Multiple inflation, cosmic string networks and the string landscape,''}
  \textcolor{purple}{JHEP {\bf 0505}, 067 (2005)
  [hep-th/0501125]}.
  
  
\bibitem{Choudhury:2013woa}
  S.~Choudhury and A.~Mazumdar,
  \textcolor{blue}{\it``Primordial blackholes and gravitational waves for an inflection-point model of inflation,''}
  \textcolor{purple}{Phys.\ Lett.\ B {\bf 733} (2014) 270
  [arXiv:1307.5119 [astro-ph.CO]]}.
 

 \bibitem{Mantz:2009fw}
  A.~Mantz, S.~W.~Allen, D.~Rapetti and H.~Ebeling,
  \textcolor{blue}{\it``The Observed Growth of Massive Galaxy Clusters I: Statistical Methods and
  Cosmological Constraints,''}
  \textcolor{purple}{Mon.\ Not.\ Roy.\ Astron.\ Soc.\  {\bf 406} (2010) 1759
  [arXiv:0909.3098 [astro-ph.CO]]}.


\bibitem{Adams:1997de}
  J.~A.~Adams, G.~G.~Ross and S.~Sarkar,
  \textcolor{blue}{\it``Multiple inflation,''}
   \textcolor{purple}{Nucl.\ Phys.\  B {\bf 503} (1997) 405
  [arXiv:hep-ph/9704286]}.

\bibitem{Carr:2009jm}
  B.~J.~Carr, K.~Kohri, Y.~Sendouda and J.~Yokoyama,
  \textcolor{blue}{\it``New cosmological constraints on primordial black holes,''}
   \textcolor{purple}{Phys.\ Rev.\  D {\bf 81}, 104019 (2010)
  [arXiv:0912.5297 [astro-ph.CO]]}.

\bibitem{Alabidi:2009bk}
  L.~Alabidi and K.~Kohri,
  \textcolor{blue}{\it``Generating Primordial Black Holes Via Hilltop-Type Inflation Models,''}
  \textcolor{purple}{Phys.\ Rev.\  D {\bf 80}, 063511 (2009)
  [arXiv:0906.1398 [astro-ph.CO]]}.

\bibitem{Drees:2011hb}
  M.~Drees, E.~Erfani,
  \textcolor{blue}{\it``Running-Mass Inflation Model and Primordial Black Holes,''}
  \textcolor{purple}{JCAP {\bf 1104 } (2011)  005
  [arXiv:1102.2340 [hep-ph]]}.
  
  
  

 \bibitem{Lyth} 
   Y.~-Z.~Ma and Y.~Wang,
  \textcolor{blue}{\it``Local Reconstruction of the Inflationary Potential with BICEP2 data,''}
  \textcolor{purple}{arXiv:1403.4585 [astro-ph.CO]}.\\
  X.~Calmet and V.~Sanz,
  \textcolor{blue}{\it``Excursion into Quantum Gravity via Inflation,''}
  \textcolor{purple}{arXiv:1403.5100 [hep-ph]}.
  
  
  

  
  \bibitem{Enqvist:2010vd} 
  K.~Enqvist, A.~Mazumdar and P.~Stephens,
  \textcolor{blue}{\it``Inflection point inflation within supersymmetry,''}
  \textcolor{purple}{JCAP {\bf 1006}, 020 (2010)
  [arXiv:1004.3724 [hep-ph]]}.
  
\bibitem{Choudhury:2014sxa}
  S.~Choudhury, A.~Mazumdar and E.~Pukartas,
  \textcolor{blue}{\it``Constraining ${\cal N}=1$ supergravity inflationary framework with non-minimal Kähler operators,''}
   \textcolor{purple}{JHEP {\bf 1404} (2014) 077
  [arXiv:1402.1227 [hep-th]]}.
  
  \bibitem{Choudhury:2014uxa}
  S.~Choudhury,
  \textcolor{blue}{\it``Constraining N = 1 supergravity inflation with non-minimal Kaehler operators using $\delta$N formalism,''}
   \textcolor{purple}{JHEP {\bf 1404} (2014) 105
  [arXiv:1402.1251 [hep-th]]}.
  
  \bibitem{Choudhury:2011jt}
  S.~Choudhury and S.~Pal,
  \textcolor{blue}{\it``Fourth level MSSM inflation from new flat directions,''}
  \textcolor{purple}{JCAP {\bf 1204} (2012) 018
  [arXiv:1111.3441 [hep-ph]]}.
  
\bibitem{camb} CAMB, \textcolor{purple}{\it Online link: \textit{http://camb.info/}}.

\bibitem{Hanson:2013hsb}
  D.~Hanson {\it et al.}  [SPTpol Collaboration],
  \textcolor{blue}{\it``Detection of B-mode Polarization in the Cosmic Microwave Background with Data from the South Pole Telescope,''}
  \textcolor{purple}{Phys.\ Rev.\ Lett.\  {\bf 111} (2013) 14,  141301
  [arXiv:1307.5830 [astro-ph.CO]]}.
  
  \bibitem{Choudhury:2015baa}
  S.~Choudhury and S.~Banerjee,
  \textcolor{blue}{\it``Hysteresis in the Sky,''}
  \textcolor{purple}{arXiv:1506.02260 [hep-th]}.
  
  
\bibitem{Bonvin:2014xia}
  C.~Bonvin, R.~Durrer and R.~Maartens,
  \textcolor{blue}{\it``Can primordial magnetic fields be the origin of the BICEP2 data?,''}
  \textcolor{purple}{Phys.\ Rev.\ Lett.\  {\bf 112} (2014) 191303
  [arXiv:1403.6768 [astro-ph.CO]]}.

\bibitem{Choudhury:2014hua}
  S.~Choudhury,
  \textcolor{blue}{\it``Inflamagnetogenesis redux: Unzipping sub-Planckian inflation via various cosmoparticle probes,''}
  \textcolor{purple}{Phys.\ Lett.\ B {\bf 735} (2014) 138
  [arXiv:1403.0676 [hep-th]]}.
  
  \bibitem{Choudhury:2015jaa}
  S.~Choudhury,
  \textcolor{blue}{\it``Braneflamagnetogenesis from Cosmoparticle Physics after Planck,''}
  \textcolor{purple}{arXiv:1504.08206 [astro-ph.CO]}.


\bibitem{Zaldarriaga:1998ar}
  M.~Zaldarriaga and U.~Seljak,
  \textcolor{blue}{\it``Gravitational lensing effect on cosmic microwave background polarization,''}
  \textcolor{purple}{Phys.\ Rev.\ D {\bf 58} (1998) 023003
  [astro-ph/9803150]}.

\bibitem{Hazra:2014aea}
  D.~K.~Hazra, A.~Shafieloo, G.~F.~Smoot and A.~A.~Starobinsky,
  \textcolor{blue}{\it``Ruling out the power-law form of the scalar primordial spectrum,''}
  \textcolor{purple}{JCAP {\bf 1406} (2014) 061
  [arXiv:1403.7786 [astro-ph.CO]]}.
 

\bibitem{Choudhury:2014hna}
S.~Choudhury, J.~Mitra and S.~SenGupta,
  \textcolor{blue}{\it``Fermion localization and flavour hierarchy in higher curvature spacetime,''}
  \textcolor{purple}{arXiv:1503.07287 [hep-th]}.\\
S.~Choudhury, J.~Mitra and S.~SenGupta,
  \textcolor{blue}{\it``Modulus stabilization in higher curvature dilaton gravity,''}
  \textcolor{purple}{JHEP {\bf 1408} (2014) 004
  [arXiv:1405.6826 [hep-th]]};\\
S.~Choudhury and S.~SenGupta,
  \textcolor{blue}{\it``A step toward exploring the features of Gravidilaton sector in Randall-Sundrum scenario via lightest Kaluza-Klein graviton mass,''}
  \textcolor{purple}{Eur.\ Phys.\ J.\ C {\bf 74} (2014) 11,  3159
  [arXiv:1311.0730 [hep-ph]]};\\
S.~Choudhury, S.~Sadhukhan and S.~SenGupta,
  \textcolor{blue}{\it``Collider constraints on Gauss-Bonnet coupling in warped geometry model,''}
  \textcolor{purple}{arXiv:1308.1477 [hep-ph]};\\
 S.~Choudhury and S.~SenGupta,
   \textcolor{blue}{\it``Thermodynamics of Charged Kalb Ramond AdS black hole in presence of Gauss-Bonnet coupling,''}
  \textcolor{purple}{arXiv:1306.0492 [hep-th]}. \\
 S.~Choudhury and S.~Sengupta,
  \textcolor{blue}{\it``Features of warped geometry in presence of Gauss-Bonnet coupling,''}
  \textcolor{purple}{JHEP {\bf 1302} (2013) 136
  [arXiv:1301.0918 [hep-th]]}.\\
  S.~Choudhury and S.~Pal,
  \textcolor{blue}{\it``Primordial non-Gaussian features from DBI Galileon inflation,''}
  \textcolor{purple}{Eur.\ Phys.\ J.\ C {\bf 75} (2015) 6,  241
  [arXiv:1210.4478 [hep-th]]}.\\
  S.~Choudhury and S.~Pal,
  \textcolor{blue}{\it``DBI Galileon inflation in background SUGRA,''}
   \textcolor{purple}{Nucl.\ Phys.\ B {\bf 874} (2013) 85
  [arXiv:1208.4433 [hep-th]]}.
  
\bibitem{Biswas:2011ar}
  T.~Biswas, E.~Gerwick, T.~Koivisto and A.~Mazumdar,
  \textcolor{blue}{\it``Towards singularity and ghost free theories of gravity,''}
  \textcolor{purple}{Phys.\ Rev.\ Lett.\  {\bf 108} (2012) 031101
  [arXiv:1110.5249 [gr-qc]]}.

\bibitem{Choudhury:2014hja}
  S.~Choudhury, B.~K.~Pal, B.~Basu and P.~Bandyopadhyay,
  \textcolor{blue}{\it``Measuring CP violation within Effective Field Theory of inflation from CMB,''}
 \textcolor{purple}{ arXiv:1409.6036 [hep-th]}.

\end{thebibliography}
\end{document}